\documentclass[sort&compress,AMA,STIX1COL]{WileyNJD-v2}

\articletype{Research article: accepted version}%

\usepackage{xcolor}
\hypersetup{colorlinks=true}

\usepackage{amsmath,amsfonts}
\usepackage{hyphenat}
\usepackage{booktabs}

\usepackage[absolute]{textpos}
\TPGrid[0pt,0pt]{20}{30}
\TPMargin{2.0mm}
\TPoptions{showboxes = true }
\textblockcolour{cyan!50!lightgray}
\textblockrulecolour{blue!90!violet}

\newcommand{\pder}[2]{\frac{\partial #1}{\partial #2}}
\newcommand{\oder}[2]{\frac{\mathrm{d} #1}{\mathrm{d} #2}}
\newcommand{\tmdder}[3]{\left(\frac{\partial #1}{\partial #2}\right)_{\!\!#3}}
\newcommand{\reff}{_\mathrm{r}}

\newcommand{\arho}{\alpha \rho}
\newcommand{\am}{\alpha m}
\newcommand{\aet}{\alpha E}

\newcommand{\ui}{u_\mathrm{I}}
\newcommand{\ppi}{P_\mathrm{I}}
\newcommand{\cci}[1]{c_{\mathrm{I},#1}^2}

\newcommand{\ph}{_\sigma}
\newcommand{\phSymbol}{\sigma}

\newcommand{\aph}{\alpha\ph}
\newcommand{\rhoph}{\rho\ph}
\newcommand{\arhoph}{\alpha\rho\ph}
\newcommand{\mph}{m\ph}
\newcommand{\amph}{\alpha m\ph}
\newcommand{\uph}{u\ph}
\newcommand{\Pph}{P\ph}
\newcommand{\eph}{e\ph}

\newcommand{\td}{\tilde{t}}
\newcommand{\xd}{\widetilde{x}}
\newcommand{\Ld}{\widetilde{L}}
\newcommand{\ud}{\widetilde{u}}
\newcommand{\Pd}{\widetilde{P}}
\newcommand{\rhod}{\widetilde{\rho}}
\newcommand{\ed}{\widetilde{e}}
\newcommand{\mud}{\widetilde{\mu}}
\newcommand{\lambdad}{\widetilde{\lambda}}

\newcommand{\uid}{\widetilde{u}_\mathrm{I}}
\newcommand{\ppid}{\widetilde{P}_\mathrm{I}}

\newcommand{\arhophd}{\alpha\widetilde{\rho}\ph}
\newcommand{\amphd}{\alpha\widetilde{m}\ph}
\newcommand{\uphd}{\widetilde{u}\ph}
\newcommand{\Pphd}{\widetilde{P}\ph}
\newcommand{\ccphd}{\widetilde{c}\ph^2}
\newcommand{\ccIphd}{\widetilde{c}_{\mathrm{I},\phSymbol}^2}

\newcommand{\tn}{^n}
\newcommand{\tns}{^{n*}}
\newcommand{\tnss}{^{n**}}
\newcommand{\tnn}{^{n+1}}
\newcommand{\thypSymbol}{\circ}
\newcommand{\tsouSymbol}{\bullet}
\newcommand{\thyp}{^{\thypSymbol}}
\newcommand{\tsou}{^{\tsouSymbol}}

\newcommand{\volc}[1]{\left\vert \mathcal{C}_{#1} \right\vert}
\newcommand{\volz}[1]{\left\vert \zeta_{#1} \right\vert}

\newcommand{\Rus}{^{\mathrm{Rus}}}
\newcommand{\iph}{_{i+\frac{1}{2}}}
\newcommand{\imh}{_{i-\frac{1}{2}}}
\newcommand{\ipo}{_{i+1}}
\newcommand{\imo}{_{i-1}}
\newcommand{\kpo}{_{k+1}}
\newcommand{\kmo}{_{k-1}}
\newcommand{\kph}{_{k+\frac{1}{2}}}
\newcommand{\kmh}{_{k-\frac{1}{2}}}
\newcommand{\jpo}{_{j+1}}
\newcommand{\jmo}{_{j-1}}
\newcommand{\KFph}{\mathrm{K}^F\ph}
\newcommand{\KHph}{\mathrm{K}^H\ph}
\newcommand{\KUph}{\mathrm{K}^U\ph}
\newcommand{\KH}[1]{\mathrm{K}^H_{#1}}
\newcommand{\KU}[1]{\mathrm{K}^U_{#1}}
\graphicspath{{FIG/}}

\newtheorem{rmk}{Remark}

\begin{document}
\hypersetup{ citecolor=teal, filecolor=magenta, urlcolor=blue!90!violet }

\title{A pressure-based method for weakly compressible two-phase flows under a Baer-Nunziato type model with generic equations of state and pressure and velocity disequilibrium}

\author[1]{Barbara Re*}
\author[2]{R\'{e}mi Abgrall}
\authormark{B. Re and R. Abgrall}

\address[1]{\orgdiv{Department of Aerospace Science and Technology}, \orgname{Politecnico di Milano}, \orgaddress{\country{Italy}}}
\address[2]{\orgdiv{Institute of Mathematics}, \orgname{University of Z\"{u}rich}, \orgaddress{\country{Switzerland}}}

\corres{*\email{barbara.re@polimi.it}}

\abstract[Summary]{
Within the framework of diffuse interface methods,
we derive a pressure-based Baer-Nunziato type model well-suited to weakly compressible multiphase flows.
The model can easily deal with different equation of states and it includes relaxation terms characterized by user-defined finite parameters, which drive the pressure and velocity of each phase toward the equilibrium. There is no clear notion of speed of sound, and thus, most of the classical low Mach approximation cannot easily be cast in this context. 
The proposed solution strategy consists of two operators: a semi-implicit finite-volume solver for the hyperbolic part and an ODE integrator for the relaxation processes.
Being the acoustic terms in the hyperbolic part integrated implicitly, the stability condition on the time step is lessened.
The discretization of non-conservative terms involving the gradient of the volume fraction fulfills by construction the non-disturbance condition on pressure and velocity to avoid oscillations across the multimaterial interfaces.
The developed simulation tool is validated through one-dimensional simulations of shock-tube and Riemann-problems, involving water-aluminum and water-air mixtures,  vapor-liquid mixture of water and of carbon dioxide, and almost pure flows.
The numerical results match analytical and reference ones, except some expected discrepancies across shocks, which however remain acceptable (errors within some percentage points).
All tests were performed with acoustic CFL numbers greater than one, and no stability issues arose, even for CFL greater than 10.
The effects of different values of relaxation parameters and of different amount equations of state---stiffened gas and  Peng-Robinson---were investigated.
}

\keywords{
Baer-Nunziato type model,
pressure formulation,
compressible two-phase flows,
pressure and velocity relaxation with finite parameters,
semi-implicit finite-volume scheme,
Peng-Robinson equation of state
}

\maketitle

\begin{textblock}{15}(2.5,25)\large
\textit{Accepted version}

\textbf{Int J Numer Meth Fluids (2022)} \normalsize
\href{https://doi.org/10.1002/fld.5087}{DOI: 10.1002/fld.5087}

The final publication is available at \href{https://onlinelibrary.wiley.com/doi/10.1002/fld.5087}{onlinelibrary.wiley.com}
\end{textblock}

\section{Introduction}

Compressible multiphase flows may manifest themselves in a variety of configurations, ranging from dispersed flows (e.g., bubble or spray flows) to interface problems involving two nearly pure fluids (e.g., liquid accumulation or sloshing of a liquid in a tank).
From a numerical point of view, a distinguishing and challenging feature of such flows is the presence of dynamic interfaces that separate immiscible fluids with different physical or chemical properties.
The several ways this challenge can be answered has led to the development of different multi-phase simulation strategies.
The first one that can come to mind is the explicit tracking of the interface, either by deforming the grid to preserve interfaces as resolved surfaces, e.g. in~\cite{Barlow2016,DumbserBoscheri2013}, or by tracking their motion indirectly by means of Lagrangian markers, e.g. in~\cite{Glimm1998,Tryggvason2001}. These methods can be very accurate in well-resolved interface problems with limited deformations, but cannot easily handle significant interface distortions or topological modifications.
A different strategy is pursued by interface capturing methods, which reconstruct the interfaces from the solution according to an indicator function. Popular instances in this class are the level-set methods (e.g., see the reviews~\cite{Sethian2003} and \cite{Saye2020}), in which an interface is described by a zero-level curve of a continuous function expressing the (signed) distance from the interface, and the jump conditions can be transferred across the interface by the ghost fluid method~\cite{Fedkiw1999}. This strategy facilitates the tracking of complex interfaces, but it may prevent mass conservation and robustness~\cite{Owkes2014}.

In this work, we focus on diffuse-interface methods (DIMs)~\cite{Saurel2018}, which are another class of interface capturing methods, initiated by the volume of fluids method of Hirt and Nichols~\cite{Hirt1981} for incompressible flows, and extended to compressible flows by Saurel and Abgrall~\cite{Saurel1999}, and Kapila et al.~\cite{Kapila2001}. DIMs rely on an augmented system of governing equations that specifically model the behavior of the continuum close to the interfaces, while they aim to recover the pure fluid behavior far from them.
In practice, DIMs assume that at least a small quantity of all fluids coexist in each computational cell, and, rather than local instantaneous realizations of the multiphase flows, they aim to describe its behavior on average (in time, space, an ensemble, or in some combination of those)~\cite{Drew1999}, which is usually the quantity of interest in industrial applications.
Finally, DIMs appear particularly suited for fluids governed by different equations of state (EOSs), since the behavior of each fluid is described through its own thermodynamic model~\cite{Saurel1999}.

\paragraph{Baer-Nunziato model}
The cornerstone of the DIM class is the Baer-Nunziato (BN) two-phase model~\cite{Baer1986}, which was originally developed for reactive granular materials and allows unequal phase pressures, velocities, and internal energies.
The BN model consists of a set of mass, momentum, and total energy equations for each phase and a topological equation for the volume fraction, so seven equations for a one-dimensional problem.
Starting from the original one, a wide set of BN-type models have been proposed~\cite{Saurel1999,Romenski2007,Ambroso2012,Saurel2014,Muller2016,Saurel2017}, according to different modeling and closure assumptions.
While using different definitions for the interface and relaxation terms, these models typically share the same homogeneous and hyperbolic part. 
Thus, they require to face similar analytical and computational challenges, which concern the presence of non-conservative terms, the large number of waves and the requirement to deal with many equations.
To mitigate the last two shortcomings, reduced models have been also proposed.

Five-equations models have been derived by means of asymptotic expansions of the BN model in the limit of stiff mechanical (i.e., pressure and velocity) equilibrium~\cite{Kapila2001,Allaire2002,Murrone2005,Saurel2008,Kreeft2010}.
Although these models are simplified, they have different difficulties, as for instance,
the discretization of a non-conservative term involving the divergence of the velocity in the transport equation and the non-monotonic behavior of the mixture sound speed with volume fraction, which may lead to an erroneous wave propagation speed through the diffuse interface~\cite{Saurel2009}.
The roots of these issues are found in the pressure-equilibrium condition, which can be thus removed, as in the pressure non-equilibrium 6-equation models~\cite{Saurel2009,Zein2010}, which however need to be augmented by an energy conservation law for the mixture to correct the predicted thermodynamic states, unless they use the formulation recently proposed by Pelanti and Shyue for simplified EOSs (i.e., stiffened gas)~\cite{Pelanti2014,Pelanti2019}.
A different choice underlies the six-equation two-fluid models~\cite{Ishii1975,Staedtke2005}, in which the fluids have same pressure, but other thermodynamic quantities are in non-equilibrium. These models are generally considered as ill-posed~\cite{Stewart1984}, but recently Hantke and co-authors~\cite{Hantke2021} have proposed some constraints on the interfacial pressure that can ensure hyperbolicity.

Even from this short and basic outline about two-phase models, it appears evident that each model has its own strengths and weaknesses, and which is the best one clearly depends on the application under investigation.
However, from a general BN-type model, a hierarchy of hyperbolic multiphase models can be derived on the basis of asymptotic analysis~\cite{Lund2012}, so it is possible to derive the simplest model involving the relevant physical effects.
Keeping into account these considerations, in this work, we propose a full non-equilibrium, BN-type model, to provide the widest applicability within the class of DIMs, and eventual reduced models will be considered in future works.
Nevertheless, the selected BN-type model includes terms for pressure and velocity relaxation determined by finite parameters, which could be tuned to manage how the mechanical equilibrium between phases is reached.

\paragraph{Pressure formulation}
Most of the literature about DIMs for BN-type models solve the governing equations for the conservative variables, that is volume fraction, density, momentum, and total energy, and contribute to the development and improvement of the so-called density-based methods.
These are the solvers of choice for flows characterized by significant compressibility, but they suffer from ill-conditioning and accuracy problems at low Mach number~\cite{Guillard1999}, that is when the flow speed is considerably lower than the speed of sound.
In these conditions, the stability constraint on the time step becomes stringent and sophisticated preconditioning techniques are required to recover the correct scaling of the pressure fluctuation with the Mach number~\cite{Guillard1999,Guillard2004}.
Because of different thermo-physical properties, two-phase flow fields, especially when involving gas and liquid mixtures, often exhibit a wide range of Mach numbers, including also the low Mach limit.
A classical way to take into account the stiffness due to low Mach effects is the dimensionless scaling of the system of partial differential equations according to a reference density, a reference speed and a reference speed of sound~\cite{Dellacherie2016,Guillard1999}.
This leads to a system that looks similar to the original one, but that is able to describe incompressible flows. However, this approach is difficult to apply to non-equilibrium multi-phase models, because it is not possible to define a unique, unambiguous reference speed of sound.
In addition, in density-based method, the pressure field is generally updated by means of an EOS, an operation that, in compressible multi-phase flows, may generate spurious oscillations at material interfaces~\cite{Karni1996,Lv2020}.

On the other hand, using the pressure rather than the density as a solution variable in the governing equations could circumvent most of the issues arising from the weak pressure-density coupling at low Mach numbers, because pressure variations are significant at all speeds.
Thus, a unique reference pressure can be easily identified for the non-dimensional scaling of the governing equations, as in~\cite{Bijl1998}.
Moreover, solving for pressure (a primitive base) rather than total energy (a conservative variable) could facilitate the achievement of mechanical equilibrium across interfaces and regions with varying thermo-physical properties~\cite{Karni1996}, and paves the wave for a straightforward implementation of arbitrary EOSs~\cite{Kawai2015}.
These features could be substantially beneficial for the simulation of compressible multiphase flows and thus have prompted us to study a pressure-based BN-type model.

Pressure-based methods have their roots in numerical methods for incompressible single-phase flows, which have been extended to compressible flows following the general idea to replace the divergence-free condition on the velocity field of standard incompressible solvers by a modified continuity equation~\cite{Karki1989,Demirdzic1995}.
This concept has been extensively applied to the semi-implicit method for pressure linked equations (SIMPLE)~\cite{Patankar1972}, to projection or fraction-step methods~\cite{Kim1985}, and to the MAC method~\cite{Harlow1965}, leading to a large variety of pressure-based formulations, e.g.~\cite{Bijl1998,Buffard2000,Munz2003,Xiao2004,Park2005,%
Degond2011,Xiao2017,Xie2017,Bermudez2020,Busto2021}.
Although the research area of pressure-based formulation has been very active for decades, most of the available techniques consider single-phase flows and, but for a few exceptions~\cite{Kwatra2009,Cordier2012,Terashima2012,Kawai2015,Dumbser2016}, they are valid only under the assumption of polytropic ideal gas.
Recently, some examples of pressure-based methods have been proposed in the framework of volume of fluid methods, e.g., \cite{Duret2018} and \cite{Denner2018},
while Zhang et al.~\cite{Zhang2019} have developed a pressure-based solver for the two-fluid six-equation model, and Abgrall et al.~\cite{Abgrall2018} have used the non-conservative pressure formulation of Kapila's model.
However, according to our knowledge, no pressure-based algorithms have been proposed for a full non-equilibrium BN-type model, except a preliminary work for the homogeneous part~\cite{ReAbgrall2019}.

Although pressure-based methods offer several potential advantages, they are non-conservative, so they are not able to correctly predict the propagation speed of shock waves.
Some techniques have been proposed to cure this inherent drawback: for instance, it would be possible to switch to a fully-conservative formulation far from material interfaces~\cite{Karni1996}, correction terms can be added to the pressure equation~\cite{VanderHeul2003,Abgrall2018}, or the pressure equation can be considered only as a predictor for the updated value to be inserted in the conservative energy equation~\cite{Zhang2019}.
However, in this work, we do not resort to any corrective measures, because we focus here on the validation of the proposed pressure-based BN-type model and on the convergence to the correct solution in the low-Mach regime using a simple numerical technique, while we leave all the numerical advancements for a further work.
Nevertheless, we solve conservatively the part of density and momentum equations related to the Euler equations, so the conservation of mass and momentum of the two-phase mixture mitigate the error in the shock propagation, unless very strong discontinuities are involved.

\paragraph{Weakly compressible multiphase flows}
Our research about a pressure-based solver well-suited for weakly compressible two-phase flows was motivated also by a specific application, the pipeline transport of pressurized carbon dioxide (CO\textsubscript{2}) within the carbon-capture and storage framework, a promising measure to mitigate climate change~\cite{Metz2005}.
In standard working conditions, CO\textsubscript{2} is transported in liquid or dense gas state, but two-phase flows may occur because of transient events such as start-up, de-pressurization, or oscillations in the supply chain. 
In these situations, the Mach number is low, but if we treated the flows as incompressible, pressure waves would generate no changes in the density. On the contrary, the capability to correctly evaluate the density and temperature variation is of paramount importance for a safe design  of the pipeline and for flow metering~\cite{Munkejord2016}.
Other examples of low-Mach multiphase problems involve sloshing phenomena, and boiling or cavitating flows, which are encountered in various applications, such as combustion engines, pumps, heat exchange, nuclear power plants, transportation and storage systems.
Here, the liquid phase is almost incompressible, but the liquid-gas mixture is highly compressible and the presence of bubbles or entrapped air, especially if close to a wall, impacts on the flow behavior and on the structural loads.
Hence, we need to take into account the compressibility of both phases to correctly evaluate the thermodynamic pressure and the wave propagation~\cite{Daru2010,LeMartelot2013,Pelanti2017}.

\paragraph{Goal and highlights}
As anticipated before, the goal of our work is to develop a pressure-based formulation of a BN-type model, well suited for the simulation of unsteady weakly-compressible multiphase flows.
The rationale behind the pressure-based formulation is to avoid 
preconditioning---required by a standard density-based approach---which could change the topology, as well as to have a clear scaling, even if more than one speed of sound characterizes the flow field.

In this paper, we describe the model and validate it in simplified test cases. Therefore, we focus only on one-dimensional problems. Nevertheless, the software implementation keeps into consideration the possibility to extend the method to two- and three-dimensions in the future.
We are particularly interested in creating a flexible and robust simulation tool, which is able to deal with different fluids and flow configurations, but, whenever possible, we pursue the strategy to combine existing tools to address specific problems.
These guidelines motivate the following modeling choices, which characterize the method proposed in this work.
\begin{itemize}
\item We adopt the BN-type model proposed by Saurel and Abgrall~\cite{Saurel1999}, but we consider finite parameters for pressure and velocity relaxation terms.
\item A general thermodynamic description is assumed during the derivation of the governing equations and the numerical method, which are thus valid for different EOSs, such as stiffened gas, Peng-Robinson, and more complex, multi-parameter EOSs.
\item Two different pressure variables are defined according to the scaling proposed by Bijl and Wesseling~\cite{Bijl1998}, so that the acoustics is filtered out from the model.
\item Staggered grids are used to prevent stability issues related to the checker-board problem.
\item A robust, but scheme-dependent, discretization of the non-conservative terms has been derived following the pressure non-disturbance condition~\cite{Abgrall1996}, to avoid spurious oscillations across multi-material interfaces.
\end{itemize}
These choices are explained and justified better in the next sections.

\paragraph{Paper structure}
The next section concisely reviews some key features of pressure-based formulations proposed for single-phase flows, which are important to support the modeling choices made in this work.
Section~\ref{s:model} begins with the presentation of the underlying BN-type model, continues with the derivation of the dimensionless pressure-based model, and ends with a short digression about the thermodynamic models.
Section~\ref{s:numMeth} presents the numerical method developed to solve the resulting BN-type model and it is split in four subsections: Sec.~\ref{ss:timedis} introduces the semi-implicit time integration of the hyperbolic operator, Sec.~\ref{ss:variablePosit} defines the organization of the variables over the grids, Sec.~\ref{ss:spacedis} details the spatial discretization of each equation of the hyperbolic part of the model, and Sec.~\ref{ss:Lrelax} explains how the relaxation terms are treated.
To have an organized framework, the results are presented in three different sections. First, Sec.~\ref{s:sphase} concerns the verification of the proposed numerical method for single-phase flows, and compares some possible alternatives in the solution strategy.
Section~\ref{s:twophase} moves to two-phase simulations but still on a verification level, as it presents some water-air problems without relaxation, to validate the behavior of the hyperbolic operator.
Finally, Sec.~\ref{s:relax} presents the results of the complete model. Sections~\ref{ss:waterAlu} and \ref{ss:waterAirRelax} give also an illustration of the effects of different values of finite relaxation parameters, Sec.~\ref{ss:almostPure} considers almost pure fluids, and Sec.~\ref{ss:coo} compares the results obtained with two different thermodynamic models.
Lastly, in Sec.~\ref{s:conclusion} we draw the conclusions of our works and we discuss future development and potential exploitation.
The manuscript includes also \ref{a:pform} and \ref{a:Hp}, which contain some passages omitted  in the derivation of the model and the numerical discretization for reason of space.

\section{Some key features of standard pressure-based approaches for single phase flows}
\label{s:pressFeat}
Many researchers have proposed diverse pressure-based formulations for the Euler equations, able to address the challenges of the low-Mach limits. 
These studies serve as a precious basis for our work, in which we attempt to blend together some key features of these previous studies and to extend them to multiphase flows.
For this reason, before explaining our work, we briefly review here, without claiming to be exhaustive, some fundamental concepts widely used in numerical methods for weakly compressible single-phase flows.

One of the most challenging aspects of low Mach flows is that the governing equations change their character: the system of equations of the compressible gas-dynamics is purely hyperbolic, while its incompressible counterpart has a mixed hyperbolic-elliptic character with infinite propagation speed.
As a consequence, pressure and density are weakly coupled, so the problem of retrieving the pressure from the density becomes ill-conditioned. This explains, at least partially, the misbehavior of standard density\hyp{}based methods at low Mach~\cite{Guillard1999} and has motivated the widespread of pressure\hyp{}based formulations. These latter strategies reflect, in general, the weak pressure-density coupling by solving the governing equations in a segregate approach: the velocity is first predicted using a pressure approximation in the momentum equation, then a correction step is carried out to update the pressure and, finally, the velocity is corrected for the new pressure~\cite{Bijl1998,Wall2002,Wenneker2002}.
Segregate solution strategies prompt the use of staggered spatial discretization where thermodynamic variables are stored at cell centers, while velocity variables are stored at cell faces~\cite{Harlow1965,Perot2000,Zhang2002}. Contrary to co-located formulations, staggered formulations filter spurious pressure modes providing an improved stability, at a similar level of efficiency and conservation properties~\cite{Perot2003}.

Asymptotic analyses have suggested the use of multiple pressure variables, which account for the different physical roles played by the different orders of the pressure in the low Mach limit~\cite{Klein1995,Munz2003}.
Performing a single time scale/multiple space scale asymptotic analysis of the Euler equations, in which the pressure for small Mach numbers ($M$) is expressed as 
\begin{equation}
\label{ep:p_exp}
P = P^{(0)} + M P^{(1)} + M^2 P^{(2)} + \mathcal{O}(M^3) \, ,
\end{equation}
Klein~\cite{Klein1995} showed that a scheme for low Mach flows should take into consideration at least two pressure variables: the leading order $P^{(0)}$ which plays the role of the thermodynamic variable, and the second-order term $P^{(2)}$ which is the ``standard pressure'' that accounts for local force balancing and, for $M\rightarrow 0$, satisfies the Poisson equation.
Instead, the first-order term $P^{(1)}$ is associated with long wave acoustics and it should be taken into account when pressure waves of order $\mathcal{O(M)}$ are important.
The pressure decomposition~\eqref{ep:p_exp} allows the compressible equations to converge toward the correct limit, namely to the solution of the incompressible ones for $M\rightarrow 0$~\cite{Wenneker2002,Munz2003}.
Standard numerical methods for compressible flows use on a non-dimensionalization based on a single reference velocity (e.g. computed from a set of reference pressure, density, and length), which introduces, in the low Mach limit, a singularity in the momentum equation, due to the presence of the term $\frac{1}{M^2}$ in front of the pressure gradient~\cite{Wenneker2002}. The pressure decomposition~\eqref{ep:p_exp} cures this singularity.
As a final remark, this asymptotic analysis highlights that the divergence condition for incompressible flows,
that is $\boldsymbol \nabla \cdot \boldsymbol {u}=0$ (where $\boldsymbol u$ is the flow velocity), results from the energy equation, not from the continuity equation, which, in the zero Mach limit, simply describes the advection of density fluctuations.

The idea of multiple pressure variables can be implemented in several ways~\cite{Bijl1998,Munz2003,Ventosa-Molina2017}. Here, we follow the strategy proposed by Bijl and Wesseling~\cite{Bijl1998,Wenneker2002}, who defined the pressure scaling as
\begin{equation}
\label{ep:scaling}
P = \frac{\Pd - \Pd\reff}{\rhod\reff \ud\reff^2} \, ,
\end{equation}
where $\rho$ is the density, $u$ is a scalar velocity, the tilde indicates dimensional variables, and the subscript $\mathrm{r}$ denotes reference quantities.
The scaling~\eqref{ep:scaling} is characterized by the parameter $M\reff$ (reference Mach number), defined as
\begin{equation}
\label{ep:refMach}
 M\reff^2= \frac{\rhod\reff \ud\reff^2}{\Pd\reff} \, ,
\end{equation}
which expresses the overall compressibility of the flow field.
This strategy does not take into account the first order pressure $P^{(1)}$.

In the low-Mach limit, the system of differential equations describing the evolution of the flow becomes  stiff. Hence, the explicit time stepping schemes, routinely used for highly non-linear compressible flows, become inefficient~\cite{Munz2003,Kwatra2009}, because the CFL (Courant-Friedrichs-Lewy) condition imposes a severe limitation of the maximum admissible time step.
To circumvents the most stringent time step limitation, the acoustic terms should be integrated implicitly, while the convective and diffusive terms can be treated explicitly, since they impose only a mild stability limitation of the time step based on the flow velocity.
This strategy is called semi-implicit time integration~\cite{Casulli1984,Klein1995,Boscarino2018}, and it is a common  feature of pressure-based schemes, shared especially by the schemes that are able to represent all-Mach numbers---i.e., from very small to Mach numbers of order one.
The concept can been easily implemented in a semi-discrete, fractional step projection method, in which the equations to be solved sequentially contain both implicit and explicit terms~\cite{Wall2002,Cordier2012,Bermudez2014,Dumbser2016,Ventosa-Molina2017,Bermudez2020}.
This is the strategy adopted in the present work, but, alternatively, a similar idea can be enforced by splitting the fluxes in two parts, advective and non-advective~\cite{Xiao2004,Kwatra2009,Lentine2011}.
Both strategies can be used to derive asymptotic preserving schemes~\cite{Degond2011,Cordier2012,Noelle2014}.

\section{A Baer and Nunziato type model for non-equilibrium multiphase flows at low Mach number}\label{s:model}
In this section, we derive the set of equations that is the basis of the proposed numerical method, explaining the starting point and the modeling choices.
We start presenting the BN-type model and the relevant notation, then we derive the pressure formulation and we apply the scaling~\eqref{ep:scaling}.

\subsection{The Baer and Nunziato type model}
The non-equilibrium  multiphase model derived by Saurel and Abgrall in~\cite{Saurel1999} assumes that each phase is compressible and evolves with its own pressure, temperature, and velocity.
The model does not consider heat or mass transfer and it tends to the Euler equations far from the interfaces.
The system of 7 governing equations can be written in the following compact non-conservative form:
\begin{gather}
\pder{\mathbf{U}}{t} + \pder{\mathbf{F}(\mathbf{U})}{x}
+ \mathbf{B}(\mathbf{U}) \pder{\alpha_1}{x}
= \mathbf{S}_P(\mathbf{U}) + \mathbf{S}_u(\mathbf{U})\label{em:bn}\\
\mathbf{U} = \begin{bmatrix}
\alpha_1\\ \arho_1 \\ \am_1 \\ \aet_1 \\  \arho_2 \\ \am_2 \\ \aet_2
\end{bmatrix}\!\!, \;
\mathbf{F} = \begin{bmatrix}
0\\ \arho_1 u_1 \\ \am_1 u_1 + \alpha P_1 \\ (\aet_1+\alpha P_1)u_1 \\ 
    \arho_2 u_2 \\ \am_2 u_2 + \alpha P_2 \\ (\aet_2+\alpha P_2)u_2
\end{bmatrix}\!\!, \;
\mathbf{B} = \begin{bmatrix}
\ui\\ 0 \\ -\ppi \\ -\ppi \ui \\  0 \\  \ppi \\ \ppi \ui
\end{bmatrix}\!\!, \;
\mathbf{S}_P = \begin{bmatrix}
 \mu (P_1 - P_2)\\ 0 \\ 0 \\ -\mu \ppi(P_1 - P_2) \\  0 \\  0 \\ \mu \ppi(P_1 - P_2)
\end{bmatrix}\!\!, \;
\mathbf{S}_u = \begin{bmatrix}
0 \\ 0 \\ -\lambda (u_1 - u_2)\\  -\lambda \ui(u_1 - u_2) \\ 
     0 \\  \lambda (u_1 - u_2)\\  \lambda \ui(u_1 - u_2)
\end{bmatrix}\notag
\end{gather}
where $\mathbf{U}$ is the vector of evolution variables, $\mathbf{F}$ is the flux function (the pure conservative part), $\mathbf{B}$ contains the non-conservative part, $\mathbf{S}_P$ and $\mathbf{S}_u$ are vectors of source terms modeling, respectively, pressure and velocity relaxation.
According to the standard notation, the variables are: $\alpha$ the volume fraction (note that $\alpha_1 + \alpha_2=1$), $\rho$ the density, $u$ the velocity, $m=\rho u$ the momentum, $P$ the pressure, and $E$ the total energy, defined as $E= e + \frac{1}{2} \rho u^2$, where $e$ is the internal energy.
The numerical subscript of each variable denotes the phase to which it refers\footnote{When a single numerical subscript is written at the end of a group of variables starting with $\alpha$, it refers to all variables in the group, e.g., $\alpha \rho_1$ means $\alpha_1 \rho_1$.}.
The pressure $\ppi$ and the velocity $\ui$ model the average interface values over the two-phase control volume, and they are estimated as
\begin{equation}
\label{em:intervar}
\ppi = \alpha P_1 + \alpha P_2 \, , \qquad
\ui = \frac{\am_1 + \am_2}{\arho_1 + \arho_2} \, ,
\end{equation}
while $\mu$ and $\lambda$ are  relaxation parameters that express, respectively, how fast the pressure and velocity equilibrium is reached~\cite{Baer1986,Saurel1999}.
These variables depends on the nature of each fluid as well as on the topology of the multiphase flow. For this reason, in this work, they are user-definite finite and positive parameters.
All variables in Eqs.~\eqref{em:bn} and \eqref{em:intervar} are dimensional.

Even if presented for two phases, this model can be adapted to three or more phases, provided a definition of interface and relaxation terms is given.
Moreover, considering $\alpha_1=1$ and $\alpha_2=0$, the model simplifies to the classical Euler equations for single-phase flows.
Furthermore, if we sum the equations per variables, we have the Euler equations for the mixture conservative variables, that is for the mixture density, i.e., $\bar{\rho} = \arho_1 + \arho_2$, the mixture momentum, and the mixture total energy.

Thermodynamic models are required to close the model. Each component obeys its own EOS as a pure material, so for each fluid or phase, we consider a generic EOS in the shape $e=e(\rho,P)$, where $e$ is the internal energy per unit of volume, namely $e = \varepsilon \rho$ with $\varepsilon$ the specific internal energy.
We remind here some thermodynamic definitions and relations which are of interest in the next sections.
First, we introduce the following thermodynamic derivatives
\begin{equation}
\label{emt:tmdder}
\chi = \tmdder{P}{\rho}{e}\, \qquad \kappa = \tmdder{P}{e}{\rho} \,.
\end{equation}
Accordingly, the definition of the speed of sound reads for each fluid
\begin{equation}
\label{emt:spsound}
\begin{aligned}
c^2 = \tmdder{P}{\rho}{s} &=
\tmdder{P}{\rho}{e} + \tmdder{P}{e}{\rho} \tmdder{e}{\rho}{s} =
\chi +\kappa \tmdder{e}{v}{s} \frac{\mathrm{d} v}{\mathrm{d} \rho} =
\chi - \kappa \dfrac{1}{\rho^2} \left[ \rho \tmdder{\varepsilon}{v}{s} + \varepsilon  \frac{\mathrm{d} \rho}{\mathrm{d} v} \right]\\
&= \chi - \kappa \dfrac{1}{\rho^2} \left[-\rho P - {\varepsilon \rho^2} \right] =
\chi + \kappa \dfrac{P +e}{\rho} \, 
\end{aligned}
\end{equation}
Definitions~\eqref{emt:tmdder} and \eqref{emt:spsound} are valid for each fluid separately, although we have omitted the subscript denoting the phase to simplify the notation.
For later convenience, we define also, for each phase, an interface speed of sound as
\begin{equation}
\label{emt:spsoundInt}
\cci{\phSymbol} = \chi\ph + \kappa\ph \dfrac{\ppi + e\ph}{\rhoph} \,.
\end{equation}

\subsection{Pressure-based formulation}
To formulate a pressure-based BN-type model, we need to derive an equation for the pressure evolution from the conservative form~\eqref{em:bn}.
Here, we describe only the main steps and the results, while the step-by-step derivation is given in~\ref{a:pform}.

The first step consists in expressing the total energy in terms of pressure, density, momentum, and energy.
Given an EOS in the form $e=e(\rho,P)$, we can express the partial derivatives of $E$ with respect to $\xi=\{t,x\}$ as
\[ \pder{E}{\xi} =  
\left[ \frac{1}{\kappa} \pder{P}{\xi} -\frac{\chi}{\kappa} \pder{\rho}{\xi} \right] + u  \pder{m}{\xi} - \frac{u^2}{2} \, \pder{\rho}{\xi} \,. \]
Thus, we insert this into the energy equation for phase $\phSymbol=\lbrace 1,2 \rbrace$ in \eqref{em:bn}, which can be re-written as\footnote{
Note that, since $\alpha_1 + \alpha_2 = 1$, $\pder{\alpha_1}{\xi} = - \pder{\alpha_2}{\xi}$, so the change of sign between phase 1 and 2 in $\mathbf{B}(\mathbf{U})$ is correctly reproduced in Eq.~\eqref{emp:tmp1} by using $\pder{\aph}{\xi}$.}
\begin{equation}
\label{emp:tmp1}
\begin{split}
\aph \left[
\frac{1}{\kappa\ph}\pder{\Pph}{t} -\frac{\chi\ph}{\kappa\ph} \pder{\rhoph}{t} + \uph  \pder{\mph}{t} -  \frac{\uph^2}{2} \, \pder{\rhoph}{t} \right]  
&+\left( e\ph + \frac{\mph^2}{2\rhoph} \right) \left[ \pder{\aph}{t}  + \pder{\aph \uph}{x} \right]\\
+ \aph \uph \left[
\frac{1}{\kappa\ph}\pder{\Pph}{x} -\frac{\chi\ph}{\kappa\ph} \pder{\rhoph}{x} + \uph  \pder{\mph}{x} -  \frac{\uph^2}{2} \, \pder{\rhoph}{x} \right]
&+ \pder{(\alpha \Pph \uph)}{x} - \ppi \ui\pder{\aph}{x} =  \ppi \mu\Delta\ph P - \ui \lambda\Delta\ph u
\end{split}\end{equation}
where we have introduced the operator $\Delta\ph$ which takes the difference between the phase $\phSymbol$ and the opposite one, i.e., $\Delta_1 P = P_1 - P_2$ and $\Delta_2 u = u_2 - u_1$.

As detailed in~\ref{a:pform},
we replace the temporal derivatives of $\rhoph$, $\mph$, and $\aph$ according to the respective equations in \eqref{em:bn}. Then, re-arranging the terms and recalling the definitions of the speed of sound \eqref{emt:spsound} and interface speed of sound \eqref{emt:spsoundInt}, we write the pressure formulation of the BN-type model~\eqref{em:bn} as
\begin{equation}
\label{emp:peq}
\aph\pder{\Pph}{t}+\alpha\uph \pder{\Pph}{x} 
+\alpha\rhoph c\ph^2 \pder{\uph}{x} - \rhoph \cci{\phSymbol}(\ui - \uph)\pder{\aph }{x} 
 = -\rhoph \cci{\phSymbol}\mu \Delta\ph P - \kappa\ph (\ui- \uph ) \lambda  \Delta\ph u \, .
\end{equation}

\subsection{Dimensionless pressure-based BN-type model}
In this section, we proceed to make the system of governing equations dimensionless, according to the pressure scaling~\eqref{ep:scaling} proposed by Bijl and Wesseling~\cite{Bijl1998}. 
For the sake of clarity, we re-write here the volume fraction, density and momentum equation of system~\eqref{em:bn}, along with the pressure equation~\eqref{emp:peq}, for one phase only, highlighting the dimensional variables with a tilde. We remind that the volume fraction $\alpha$ is, by definition, a dimensionless variable.
\begin{align}
\pder{\aph}{\td}    + \uid\pder{\aph}{\xd}   &= \mud\Delta\ph \Pd \label{emd:1alpha}\\
\pder{\arhophd}{\td}+ \pder{(\arhophd \uphd)}{\xd}  &= 0\\
\pder{\amphd}{\td}+ \pder{(\amphd \uphd + \alpha\Pphd)}{\xd} - \ppid \pder{\aph}{\xd} &= - \lambdad \Delta\ph \ud \\
\aph\pder{\Pphd}{\td}+ \alpha\uphd\pder{\Pphd}{\xd} + \arhophd \ccphd \pder{\uphd}{\xd}
 - \rhod\ph \ccIphd(\uid - \uphd)\pder{\aph }{\xd} 
 &= - \rhod\ph \ccIphd \mud\Delta\ph \Pd - \widetilde{\kappa}\ph (\uid - \uphd) \lambdad \Delta\ph \ud \label{emd:1P}
\end{align}

The scaling procedure requires the definition of the set of (dimensional) reference variables.
The first entries in this set are: a density $\rhod\reff$, a length $\tilde{L}\reff$, and a velocity $\ud\reff$. Conventionally, we define dimensionless density, length, and velocity as
\begin{equation*}
 \rho = \frac{\rhod}{\rhod\reff}, \qquad
 x = \frac{\xd}{\tilde{L}\reff} \qquad
 u = \frac{\ud}{\ud\reff}  \,.
\end{equation*}
Combinations of these three reference variables are sufficient to make dimensionless all the variables in Eqs.~\eqref{emd:1alpha}--\eqref{emd:1P}, as shown in Tab.~\ref{tmd:dimAnalys}.
However, as anticipated in Sec.~\ref{s:pressFeat}, we adopt a special scaling for the pressure to filter out the long-wave acoustics and to cure the singularity in the momentum equations in the zero Mach limit. Indeed, we introduce also a pressure reference variable $\Pd\reff$, and we define the dimensionless pressure as $P= \frac{\Pd - \Pd\reff}{\rhod\reff \ud\reff^2}$.

Let us illustrate how this choice influences the scaling of the thermodynamic variables.
To preserve the relation between the internal and total energy at dimensionless level, we define 
\begin{equation*}
e = \frac{\ed}{\rhod\reff \ud\reff^2} , \qquad \text{and}\qquad 
E = \frac{\widetilde{E}}{\rhod\reff \ud\reff^2} = \frac{\ed}{\rhod\reff \ud\reff^2} +\frac{1}{2}\frac{\rhod\ud^2}{\rhod\reff \ud\reff^2} =  e +\frac{1}{2}\rho u^2 \, .
\end{equation*}
Consequently, the pressure derivatives defined in~\eqref{emt:tmdder} are scaled as
\begin{equation}
\label{emd:tmdder}
\widetilde{\chi} = \tmdder{\Pd}{\rhod}{\ed} = \tmdder{P}{\rho}{e} \ud\reff^2 = \chi\ud\reff^2 \,, \qquad 
\widetilde{\kappa} =  \tmdder{\Pd}{\ed}{\rhod} = \tmdder{P}{e}{\rho} = \kappa\,.
\end{equation}
More care is required for the speed of sound, which depends explicitly on the pressure. We want to preserve the definitions~\eqref{emt:spsound} and \eqref{emt:spsoundInt} also at dimensionless level, so we define
\begin{equation}
\label{emd:spsounds}
c^2 = \chi + \kappa \dfrac{P +e}{\rho} \,  \qquad \text{and} \qquad 
c_\mathrm{I}^2= \chi + \kappa \dfrac{\ppi + e}{\rho} \,.
\end{equation}
But, with this choice, the dimensional speed of sound in terms of dimensionless variables reads
\begin{equation}
\widetilde{c}^2 = \widetilde{\chi} + \widetilde{\kappa} \dfrac{\Pd +\ed}{\rhod} =
\left[ \chi + \kappa \dfrac{(P +e)\rhod\reff}{\rho\rhod\reff} \right] \ud\reff^2 + \kappa \frac{\Pd\reff}{\rho\rhod\reff} =
c^2\ud\reff^2   + \frac{\kappa}{\rho} \frac{\Pd\reff}{\rhod\reff \ud\reff^2}\ud\reff^2 =
\left[ c^2 + \frac{1}{M\reff^2}\frac{\kappa}{\rho} \right]\ud\reff^2 
\end{equation}
where we have introduced the reference Mach number defined in~\eqref{ep:refMach}. A similar expression is found for the interface speed of sound, which is reported in Tab.~\ref{tmd:dimAnalys}.
As we show in the next paragraph, the additional term $\frac{1}{M\reff^2}\frac{\kappa}{\rho}$ plays a fundamental role in the scaling of the pressure equation~\eqref{emp:peq}.

\begin{table}
\caption{This table summaries the dimensionless scaling of the variables and of some operators in equations~\eqref{emd:1alpha}--\eqref{emd:1P}.
In particular, we show the combination of reference quantities ($\rhod\reff$, $\ud\reff$, $\Ld\reff$, and $\Pd\reff$) required to express each dimensional variable in terms of its dimensional counterpart. The first two columns report the variables that are not affected by the pressure, whose scaling is standard. The last column refers to the thermodynamic variables that require particular care because of the presence of the pressure in their definition.}
\label{tmd:dimAnalys}
\begin{equation*}\begin{array}{*{3}{r@{\,=\,}l}}
\toprule
\widetilde{m} & m \, \rhod\reff \ud\reff              &   \ed & e \,\rhod\reff \ud\reff^2   &
      \Pd & P \, \rhod\reff \ud\reff^2 + \Pd\reff \\
\mud & \mu \, \frac{1}{\Ld\reff \rhod\reff \ud\reff}   &
  \lambdad & \lambda \,\frac{\rhod\reff \ud\reff}{\Ld\reff}  &  
      \widetilde{c}^2& \left(c^2 + \frac{1}{M\reff^2}\frac{\kappa}{\rho} \right)\ud\reff^2 \\
\pder{}{\td} & \frac{\ud\reff}{\Ld\reff} \,\pder{}{t} &
  \pder{}{\xd} & \frac{1}{\Ld\reff} \, \pder{}{x}        & 
      \widetilde{c}_\mathrm{I}^2& \left( c_\mathrm{I}^2 + \frac{1}{M\reff^2}\frac{\kappa}{\rho} \right)\ud\reff^2 \\
\Delta\ph \ud & \ud\reff \, \Delta\ph u              & \multicolumn{2}{c}{\quad}  & 
      \Delta\ph \Pd & \rhod\reff \ud\reff^2 \, \Delta\ph P \\
\bottomrule
\end{array}\end{equation*}
\end{table}

Following the definitions given above and in Tab.~\ref{tmd:dimAnalys}, we express all variables in Eqs.~\eqref{emd:1alpha}--\eqref{emd:1P} in terms of their dimensionless counterpart. By using a verbose notation to show all substitutions, we obtain
\begin{align}
\frac{\ud\reff}{\Ld\reff}\pder{\aph}{t}  + \frac{\ud\reff}{\Ld\reff}\ui\pder{\aph}{x} 
   &= \frac{\rhod\reff \ud\reff^2}{\Ld\reff \rhod\reff \ud\reff}  \mu\Delta\ph P \label{emd:tmp_alpha} \\
\frac{\rhod\reff \ud\reff}{\Ld\reff}\pder{\arhoph}{t} + \frac{\rhod\reff \ud\reff}{\Ld\reff}\pder{(\arhoph \uph)}{x}  &= 0\\
\frac{\rhod\reff \ud\reff^2}{\Ld\reff}\pder{\amph}{t} + \frac{\rhod\reff \ud\reff^2}{\Ld\reff}\pder{(\amph \uph + \alpha\Pph)}{x}
 + \frac{\Pd\reff}{\Ld\reff}\pder{\aph}{x} - \frac{\ppi\rhod\reff \ud\reff^2 + \Pd\reff}{\Ld\reff} \pder{\aph}{x} &= 
   - \frac{\rhod\reff \ud\reff^2}{\Ld\reff} \lambda \Delta\ph u \label{emd:tmp_mom}\\
\frac{\rhod\reff \ud\reff^3}{\Ld\reff} \left[ \aph\pder{\Pph}{t}+\alpha\uph\pder{\Pph}{x} \right]
   + \frac{\rhod\reff \ud\reff^3}{\Ld\reff} \left[ \arhoph  \left(c\ph^2 + \frac{1}{M\reff^2}\frac{\kappa\ph}{\rho\ph} \right)  \pder{\uph}{x} \right.
 - \rho\ph &\left. \left(\cci{\phSymbol} + \frac{1}{M\reff^2}\frac{\kappa\ph}{\rho\ph} \right)(\ui - \uph)\pder{\aph }{x}\right] \notag\\ 
 = - \frac{\rhod\reff \ud\reff^2\, \rhod\reff \ud\reff^2}{\Ld\reff \rhod\reff \ud\reff} \rho\ph
     \left(\cci{\phSymbol} + \frac{1}{M\reff^2}\frac{\kappa\ph}{\rho\ph} \right) \mu \Delta\ph P 
   &- \frac{\rhod\reff \ud\reff^3}{\Ld\reff } \kappa\ph (\ui - \uph) \lambda \Delta\ph u \,. \label{emd:tmp_P}
\end{align}
Clearly, the previous equations can be simplified. Noting that in Eq.~\eqref{emd:tmp_mom} the two terms involving $\Pd\reff$ cancel out, we can immediately simplify Eqs.~\eqref{emd:tmp_alpha}--\eqref{emd:tmp_P} by deleting the factors comprising $\rhod\reff$, $\ud\reff$, and $\Ld\reff$. 
Then, we multiply  Eq.~\eqref{emd:tmp_P} by $M\reff^2$ and we re-arrange the terms, gathering those with this factor together.
In summary, the final system of equations expressing the pressure-based formulation of the BN-type model defined in Eq.~\eqref{em:bn} reads
\begin{align}
\pder{\aph}{t}  + \ui\pder{\aph}{x} 
   &=   \mu\Delta\ph P \label{emd:fin_alpha}\\
\pder{\arhoph}{t} + \pder{(\arhoph \uph)}{x}  &= 0\\
\pder{\amph}{t} + \pder{(\amph \uph + \alpha\Pph)}{x} - \ppi  \pder{\aph}{x} &= 
   -  \lambda \Delta\ph u \\
 M\reff^2 \left[ \alpha\ph\pder{\Pph}{t}+\alpha\uph\pder{\Pph}{x} 
   +  \arhoph c\ph^2  \pder{\uph}{x} 
 - \rho\ph  \cci{\phSymbol} (\ui - \uph)\pder{\aph }{x}\right] &+
 \kappa\ph \left[ \aph \pder{\uph}{x} - (\ui - \uph)\pder{\aph }{x} \right] \notag\\ 
 = - M\reff^2 \left[ \rho\ph \cci{\phSymbol} \mu \Delta\ph P 
   + \kappa\ph (\ui - \uph) \lambda \Delta\ph u \right] &-  \kappa\ph \mu \Delta\ph P \,. \label{emd:fin_P}
\end{align}
We highlight that the adopted pressure scaling, by means of the additional term proportional to $\kappa$ in the definition of $c^2\ph$ and $\cci{\phSymbol}$, is directly responsible for the peculiar expression of Eq.~\eqref{emd:fin_P}, in which terms proportional to $M\reff^0$ and $M\reff^2$ coexist.
The fundamental benefit of this choice is expressed by the following Remark.
\begin{rmk}[Multiphase incompressibility constraint]\label{rmk:incompressibility}
From Eq.~\eqref{emd:fin_P}, we can derive the multiphase counterpart of the kinematic constraint for incompressible flows, which for a 1D single-phase flow reads $\pder{u}{x} =0$. In the limit for $M\reff \rightarrow 0$, Eq.~\eqref{emd:fin_P} simplifies to
$   \aph \pder{\uph}{x} - (\ui - \uph)\pder{\aph }{x} = - \mu \Delta\ph P $,
which, exploiting Eq.~\eqref{emd:fin_alpha}, can be re-written as
\begin{equation}
\label{emd:incomp}
   \pder{\aph}{t} + \pder{\alpha \uph}{x}  = 0  \,.
\end{equation}
Equation~\eqref{emd:incomp} can be considered the multiphase incompressibility condition, since, if we sum Eq.~\eqref{emd:incomp} for $\phSymbol =1$ and $\phSymbol =2$, we have $ \pder{\alpha u_1 + \alpha u_2 }{x} = \pder{\bar{u}}{x}  = 0  $, where $\bar{u}$ is the mixture velocity.
This result reminds us that the incompressibility condition comes from the energy equation, and not from the mass equation.
\end{rmk}

\begin{rmk}[Symmetry]
The model expressed by Eqs.\eqref{emd:fin_alpha}--\eqref{emd:fin_P} is symmetric, in the sense that an exchange in the phase index $\phSymbol$ does not change the set of governing equations.
\end{rmk}

\begin{rmk}[Uniform pressure and velocity field]
\label{rmk:uniform}
From Eqs.\eqref{emd:fin_alpha}--\eqref{emd:fin_P}, we can see that, if the initial state is (spatially) uniform in pressure and velocity, this condition is preserved in time. Actually, if $u_1=u_2=\ui=u$ and $P_1 = P_2 =\ppi=P$, we have:
\begin{equation*}
\pder{\aph}{t}  + u\pder{\aph}{x} = 0\,; \qquad  \pder{\rhoph}{t} + u\pder{\rhoph}{x}  = 0\,; \qquad
 \pder{ \uph}{t} = 0\,;\qquad \pder{\Pph}{t}  = 0 \,,
\end{equation*}
so no pressure or velocity variations are generated~\cite{Saurel1999}.
\end{rmk}

\subsection{Thermodynamic models used in this work}
The system of governing equations presented above and the numerical method described in the next section are derived without any specific assumption on the thermodynamic models, as long as the EOS of fluid can be expressed as $e=e(\rho,P)$.
Since this requirement is pretty easy to meet, most of the EOSs used for academic and industrial purposes can be adopted to model the behavior of each component within the proposed pressure-based BN-type model.
To introduce the nomenclature used in the following sections, we briefly introduce here the models used in for the simulations presented in result sections~\ref{s:sphase}, \ref{s:twophase} and \ref{s:relax}, namely the stiffened gas model~\cite{Harlow1971,Menikoff1989} and the polytropic Peng-Robinson EOS~\cite{PengRobinson1976}.

A complete thermodynamic model of a pure fluid at equilibrium can be obtained from two independent~EOSs, the thermal and caloric one.
For the stiffened gas, their dimensional expressions read (omitting the tilde to lighten the notation)
\begin{equation}\label{emd:stiff}
e\ph(\rhoph, \Pph)=\frac{\Pph + \gamma\ph P_{\infty, \phSymbol} } {\gamma\ph - 1} + \rhoph q\ph 
\qquad \text{and} \qquad
T\ph(\rhoph, \Pph)=\frac{\Pph + P_{\infty, \phSymbol} } {c_{v,\phSymbol} \rhoph (\gamma\ph - 1)} \,,
\end{equation}
where $T$ is the temperature and $c_v= \tmdder{\varepsilon}{T}{v}$ is the specific heat capacity at constant volume, which is constant in the stiffened gas approximation.
The parameters  $\gamma$ (ratio of specific heat capacity), $P_\infty$, and $q$  depend on the  material and can be determined by fitting experimental data, e.g., the saturation curve~\cite{Saurel2008,LeMetayer2016}.
The expressions of other thermodynamic variables can be found in~\cite{Rodio2014,Han2017}.

Stiffened gas model can be considered an extension of the polytropic ideal gas (which is recovered when $P_\infty=q=0$) able to take into consideration the repulsive effects present in all states of matter (modeled by the term $\frac{e - \rho q}{\gamma - 1}$) and 
the cohesive forces typical of liquid and solid states (thanks to the term $\gamma P_{\infty}$)~\cite{LeMetayer2016}.
This capability together with its simplicity accounts for its wide use in the research activities focused on the development of models and numerical tools for two-phase flows, as the present one.
When the focus is the study of complex two-phase flow behavior, e.g., for the investigation of water cavitation problems or in process simulation of renewable energy technologies, more accurate EOSs may be required.
An answer to this demand may come from cubic EOSs, which are widely used also in industrial applications because they combine a decent accuracy with computational efficiency.
A popular instance in this class is the Peng-Robinson~\cite{PengRobinson1976}~EOS, which is used in this work to model liquid and vapor CO\textsubscript{2} in Sec.~\ref{ss:coo}.
The expression of thermal and caloric EOSs for this model can be found in~\cite{Rodio2014,Re2019}.

All thermodynamic variables are made dimensionless following the scaling rules defined in Tab.~\ref{tmd:dimAnalys} and a standard scaling for the temperature according to a reference dimensional value $\widetilde{T}\reff$, that is $\widetilde{T}= T \, \widetilde{T}\reff$.

\section{Numerical method}\label{s:numMeth}
For the aim of this work, a numerical method that is first order accurate, both in time and in space, is considered.
The system of governing equations \eqref{emd:fin_alpha}--\eqref{emd:fin_P} is solved according to the Strang splitting approach, as in~\cite{Saurel1999,Zein2010,Herard2012,Pelanti2014}.
Hence, given the solution  $\mathbf{U}\tn$ at a initial time $t\tn$, the solution $\mathbf{U}\tnn$ after a time interval $\Delta t$ is obtained by the sequence of operators
\begin{equation}
\mathbf{U}\tnn = L_\mathrm{relax} \,L_\mathrm{hyp}(\mathbf{U}\tn) \label{en:split} \,,
\end{equation}
where $L_\mathrm{hyp}$ is the operator that solves the hyperbolic part of the system over a time step $\Delta t$, while $L_\mathrm{relax}$  is the relaxation operator that solves the system of ordinary differential equations (ODEs) considering only the relaxation terms for the velocity and the pressure.
We describe $L_\mathrm{hyp}$ in the sec.~\ref{ss:timedis}--\ref{ss:spacedis},
while $L_\mathrm{relax}$ in sec.~\ref{ss:Lrelax}.

\subsection{Temporal discretization of the hyperbolic operator}\label{ss:timedis}
For the numerical discretization of the hyperbolic operator $L_\mathrm{hyp}$, we start from the time integration, keeping the spatial derivatives continuous.
To mitigate the time step restriction imposed by the CFL constraint, we use a semi-implicit temporal discretization where the acoustic effects are treated implicitly.
This requires to integrate implicitly the pressure gradient in the momentum equations and the divergence of the velocity in the pressure equations.
To easily handle the first task, we adopt a time splitting in which the momenta (and the velocities) are first estimated by treating explicitly the pressure gradient, then they are corrected according to the updated pressure values, as done for instance in~\cite{Xiao2004,Kwatra2009}.
Moreover, at the end of the time step, we recompute the density with the current advection velocity, to have a better accuracy when the Mach number is particularly low. 
The semi-discretization of the governing equations per each phase reads
\begin{align}
\frac{\aph\tnn - \aph\tn}{\Delta t}  + \ui\tn \pder{\aph\tnn}{x}    &=   0\label{ent:a} \\
\frac{\arhoph\tns - \arhoph\tn}{\Delta t} + \pder{(\arhoph\tns \uph\tn)}{x}  &= 0 \label{ent:dF} \\
\frac{\amph\tns - \amph\tn}{\Delta t} + \pder{(\amph\tns \uph\tn + \aph\tnn\Pph\tn)}{x} - \ppi\tn  \pder{\aph\tnn}{x} &= 0 \label{ent:m}\\
 M\reff^2  \aph\tnn \left[\frac{\Pph\tnn - \Pph\tn}{\Delta t}+ \uph\tns\pder{\Pph\tnn}{x} \right] 
  +  \left( M\reff^2  \rhoph c\ph^2 + \kappa\ph \right)\tn \aph\tnn \pder{\uph\tnn}{x}& \notag\\ 
 - \left( M\reff^2 \rho\ph  \cci{\phSymbol} + \kappa\ph\right)\tn (\ui - \uph)\tns\pder{\aph\tnn}{x} &= 0  \label{ent:P}\\
\frac{\amph\tnss - \amph\tns}{\Delta t} + \pder{\left[(\amph\tnss - \amph\tns)\uph\tn\right]}{x}
 + \pder{\left[\aph\tnn(\Pph\tnn- \Pph\tn)\right]}{x} - (\ppi\tnn- \ppi\tn) \pder{\aph\tnn}{x} &= 0\label{ent:dm} \\
\frac{\arhoph\tnn - \arhoph\tn}{\Delta t} + \pder{(\arhoph\tnn \uph\tnn)}{x}  &= 0 \label{ent:dL} \,,
\end{align}
where $\alpha \rho\tns$, $\alpha m\tns = (\alpha \rho)\tns u\tns$ and $u\tns$ are the predicted density, momentum and velocity.
The superscripts $n$ and $n+1$ indicate, as usual, variables at the previous time step, $t\tn$, and at the end of the hyperbolic operator, $t\tnn$. The double star in the momentum correction equation \eqref{ent:dm} highlight that it is related to the predicted density, i.e. $\amph\tnss = (\arhoph)\tns u\tnn$. 

We can interpret the previous set of equations also in the framework of multiple-pressure variables. Indeed, this semi-implicit discretization computes the convective and thermodynamic effects in a predictor step, composed by Eqs.~\eqref{ent:a}--\eqref{ent:m}, while the high-order pressure effects, that is the ones due to the term $P^{(2)}$ in Eq.~\eqref{ep:p_exp}, are corrected implicitly through the Eqs.~\eqref{ent:P} and \eqref{ent:dm}~\cite{Klein1995}.
Moreover, if we sum \eqref{ent:m} and \eqref{ent:dm}, we get the equation for the momentum $\alpha m\tnss$ with implicit pressure gradient, i.e.,
\begin{equation*}
\frac{\amph\tnss - \amph\tn  }{\Delta t} + \pder{(\amph\tnss \uph\tn + \aph\tnn\Pph\tnn)}{x} - \ppi\tnn  \pder{\aph\tnn}{x} = 0 \,.
\end{equation*}
The final momentum is computed after solving \eqref{ent:dL} as
\begin{equation}
\label{ent:mtnn}
\amph\tnn = (\arhoph)\tnn u\tnn =  (\arhoph)\tnn \frac{\amph\tnss}{\arhoph\tns}
\end{equation}

On the other hand, we use a different approach for the density equations \eqref{ent:dF} and \eqref{ent:dL}: the results of the former one, i.e., $\arhoph\tns$, are used only while solving \eqref{ent:m}--\eqref{ent:dm}; but, at the end of the time step, while solving \eqref{ent:dL}, the densities $\arhoph\tnn$ are computed starting from $\arhoph\tn$, discharging $\arhoph\tns$. This re-computation allows the use of the most updated advection velocity, $\uph\tnn$, which is particularly important in flow problems close to the incompressibility limit, where the density equations simplify to transport equations. 
We compare the results obtained with and without density re-computation in the first numerical test, in Sec.~\ref{ss:lmair}.

A final remark concerns the divergence of the velocity in \eqref{ent:P}, which needs to be treated implicitly to overcome acoustic CFL limitations~\cite{Kwatra2009}.
However, since Eqs.~\eqref{ent:a}--\eqref{ent:dL} are solved in a segregate approach in the order they appear, the value of the velocity $u\tnn$ is not known while solving the pressure equation.
Inspired by the use of the momentum equation to derive an implicit pressure equation~\cite{Kwatra2009,Cordier2012}, we use Eq.~\eqref{ent:dm} to approximate the value of $u\tnn$. Indeed, to a first approximation, the difference in the convective terms can be neglected~\cite{Bijl1998}, so the final velocity can be approximated as
\begin{equation}
\uph\tnn =  \uph\tns + \frac{\Delta t}{\arhoph\tns}\left[- \pder{\left[\aph\tnn(\Pph\tnn- \Pph\tn)\right]}{x} + (\ppi\tnn- \ppi\tn) \pder{\aph\tnn}{x}\right] \label{ent:unn} \,.
\end{equation}
As explain better in the following, this choice, and in particular the term $\ppi\tnn = \sum_\phSymbol \aph \Pph\tnn$, couples the pressure equations for all phases together. On the contrary, Eqs.~\eqref{ent:dF},\eqref{ent:m}, \eqref{ent:dm}, and \eqref{ent:dL}, are solved per each phase independently.

\begin{rmk}[Alternative formulation]
\label{rmk:uform}
Considering the definition of $\amph\tnss$ and that $(\arhoph)\tns$ is already known, instead of the momentum correction \eqref{ent:dm}, we could also correct directly the velocity by solving
\begin{equation}
\frac{\uph\tnn - \uph\tns}{\Delta t} + \pder{\left[\uph\tnn - \uph\tns)\uph\tn\right]}{x}
 + \frac{1}{\arhoph\tns}\pder{\left[\aph\tnn(\Pph\tnn- \Pph\tn)\right]}{x} -  \frac{1}{\arhoph\tns}(\ppi\tnn- \ppi\tn) \pder{\aph\tnn}{x} = 0 \,. \label{ent:du} 
\end{equation}
\end{rmk}

\subsection{Variables positioning: primary and staggered grids}\label{ss:variablePosit}
\begin{figure}
\centering\includegraphics[scale=1.2]{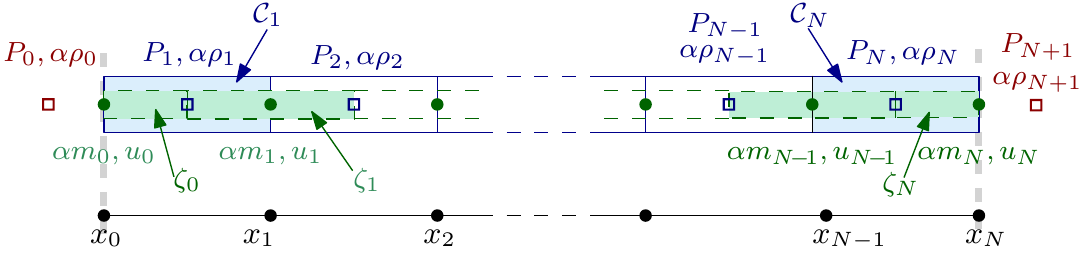}
\caption{Spatial discretization and variables positioning.
At the bottom, the computational domain $\Omega=[x_0,x_N]$ is drawn, along with the position of the grid nodes $x_i$ and the boundary domains sketched with grey dashed lines. The upper part of the picture illustrates the primary and staggered grids. They are drawn with a vertical development and separately from the computational domain only for greater clarity, but all grids and cells here defined should be considered as one-dimensional.
In the upper part of the picture, blue color refers to the primary cells $\mathcal{C}_i$ and to the quantities $\alpha$, $\alpha \rho$, and $P$, which are stored at their centers, represented by blue square marks. Green color refers to the staggered cells, at the center of which the kinetic variables $\alpha m$ and $u$ are stored; these cells are sketched by dashed lines and their centers are represented by green circle marks.
In red, the boundary values for the primary grid complement the picture.
To lighten the notation of variables, phase indication is omitted and the numerical subscripts refer simply to the spatial cells.
}
\label{fns:grids}
\end{figure}

For the spatial discretization of the pressure-based BN-type model, we consider two finite-volume schemes based on staggered grids: one for the thermodynamic variables, and one for the kinematic variables.
\figurename~\ref{fns:grids} shows how the staggered grids are defined. We split the computational domain $\Omega=[x_0,x_N]$ in $N$ intervals, defined by the equidistant grid nodes $x_i$, with $i=0, \dots, N$. Then, the grid for the finite-volume discretization of the thermodynamic quantities (hereafter called \textit{primary grid}) is built by defining each cell $\mathcal{C}_i$ corresponding to the grid element $[x_{i-1},x_i]$.
Conversely, the grid for the finite-volume discretization of the kinematic variables (hereafter called \textit{staggered grid}) is built by centering each cell $\zeta_k$ around the grid nodes $x_k$, between the centroids of the adjacent grid elements (or the boundary). 
In summary, the cells on primary and staggered grids are defined as
\begin{equation*}\begin{array}{ll@{\,=\,}lll}
\text{primary:} & \mathcal{C}_i & \,[x_{i-1},x_i] & \forall i =1,\dots, N; & \\
\text{staggered:} & \zeta_k & \left[\dfrac{x_k+x_{k-1}}{2},\dfrac{x_{k+1}+x_{k}}{2}\right] & \forall k = 1,\dots, N-1;\\
&\zeta_0&\multicolumn{2}{l}{\left[x_0,\dfrac{x_{1}+x_{0}}{2}\right]\;;   \hfill \zeta_N=\left[\dfrac{x_{N}+x_{N-1}}{2}, x_N\right]}.
\end{array}\end{equation*}
As it appears from the given definitions, starting from grid nodes equally spaced by a distance $\Delta x$, all primary cells have the same size $\volc{i} = \Delta x$, while the staggered cells have the same size  $\volz{k}= \Delta x$ only far from boundary. Indeed, the first and last staggered cells are half the size, i.e., $\volz{0} = \volz{N} =\Delta x / 2$.

A cell-centered finite-volume discretization over the primary grid is used to solve the volume fraction, density, and pressure equations, that is Eqs.~\eqref{ent:a}, \eqref{ent:dF}, \eqref{ent:dL}, and \eqref{ent:P}. So, the thermodynamic variables (sometimes called ``scalar'' in contrast to the kinematic, vectorial variables) are approximated over the cell $\mathcal{C}_i$ as
\begin{equation*}
(\xi\ph)_i\tn = \frac{1}{\volc{i}} \int_{\mathcal{C}_i} \xi\ph(x,t\tn) \mathrm{d}x \,, \qquad \text{with}\quad \xi\in \lbrace \aph, \arho\ph, \Pph \rbrace \,.
\end{equation*}
On the other hand, the momentum and momentum update equations, \eqref{ent:m} and \eqref{ent:dm}, are discretized over the staggered grid. Using a finite-volume scheme, we define the cell value of the momentum as
\begin{equation*}
(\amph)_k\tn = \frac{1}{\volz{k}} \int_{\zeta_k} \amph(x,t\tn) \mathrm{d}x \,.
\end{equation*}
These are the finite-volume cell values, illustrated also in Fig.~\ref{fns:grids}. However, it is often required to map variables from ``their'' grid to the other.
In this case, we perform a weighted average, which, being the grid nodes equidistant, simply results in an arithmetic mean.

\begin{rmk}[Notation and mapping]\label{rmk:mapping}
To have a clear notation, we use the subscript $i$ for quantities over the primary grid, and $k$ for quantities over the staggered one.
Accordingly, a thermodynamic variable with a subscript $k$ refers to its mapped value over the staggered cell $k$, and vice versa.
To clarify this point, consider the following example. The notation $(\arho\ph)_k$ indicates the mapped density over the cell $\zeta_k$, computed as $(\arhoph)_k = [(\arhoph)_{i} + (\arhoph)_{i+1} ]/2$ for $i=k$.
We can then use this mapped density to estimate the velocity in the cell $\zeta_k$ as $(\uph)_k = (\amph)_k / (\arhoph)_k $.
\end{rmk}

\subsection{Spatial discretization of the hyperbolic operator}\label{ss:spacedis}
Each hyperbolic differential equation in the model is integrated in time for the interval $\Delta t = t\tnn - t\tn$ and in space over all cells ($\mathcal{C}_j$ or $\zeta_k$).
The spatial derivative of the convective fluxes is approximated through numerical evaluations of the fluxes at the cell interfaces. In particular, we use a first-order approximation based on the Rusanov flux, as in~\cite{Saurel1999,Saurel2017}. This choice is motivated by simplicity, as it avoids the complexities related to the solution of local Riemann problems with several waves~\cite{Andrianov2004,Schwendeman2006,Deledicque2007}.

The use of staggered grids makes the discretization of some specific terms easy and natural. 
\begin{itemize}
\item The convective velocity to be used in the flux computation on the primary grid is directly the velocity defined over the staggered grid. For instance, the Rusanov flux for $\arhoph$ at the interface between $\mathcal{C}_i$ and $\mathcal{C}_{i+1}$ is computed as
\begin{equation}
\label{ens:rus_ad}
F\Rus\iph(\arhoph, \uph)
= \frac{1}{2} (\uph)_k \left[ (\arhoph)\ipo + (\arhoph)_i  \right] 
- \frac{1}{2} \vert (\uph)_k \vert \left[ (\arhoph)\ipo - (\arhoph)_i  \right] \,.
\end{equation}
\item The pressure gradient in the momentum equation is readily approximated by a centered difference scheme, since the values of the pressure at the faces of the staggered cells are available:
\begin{equation}\label{ens:p_grad}
\int_{\zeta_k} \pder{\Pph}{x} \mathrm{d}x \approx (\Pph)\ipo - (\Pph)_i \,.
\end{equation}
\item The divergence of the velocity in the pressure equation is easily discretized through a centered difference scheme:
\begin{equation}\label{ens:u_div}
\int_{\mathcal{C}_i} \pder{\uph}{x} \mathrm{d}x \approx (\uph)_k - (\uph)\kmo \,.
\end{equation}
\end{itemize}

A major complexity of the spatial discretization of Eqs.~\eqref{ent:a}--\eqref{ent:dL} concerns the presence of non\hyp{}conservative terms involving the gradient of the volume fraction. This is a challenge common to all BN-type models, which include the term $\mathbf{B}(\mathbf{U}) \pder{\alpha_1}{x}$ that models the momentum and energy transfer among phases but prevents to write Eq.~\eqref{em:bn} in divergence form.
This means that it is not possible to define weak solutions in the standard sense of distribution and to determine unique wave speeds.
From a numerical point of view, these non-conservative products have to be integrated as source terms, rather than as fluxes.
Since a naive discretization may introduce spurious oscillations across material interfaces between phases with different specific heat ratios,
we seek a robust discretization of non-conservative terms involving the volume fraction gradient by explicitly enforcing that uniform velocity and pressure profiles are maintained~\cite{Abgrall1996}.
Honestly, different strategies can be followed to integrate the non-conservative terms associated to the linearly degenerate fields, as, in particular, path conservative schemes~\cite{Pares2006}. However, this approach does not guarantee to always converge to the correct weak solution of non-conservative hyperbolic problems~\cite{Abgrall2010}. In addition, a primitive formulation of the governing equations, such as the pressure one here considered, facilitates preserving pressure equilibrium near material interfaces~\cite{Karni1996}.
All in all, for weak discontinuities, as the ones considered in the framework of weakly compressible flows, any consistent and accurate enough method would be adequate to achieve a satisfactory solution~\cite{Toro2003}.

\subsubsection{Volume fraction and density equations}
Without any non-conservative term, the density equations~\eqref{ent:dF} and \eqref{ent:dL} are easily discretized in space as
\begin{equation}\label{ens:d} 
\frac{\volc{i}}{\Delta t} \left[(\arhoph)_i^\diamond - (\arhoph)_i\tn \right]
 = - \left[ F\Rus\iph(\arhoph^\diamond, \uph\tn) - F\Rus\imh(\arhoph^\diamond, \uph\tn) \right]
\end{equation}
where the expression for the Rusanov fluxes is given in Eq.~\eqref{ens:rus_ad}, and the superscript $\diamond$ corresponds to ${n*}$ and to ${n+1}$ in the spatial discretization of \eqref{ent:dF} and \eqref{ent:dL}, respectively.

The discrete volume fraction equation is
\begin{equation}\label{ens:alpha}
\frac{\volc{i}}{\Delta t} \left[(\aph)_i\tnn - (\aph)_i\tn \right] = - H_u(\aph\tnn, \ui\tn)_i \,,
\end{equation}
where $ H_u(\aph\tnn, \ui\tn)_i \approx \int_{\mathcal{C}_i} \ui\tn \pder{\aph\tnn}{x} \mathrm{d}x $ is a suitable approximation of the non-conservative term.
To define this operator, we follow the idea that starting from a uniform pressure and velocity, no variations in these variables should be generated~\cite{Abgrall1996,Saurel1999,Saurel2017}, see also Remark~\ref{rmk:uniform}.

If we assume a uniform velocity field, e.g. $(\uph)_k = (\uph)\kpo = (\ui) = u$, the discrete mass equation reads 
\begin{equation}\label{ens:ad_uunif}
\frac{\volc{i}}{\Delta t} \left[(\arhoph)_i\tnn - (\arhoph)_i\tn \right]
 = - \frac{1}{2}u\left[ (\arhoph)\ipo - (\arhoph)\imo  \right] 
 +\frac{1}{2}\vert u \vert \left[ (\arhoph)\ipo - 2(\arhoph)_i   + (\arhoph)\imo \right] \,,
\end{equation}
where we have dropped the superscripts in the left hand side to lighten the notation.
Let us consider now the special case when also the density field is uniform~\cite{Saurel2017}. If $(\rhoph)_i = (\rhoph)\imo = (\rhoph)\ipo$, the mass equation reads
\begin{equation}\label{ens:ad_dunif}
\frac{\volc{i}}{\Delta t} \bigg[(\arhoph)_i\tnn - (\arhoph)_i\tn \bigg]=
 - (\rhoph) \frac{u}{2} \bigg[ (\aph)\ipo - (\aph)\imo  \bigg]
 + (\rhoph) \frac{\vert u \vert }{2} \bigg[ (\aph)\ipo - 2(\aph)_i   + (\aph)\imo \bigg] \,.
\end{equation}
If velocity and density are uniform, the density should remain constant, i.e., $(\rhoph)_i\tnn=(\rhoph)_i\tn$. So, in order to make Eq.~\eqref{ens:alpha} compatible with Eq.~\eqref{ens:ad_dunif} in this specific case, we need that
\[ H_u(\aph\tnn, \ui\tn)_i =
 \frac{u}{2} \bigg[ (\aph)\ipo - (\aph)\imo  \bigg]
 - \frac{\vert u \vert }{2} \bigg[ (\aph)\ipo - 2(\aph)_i   + (\aph)\imo \bigg] \,.
\]
From this, we define the following non-conservative operator $H_u$:
\begin{equation}\label{ens:Hu}
H_u(\aph\tnn, \ui\tn)_i =
     \hat{F}\Rus\iph\left(\aph\tnn, (\ui)_i\tn\right) - \hat{F}\Rus\imh\left(\aph\tnn, (\ui)_i\tn\right) \,, 
\end{equation}
where $ \; \hat{F}\Rus\iph\left(\aph\tnn, (\ui)_i\tn\right) =  \frac{1}{2} (\ui)_i\tn \left[ (\aph)\ipo\tnn + (\aph)_i\tnn  \right] 
- \frac{1}{2} \vert (\ui)_i\tn\vert \left[ (\aph)\ipo\tnn - (\aph)_i\tnn  \right] $.
We use the  notation $\hat{F}\Rus$ to highlight that that the resulting discretization of $H_u$ depends on the discretization for the convective flux in the mass equation, but, at the same time, the $\hat{\cdot}$ indicates that it is not a proper flux, as $(\ui)_i$ is the mapping of the interface velocity over the primary cell $\mathcal{C}_i$, not an interface velocity. This choice guarantees that $H_u =0 $, if the volume fraction is uniform, as expected by the integration of $\ui \pder{\aph}{x}$.

\subsubsection{Momentum equations}
The spatial discretization of the momentum equations \eqref{ent:m} and \eqref{ent:dm} requires the integration of three terms: the convective flux, for which we adopt a Rusanov flux; the pressure gradient, discretized by the central finite difference defined in~\eqref{ens:p_grad}; and the non-conservative term, for which we define the operator $ H_P(\aph\tnn, \ppi\tn)_k \approx \int_{\zeta_k} \ppi\tn \pder{\aph\tnn}{x} \mathrm{d}x $ exploiting the non-disturbance pressure and velocity condition, as explained in the following.
Accordingly, the discrete equations of the predicted momentum and of the corrected momentum read
\begin{align}\begin{split}
\frac{\volz{k}}{\Delta t} \bigg[(\amph)_k\tns - (\amph)_k\tn \bigg]
 = &- \left[ F\Rus\kph\left( \amph\tns, \uph\tn \right) - F\Rus\kmh\left( \amph\tns, \uph\tn \right)\right]\\
   &- \left[(\aph)\ipo\tnn (\Pph)\ipo\tn - (\aph)_i\tnn (\Pph)_i\tn \right] + H_P(\aph\tnn, \ppi\tn)_k \label{ens:m}
\end{split}\\
\begin{split}
\frac{\volz{k}}{\Delta t} \bigg[ (\amph)_k\tnss - (\amph)_k\tns \bigg]
 = &- \left[ F\Rus\kph\left( \delta \amph\tnss, \uph\tn \right) - F\Rus\kmh\left( \delta \amph\tnss, \uph\tn \right)\right]\\
   &- \left[(\aph)\ipo\tnn (\delta \Pph)\ipo\tnn - (\aph)_i\tnn (\delta \Pph)_i\tnn \right] + H_P(\aph\tnn, \delta\ppi\tnn)_k \label{ens:dm}
\end{split}\end{align}
where we have used the operator $\delta$ to identify the jump in the kinetic and pressure variables between the prediction and correction step. More precisely,
$$\delta \amph\tnss = \amph\tnss - \amph\tns = \arhoph\tns \delta u\tnn = \arhoph\tnn (\uph\tn -\uph\tns)\,,$$
$$\delta \Pph\tnn = \Pph\tnn - \Pph\tn\,, \qquad \text{and} \qquad
\delta \ppi\tnn = \ppi\tnn - \ppi\tn\,.$$
The Rusanov fluxes are defined, as usual, as
\begin{equation}\label{ens:rus_m}
F\Rus\kph\left( \amph\tns, \uph\tn \right)  = \frac{1}{2} \left[ (\amph)\kpo\tns (\uph)\kpo\tn +  (\amph)_k\tns (\uph)_k\tn \right]  
- \frac{1}{2} S\kph\tn     \left[ (\amph)\kpo\tns -  (\amph)_k\tns  \right] 
\end{equation}
where $S\kph\tn = \max \left( \left\vert (\uph)\kpo\tn \right\vert, \left\vert (\uph)_k\tn  \right\vert \right)$.
The same expression but with $(\delta \amph)\tnss$ instead of $(\amph)\tns$ is used for $F\Rus\kph\left( \delta \amph\tnss, \uph\tn \right)$.

The discretization of the non-conservative term $ H_P(\aph, \ppi)_k$ is derived, similarly to $H_u(\aph, \ui)_i$,  by imposing the non-disturbance  pressure and velocity constraint. The whole process is detailed in~\ref{a:Hp}. For conciseness, we report here only the final definition of the non-conservative operator $H_P$:
\begin{align}
 H_P(\aph\tnn, \ppi\tn)_k &= (\ppi)_k\tn \left[(\aph)\ipo\tnn - (\aph)_i\tnn  \right] \,, \\
 H_P(\aph\tnn, \delta\ppi\tn)_k &= \left[ (\ppi)_k\tnn - (\ppi)_k\tn \right] \left[(\aph)\ipo\tnn - (\aph)_i\tnn \right] \,,
\end{align}
where $(\ppi)_k= \frac{1}{2}\left[ (\ppi)_i + (\ppi)\ipo \right]$ is the interface pressure mapped at the staggered cell $\zeta{k}$.

Finally, we highlight that Eqs.~\eqref{ens:m} and \eqref{ens:dm} are solved only over the internal cells $\zeta_k$, with $k=1,\dots,N-1$; while the values of the predicted and updated momentum on $\zeta_0$ and $\zeta_N$ are imposed through the boundary treatment described in Sec.~\ref{s:boundarycondition}.

\subsubsection{Velocity correction equation}
If we consider the velocity correction equation \eqref{ent:du},
its discretization is straightforwardly derived from the spatially discrete equation of momentum correction \eqref{ens:dm} and it reads
\begin{multline}\label{ens:du}
\frac{\volz{k}}{\Delta t} \bigg[ (\uph)_k\tnn - (\uph)_k\tns \bigg]
 = - \left[ F\Rus\kph\left( \delta \uph\tnss, \uph\tn \right) - F\Rus\kmh\left( \delta \uph\tnss, \uph\tn \right)\right]\\
   - \dfrac{1}{(\arhoph)_k\tns}\left[(\aph)\ipo\tnn (\delta \Pph)\ipo\tnn 
   - (\aph)_i\tnn (\delta \Pph)_i\tnn \right] 
   + \dfrac{1}{(\arhoph)_k\tns} H_P(\aph\tnn, \delta\ppi\tnn)_k \;, \quad
\end{multline}
where the Rusanov fluxes are defined, similarly to \eqref{ens:rus_m}, as
\[ F\Rus\kph\left( \uph\tnn, \uph\tn \right)  = \frac{1}{2} \left[ (\uph)\kpo\tnn (\uph)\kpo\tn +  (\uph)_k\tnn (\uph)_k\tn \right]  
- \frac{1}{2} S\kph\tn  \left[ (\amph)\kpo\tns -  (\amph)_k\tns  \right] \,. \]

\subsubsection{Pressure equation}
We develop now the discrete version of the non-conservative pressure equation~\eqref{ent:P}. First, we observe that, in the considered finite volume context, the thermodynamic variables and the volume fraction are constant within the primary cell, as in~\cite{Bijl1998}. So, integrating Eq.~\eqref{ent:P} over a cell $\mathcal{C}_i$, we can write
\begin{multline}\label{ens:P1}
M\reff^2  (\aph)_i\tnn \frac{\volc{i}}{\Delta t} \left[ (\Pph)_i\tnn - (\Pph)_i\tn\right] 
+ M\reff^2  (\aph)_i\tnn \int_{\mathcal{C}_i}\!\!\uph\tns\pder{\Pph\tnn}{x} \mathrm{d}x \\
+  (\KFph)_i\tn (\aph)_i\tnn \int_{\mathcal{C}_i}\!\!\pder{\uph\tnn}{x} \mathrm{d}x  
 \; - \;(\KHph)_i\tn \int_{\mathcal{C}_i}\!\! (\ui - \uph)\tns\pder{\aph\tnn}{x} \mathrm{d}x = 0\,, \quad
\end{multline}
where, to have a more compact expression, we have introduced the two coefficients
\[  (\KFph)_i\tn =  M\reff^2  (\rhoph)_i\tn (c\ph^2)_i\tn + (\kappa\ph)_i\tn \, , \qquad\qquad
 (\KHph)_i\tn =  M\reff^2  (\rhoph)_i\tn (\cci{\phSymbol})_i\tn + (\kappa\ph)_i\tn  \,,
\]
which are known, because the variables $(c\ph^2)_i\tn$, $(\cci{\phSymbol})_i\tn$, and $(\kappa\ph)_i\tn $ are  computed using the thermodynamic state at cell $\mathcal{C}_i$ and at time $t\tn$.
For instance, $(\kappa\ph)_i\tn = \kappa\big( (\rhoph)_i\tn, (\eph)_i\tn \big)$, according to definition~\eqref{emd:tmdder}.

The second step concerns the discretization of the first integral term, which, thanks to the product rule, is re-written as
\[ \int_{\mathcal{C}_i}\!\!\uph\tns\pder{\Pph\tnn}{x} \mathrm{d}x 
= \int_{\mathcal{C}_i}\!\!\pder{(\Pph)\tnn (\uph)\tns}{x} \mathrm{d}x  - \int_{\mathcal{C}_i}\!\!(\Pph)\tnn\pder{\uph\tns}{x} \mathrm{d}x  \,.\]
To approximate the first term in the previous expression, we define the following flux (similar to~\eqref{ens:rus_ad})
\[ F\Rus\iph\left( \Pph\tnn, \uph\tns \right) = \frac{1}{2} (\uph)_k\tns \left[ (\Pph)\ipo\tnn + (\Pph)_i\tnn \right]
 - \frac{1}{2} \big\vert (\uph)_k\tns \big\vert \left[ (\Pph)\ipo\tnn -  (\Pph)_i\tnn \right] \,,\]
while for the second one, we rely on the central approximation scheme for the divergence of the velocity given in~\eqref{ens:u_div}. We obtain:
\begin{multline*}
\int_{\mathcal{C}_i}\!\!\uph\tns\pder{\Pph\tnn}{x} \mathrm{d}x 
\approx F\Rus\iph\left( \Pph\tnn, \uph\tns \right) - F\Rus\imh\left( \Pph\tnn, \uph\tns \right)
 - (\Pph)_i\tnn \big[ (\uph)_k\tns - (\uph)\kmo\tns\big]\\
= \frac{1}{2} \left[ (\Pph)\ipo\tnn - (\Pph)_i\tnn \right]   \left[(\uph)_k\tns - \big\vert (\uph)_k\tns \big\vert \right] 
 + \frac{1}{2}  \left[ (\Pph)_i\tnn - (\Pph)\imo\tnn \right] \left[ (\uph)\kmo\tns + \big\vert (\uph)\kmo\tns \big\vert \right] \,.
\end{multline*} 

A third aspect to be considered is the approximation of the non-conservative term involving the gradient of the volume fraction. Given the similarities with the non-conservative term in the volume fraction equation, we adopt the same operator $H_u$ defined in~\eqref{ens:Hu}, but for the velocity jump.
Thus,
\[ \int_{\mathcal{C}_i}\!\! (\ui - \uph)\tns\pder{\aph\tnn}{x} \mathrm{d}x  \approx H_u\big(\aph\tnn, (\ui - \uph)\tns\big)_i \,,\]
with $ \displaystyle 
H_u\big(\aph\tnn, (\ui - \uph)\tns\big)_i =
     \hat{F}\Rus\iph\left(\aph\tnn, \left((\ui)_i\tns - (\uph)_i\tns\right) \right) - \hat{F}\Rus\imh\left(\aph\tnn, \big((\ui)_i\tns - (\uph)_i\tns\big) \right) $.

The remaining integral term in Eq.~\eqref{ens:P1} is easily approximated by a central difference scheme, but it requires an expression for the  velocities at the time step $t\tnn$. This latter is derived from the discretization of the velocity update, Eq.~\eqref{ens:du}, discharging the differences in the convective terms. It reads
\begin{equation} \label{ens:unn}
(\uph)_k\tnn =  (\uph)_k\tns + \frac{\Delta t}{\volz{k} \, (\arhoph)_k\tns} \Big[- \left[(\aph)\ipo\tnn (\delta \Pph)\ipo\tnn - (\aph)_i\tnn (\delta \Pph)_i\tnn \right] + H_P(\aph\tnn, \delta\ppi\tnn)_k \Big]\,.
\end{equation}

In conclusion, the discrete version of the pressure equation is
\begin{multline}\label{ens:P}
M\reff^2  (\aph)_i\tnn \frac{\volc{i}}{\Delta t} \left[ (\Pph)_i\tnn - (\Pph)_i\tn\right] \\
= - M\reff^2  (\aph)_i\tnn \frac{1}{2} \Bigl\lbrace 
\left[ (\Pph)\ipo\tnn - (\Pph)_i\tnn \right]   \left[(\uph)_k\tns - \big\vert (\uph)_k\tns \big\vert \right] 
 +  \left[ (\Pph)_i\tnn - (\Pph)\imo\tnn \right] \left[ (\uph)\kmo\tns + \big\vert (\uph)\kmo\tns \big\vert \right] \Bigr\rbrace\\
-  (\KFph)_i\tn (\aph)_i\tnn \Big[(\uph)_k\tnn -(\uph)\kmo\tnn \Big]
 +(\KHph)_i\tn H_u\big(\aph\tnn, (\ui - \uph)\tns\big)_i \,.
\end{multline}

\begin{rmk}[Equation coupling]
The implicit treatment of the velocity divergence in the pressure equation determines the coupling of the discrete pressure equations for both phases.
In~\eqref{ens:P}, the velocities $(\uph)_k\tnn$ and $(\uph)\kmo\tnn$  depend on $(\delta\ppi)_k\tnn $ and $(\delta\ppi)\kmo\tnn $ (cfr. Eqs.~\eqref{ent:unn} and\eqref{ens:unn}).
Recalling the definition of $(\ppi)$ and the mapping from the primary to the staggered, we have
\[ (\ppi)_k\tnn 
= \frac{1}{2}\bigl[  (\ppi)\ipo\tnn +  (\ppi)_i\tnn \big]
= \frac{1}{2} \sum_\phSymbol \left[ (\aph)\ipo\tnn (\Pph)\ipo\tnn +  (\aph)_i\tnn (\Pph)_i\tnn\right]\,, \]
from which it appears evident the involvement of the pressure of both phases in the definition of the velocity $(\uph)\tnn$.
Consequently, we need to solve the pressure equations~\eqref{ens:P} for both phase together, i.e., in a coupled way.
\end{rmk}

\subsubsection{Boundary conditions}\label{s:boundarycondition}
To impose boundary conditions, we distinguish between primary and staggered grid.
For the primary grid, we use a standard method based on two  ghost states defined outside the computational domain. With reference to Fig.~\ref{fns:grids}, these states are denoted by subscripts $0$ and $N+1$, on the left and right boundary, respectively, and are defined as
\[ \left(\mathbf{W}\ph\right)_\mathrm{B}\tnn = \begin{bmatrix}
(\aph)_\mathrm{B}\tnn \\ 
(\arhoph)_\mathrm{B}\tnn \\ 
(\Pph)_\mathrm{B}\tnn
\end{bmatrix} \, \qquad \text{for} \; \phSymbol=\{1,2\} \,, \; \text{and} \; \mathrm{B}=\{0,N+1\} \,.
\]
According to the physical boundary condition we need to model, the value of the variables in $\left(\mathbf{W}\ph\right)_\mathrm{B}\tnn$ mirrors the state of the adjacent internal cell ($\mathcal{C}_1$ or $\mathcal{C}_N$), or it is directly imposed as boundary value (for the details about this selection process, see for instance~\cite{Bermudez2016}).
The boundary state $\left(\mathbf{W}\ph\right)_\mathrm{B}\tnn$ is then used in the discrete equations~\eqref{ens:d}, \eqref{ens:alpha}, \eqref{ens:m},\eqref{ens:P}, and \eqref{ens:dm} to evaluate the fluxes, the non-conservative terms, and the central difference schemes at the boundary interfaces.

For the staggered grid, we use a different strategy, because the first and the last staggered cells ($\zeta_0$ and $\zeta_N$) are boundary cells.
In addition, the velocities $(\uph)_0$ and $(\uph)_N$  are already stored at the boundary interfaces (see Fig.~\ref{fns:grids}).
Thus, the momentum and velocity in these two cells are not computed by solving Eqs.~\eqref{ens:m} and \eqref{ens:dm}, but they are computing according to the physical boundary condition. In particular, we distinguish two cases: if the boundary velocity $(\uph)_\mathrm{B}$ is known, its value is imposed; otherwise the velocity is extrapolated from the two closest internal cells. For instance, considering the left boundary:
\[
(\uph)_0 = \begin{cases}
(\uph)_\mathrm{B} & \text{if}\; \uph \; \text{known at}\; x_0\,,\\
2(\uph)_1 - (\uph)_2 & \text{otherwise.}
\end{cases} \quad \text{Then} \; 
(\amph)_0 = \frac{(\arhoph)_0 +(\arhoph)_1}{2}(\uph)_0 \,,
\]
where $(\arhoph)_0$, $(\arhoph)_1$ are the density values on the primary cells. This definition applies also to the implicit velocity in the pressure equations, i.e., while solving the equation~\eqref{ens:P} for $(\Pph)_1\tnn$ and $(\Pph)_N)$ the expressions for the velocity $(\uph)_0\tnn$ and $(\uph)_{N}\tnn$ are the ones given above, instead of~\eqref{ens:unn}.

\subsubsection{Solution of implicit system}
The implicit treatment of some terms in the discretization of the hyperbolic operator makes the equations coupled between adjacent cells. Indeed, the structure of mass, volume fraction, and momentum equations, e.g.\eqref{ens:d}, \eqref{ens:alpha}, \eqref{ens:m}, and \eqref{ens:dm}, can be approximately represented as
\begin{equation}
\frac{\Delta x_j}{\Delta t} \bigg[ w_j\tnn - w_j\tn \bigg]
 = - \left[ F_\mathrm{R}\left(w_j\tnn, w\jpo\tnn\right) - F_\mathrm{L}\left(w_j\tnn,  w\jmo\tnn \right) \right] 
  - \left[ D_\mathrm{R} - D_\mathrm{L} \right] 
   + H(\alpha\tnn)\;,
\end{equation}
where $w$ is the unknown of a specific phase in the cell $j$ (over the primary or staggered grid), $\Delta x_j$ is the cell volume, $F_\mathrm{L}$ and $F_\mathrm{R}$ are fluxes across the left and right cell face,  $D_\mathrm{R}$ and $D_\mathrm{L}$ refer to the cell centered discretization and the term $H$ represents the discretization of the non-conservative terms (involving different values of $\alpha\tnn$ according to the equation we are considering). The superscripts $n$ and $n+1$ indicates, generally, known and unknown values, respectively. 
Obviously, not all right hand side terms are present in every equation, but there is always at least one term that generates the cross-coupling.

The previous expression can be further simplified as
\begin{equation}
\frac{\Delta x_j}{\Delta t} \bigg[ w_j\tnn - w_j\tn \bigg]
 = - \Phi_j\left(w_j\tnn, w\jpo\tnn, w\jmo\tnn \right) + \mathrm{R}_j(w)
\end{equation}
where $\Phi$ includes the fluxes or the non-conservative terms that are function of the unknowns themselves and $\mathrm{R}$ includes all the known terms.
We use a first order Taylor expansion to approximate $\Phi$ as
\[
 \Phi\tnn =  \Phi\left(w_j\tnn, w\jpo\tnn, w\jmo\tnn \right) \approx
 \Phi\tn + \pder{\Phi}{w_j} \delta w_j\tnn 
  + \pder{\Phi}{w\jpo} \delta w\jpo\tnn  
  + \pder{\Phi}{w\jmo} \delta w\jmo\tnn \,,
\]
where $\delta w_j\tnn = w_j\tnn - w_j\tn$. Since we use Rusanov fluxes, the derivatives of fluxes and the non-conservative term (required only in the volume fraction equation) can be easily computed analytically.
Hence, for Eqs.~\eqref{ens:d}, \eqref{ens:alpha}, \eqref{ens:m}, and \eqref{ens:dm}, for each phase separately, we need to solve a set of equations in the form
\begin{equation}
\left[ \frac{\Delta x_j}{\Delta t} + \pder{\Phi}{w_j} \right] \delta w_j\tnn 
+ \pder{\Phi}{w\jpo} \delta w\jpo\tnn  
+ \pder{\Phi}{w\jmo} \delta w\jmo\tnn
 = \Phi_j\tn + \mathrm{R}_j(w) \, ,
\end{equation}
which is comprised of $N$ or $N+1$ equations, depending on whether we are considering the primary or the staggered grid.
The resulting systems are linear and they can be written, in a compact form, as $[A] \boldsymbol \delta \boldsymbol w = \mathbf{R}$, where $[A]$ is a tridiagonal matrix including the derivatives of $\Phi$, $ \delta \boldsymbol w $ is the vector of unknown, and $\mathbf{R}$ is the known term.
These systems are solved through the Generalized Minimal Residual (GMRES) algorithm provided by the PETSc library~\cite{petsc-web-page}.

Similar observations can be drawn also for the pressure equation~\eqref{ens:P}, which however is solved for both phases together. In this case, the generalized term $\Phi_P$ includes also the unknown terms deriving from \eqref{ens:unn}, that is
\[ (\Phi_P)_i\tnn =  \Phi\left((P_1)_i\tnn, (P_1)\ipo\tnn, (P_1)\imo\tnn, (P_2)_i\tnn, (P_2)\ipo\tnn, (P_2)\imo\tnn \right)\,. \]
Consequently, the first-order Taylor expansion involves six different unknowns, which can be organized in a vector in this order: $\left[ \dots, (\delta P_1)\imo, (\delta P_2)\imo, (\delta P_1)_i, (\delta P_2)_i, (\delta P_1)\ipo, (\delta P_2)\ipo  ,\dots\right]$, so that the resulting final linear system can be written as  $[A] \boldsymbol \delta \boldsymbol w = \mathbf{R}$, where $[A]$ is now a banded matrix with an upper bandwidth of 3 and a lower bandwidth of 2.

\subsection{Relaxation operator}\label{ss:Lrelax}
According to the Strang splitting introduced in \eqref{en:split}, 
the solution of the hyperbolic operator described in Secs.~\ref{ss:timedis}--\ref{ss:spacedis} provides a known set of variables that are used as initial data to solve the system of ODEs associated with the relaxation terms.
For this reason, in this section, we re-define the notation to distinguish the intermediate solutions after the hyperbolic operator $L_\mathrm{hyp}$, and the relaxation operator $L_\mathrm{relax}$ as follows
\begin{equation}
\label{enss:notation}
\mathbf{U}\thyp = L_\mathrm{hyp}(\mathbf{U}\tn) \quad \text{and}\qquad
\mathbf{U}\tsou = L_\mathrm{relax}(\mathbf{U}\thyp)\,.
\end{equation}
In practice, in this subsection the superscript $\thypSymbol$ denotes what in subsections~\ref{ss:timedis} and \ref{ss:spacedis} was denoted by $n+1$, and the  superscript $\tsouSymbol$ refers to the variables computed during the relaxation processes.

The relaxation operator $L_\mathrm{relax}$ plays a fundamental role in driving phasic velocities and pressures toward the equilibrium, close to interfaces.
The characteristic time of these processes depend on many factors, as the fluids features and the multiphase flow topology. For instance, the parameter $\mu$, which expresses the velocity of the pressure relaxation, may depend on the compressibility of the fluid and the parameter $\lambda$, which governs the rate of the velocity homogenization, may depend on fluid viscosity~\cite{Saurel1999}.
In general, pressure and velocity relaxation are much faster than the dynamics associated to the wave propagation, to the point that they are something modeled as instantaneous phenomena, by assuming infinite $\mu$ and $\lambda$~\cite{Saurel1999,Zein2010,Pelanti2019}.
However, in this work, we use finite relaxation parameters, as in~\cite{Saurel2007,DumbserBoscheri2013}, to allow wider modeling possibilities.
Indeed, we could define the relaxation parameters in terms of the average interfacial area of bubbles~\cite{Abgrall2003}, or, if we had experimental data about different multiphase flow topologies, we could tune the relaxation parameters in our model to match the data.

Assuming a characteristic time much shorter than the one characterizing the hyperbolic operator,
the ODE system associated to the relaxation operator is derive
from the continuous governing equations~\eqref{emd:fin_alpha}--\eqref{emd:fin_P} neglecting convective and transport terms.
It reads
\begin{align}
\oder{\alpha_1}{t}  &=   \mu\Delta_1 P \label{enns:alpha}\\
\oder{\arhoph}{t}  &= 0  && \text{for}\; \phSymbol = \left\lbrace 1,2 \right\rbrace \label{enns:rho}\\
\oder{\amph}{t}  &=  -  \lambda \Delta\ph u && \text{for}\; \phSymbol = \left\lbrace 1,2 \right\rbrace \label{enns:m} \\
 M\reff^2  \alpha\ph\oder{\Pph}{t} &=
   - \left[ M\reff^2  \rho\ph \cci{\phSymbol}  +  \kappa\ph \right]\mu \Delta\ph P 
   -  M\reff^2 \kappa\ph (\ui - \uph) \lambda \Delta\ph u  && \text{for}\; \phSymbol = \left\lbrace 1,2 \right\rbrace \,.\label{enns:P}
\end{align}
This system is characterized by a high degree of stiffness, so we use the implicit Backward Euler scheme for the time integration.
Equations~\eqref{enns:rho} give immediately $\arhoph\tnn = \arhoph\thyp$. If we use this result in~\eqref{enns:m} and we integrate in time, we have
\begin{align*}
\arho_1\tsou\frac{u_1\tsou - u_1\thyp}{\Delta t}  &=  -  \lambda \left( u_1\tsou -u_2\tsou  \right)\\
\arho_2\tsou\frac{u_2\tsou - u_2\thyp}{\Delta t}  &=  +  \lambda \left( u_1\tsou -u_2\tsou  \right) \,,
\end{align*}
where the only unknowns are the velocities. The solution of this system, expressed in term of $\Delta u$, is
\begin{equation}\label{enns:u}
\Delta u = \left( u_1\tsou -u_2\tsou  \right) = \left( u_1\thyp -u_2\thyp  \right)\Big / 
\left[ 1 + \lambda \Delta t \dfrac{\arho_1\tsou + \arho_2\tsou}{\arho_1\tsou  \arho_2\tsou} \right]
\end{equation}
which, as expected, gives $\Delta u  \rightarrow 0$ when $\lambda \rightarrow \infty$, so that $u_1\tsou=u_2\tsou= \frac{\arho_1\tsou u_1\thyp + \arho_2\tsou u_2\thyp}{\arho_1\tsou +  \arho_2\tsou} = \ui\tsou$.
In the opposite case, for $\lambda = 0$, we have  $\uph\tsou=\uph\thyp$.

The remaining part in the ODE system comprises the volume fraction equation \eqref{enns:alpha} and the two pressure equations \eqref{enns:P}. After the discretization of the time derivatives, these equations can be re-written as
\begin{align}
\alpha_1\tsou - \alpha_1\thyp  +  \mu\Delta t  \, \Delta P &=0 \label{enns:Prelax1}\\
 M\reff^2  \alpha_1\tsou (P_1\tsou - P_1\thyp) 
   +  \KH{1} \, \mu \Delta t\Delta P + \KU{1} \lambda \Delta t \Delta u &=0  \label{enns:Prelax2}\\
 M\reff^2 (P_1\tsou - \Delta P - P_2\thyp) +\alpha_1\tsou (\Delta P + P_2\thyp - P_1\thyp) 
  + \left( \KH{1} - \KH{2} \right)  \mu\Delta t \Delta P 
  + \left( \KU{1} - \KU{2} \right) \lambda \Delta t  \Delta u  &=0 \label{enns:Prelax3}
\end{align}
where $\KHph =  \left[ M\reff^2  \rhoph \cci{\phSymbol}  +  \kappa\ph \right]$, as in \eqref{ens:P1}, and 
$\KUph = M\reff^2 \kappa\ph (\ui - \uph)$.
Reminding that the velocities $\uph\tsou$ are given by~\eqref{enns:u}, the last terms in \eqref{enns:Prelax2} and \eqref{enns:Prelax3} are known.
For the discretization of the coefficient $\KHph$, we approximate the thermodynamic variables and the interface pressure by using the values at the end of the hyperbolic operator. 
This choice is a simplifying assumption, which slightly mitigates the non-linearity of the system  \eqref{enns:Prelax1}--\eqref{enns:Prelax3}, and it is motivated by the absence of differences noted in  \cite{Saurel2009} while solving the pressure relaxation system approximating the integral value of the interface pressure by $P_I\tsou$ or $P_I\thyp$.

From a numerical point of view, the non-linear system~\eqref{enns:Prelax1}--\eqref{enns:Prelax3} presents some unfavorable features, such as the simultaneous presence of very small and very large terms, which could cause a loss of accuracy, the stiffness and the non-linearity.
To tackle these aspects, we rely also for the relaxation operator on the PETSc non-linear solver~\cite{petsc-web-page} and, in particular, on the trust-region Newton-based solver.

\section{Verification for single-phase flows}\label{s:sphase}
Since both the model and the numerical method proposed in this work are new, before focusing on two-phase simulations, we present in this section some single-phase tests, to verify the numerical method and, in particular, the low-Mach treatment and the core part of the hyperbolic operator. 
The governing equations \eqref{ent:a}--\eqref{ent:dL} (in their fully discrete versions given in Sec.~\ref{ss:spacedis}) are here solved only for one phase, but the numerical solution algorithm is kept unaltered.
And by that, we mean that the volume fraction equation is solved and  the non-conservative terms are included while building the system matrix $[A]$, even if we expect them to be identically null.
However, in single phase simulations, the relaxation operator is not applied, or, in other words, $\lambda=0$ and $\mu=0$.

\begin{table}
\caption{Stiffened gas parameters for the fluids (air and water) used in the numerical tests for single-fluid flows of Sec.~\ref{s:sphase} and for two-phase flows without relaxation of Sec.~\ref{s:twophase}}
\label{t:tmd_data}
\centering
\begin{tabular}{l*{4}{c}}\toprule
        &  $\gamma$ \small $[-]$& $P_\infty$ \small $[\mathrm{Pa}]$     &  $c_v$ \small $[\mathrm{J/kg\,K} ]$ &
$q$ \small $[\mathrm{J/kg} ]$  \\ \midrule
 Air:   &     1.4      &  0          &  717.6  & 0 \\
 Water: &     4.4      &  $6.8 \cdot 10^8$ &  4178.0 & 0\\\bottomrule
\end{tabular}
\end{table}

\subsection{Low Mach Riemann problem for a perfect gas}\label{ss:lmair}
We start with a Riemann problem test at particularly low Mach number, presented in~\cite{Abbate2017}.
The pipe is filled with a perfect gas, i.e. air with the parameters given in Tab.~\ref{t:tmd_data}, at very low pressures and the left and right chambers features weak pressure and velocity jumps, according to the data given in the row \textsf{lmAir} in Tab.~\ref{t:sp_initCond}.
The solution is represented by two rarefaction waves, plus a central contact discontinuity which moves at 
$u_s = 4.7~10^{-3}~\mathrm{m/s}$. The Mach number is lower than 0.012 all over the domain.

\begin{figure}
\centering\includegraphics[width=0.75\textwidth]{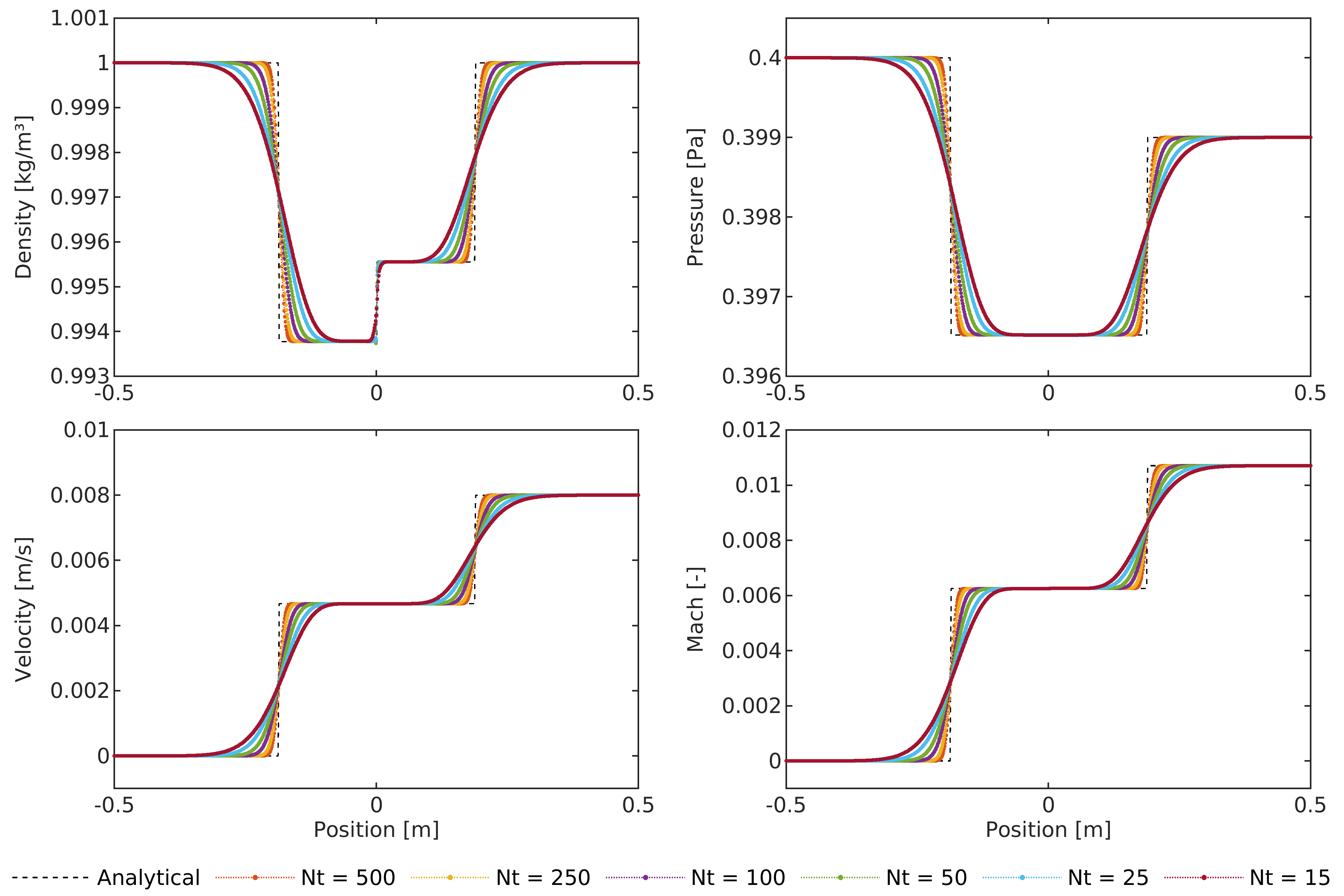}
\caption{Low Mach Air: results at $t_F=0.25~\mathrm{s}$, obtained in six simulations, each one considering a different time step $\Delta t= t_F/N_t$, with $N_t$ the number of steps as indicated in the legend. The standard formulation with density re-computation and momentum correction is used.
The analytical solution of the Riemann problem (initial conditions given in Tab.~\ref{t:sp_initCond}) is shown as a dashed line.
The six numbers of steps $N_t$ correspond to the following CFL numbers.}\footnotesize

\vspace*{1ex}
\begin{tabular}{*{7}{l}}\toprule
$N_t$ &  500 & 250 & 100 & 50 & 25 & 15\\
$\max \mathrm{CFL}(|u|+c)$ &   0.4 &  0.8 &  1.9 &  3.8 &  7.6 & 12.6\\
$\max \mathrm{CFL}(|u|)$   &   0.004 &  0.008 &  0.02 &  0.04 & 0.08 &  0.13 \\ \bottomrule
\end{tabular}
\label{fsp:lmAir_mmrC}
\end{figure}

Figure~\ref{fsp:lmAir_mmrC} displays the results at the final time obtained with the standard formulation described by \eqref{ent:a}--\eqref{ent:dL} considering only one phase.
Six simulations are run imposing six different time steps $\Delta t$, defined as
$\Delta t=t_F/N_t$ where $t_F$ is the final time and $N_t$ is the number  of integration steps used to reach the final time $t_F$.
The simulations are labeled in the picture according to the number $N_t$, which goes from $500$ to $15$ (from the smallest to the largest time step).
As reported in the caption, some of them lead to an acoustic CFL greater than one, in short, $\mathrm{CFL}(|u|+c)>1$.
The contact discontinuity, which has moved by only one cell, is sharply represented, while the rarefaction waves are smeared because of the first-order accuracy of the Rusanov flux.

We use this test also to show the role played by the density correction and the velocity formulation.
Figure~\ref{fsp:lmAir_mmrC} compares the results obtained with and without density re-computation. 
In particular, we compare three formulations:
\begin{description}
\item[$\rho \;  \mathsf{as\;first}$:] solves only the mass equation \eqref{ent:dF} at the beginning of the time step, i.e. $\arhoph\tnn \gets \arhoph\tns$ and \eqref{ent:dL} is skipped;
\item[$\rho \;  \mathsf{as\;last}$:] solves only the mass equation \eqref{ent:dL}  at the end of the time step, i.e. $\arhoph\tns \gets \arhoph\tn$   and \eqref{ent:dF} is skipped;
\item[$\rho \;  \mathsf{corr}$:] solves the mass equation \eqref{ent:dF} at the beginning of the time step and then re-computes density at the end through \eqref{ent:dF}, i.e. the standard formulation.
\end{description} 
From the density profile, we can notice that solving the mass equation only at the beginning of the time step, so that using the convective velocity $\uph\tn$, leads to some oscillations across the contact discontinuity, which are amplified if the CFL number increases.
The old value of the velocity $\uph\tn$ does not account for the pressure correction, which is responsible for enforcing the incompressibility condition (see Remark~\ref{rmk:incompressibility}).
On the other hand, from the velocity profile, we can notice that the computation of the density only at the end of the time produces slightly worse results than the standard formulation with density re-computation.
Since the density equation~\eqref{ens:d} does not present numerical difficulties, e.g. it does not include non-conservative terms, and its computational effort is almost negligible with respect to the solution of the other equations, we adopt the re-computation as the standard formulation.

\begin{figure}
\centering\includegraphics[width=0.75\textwidth]{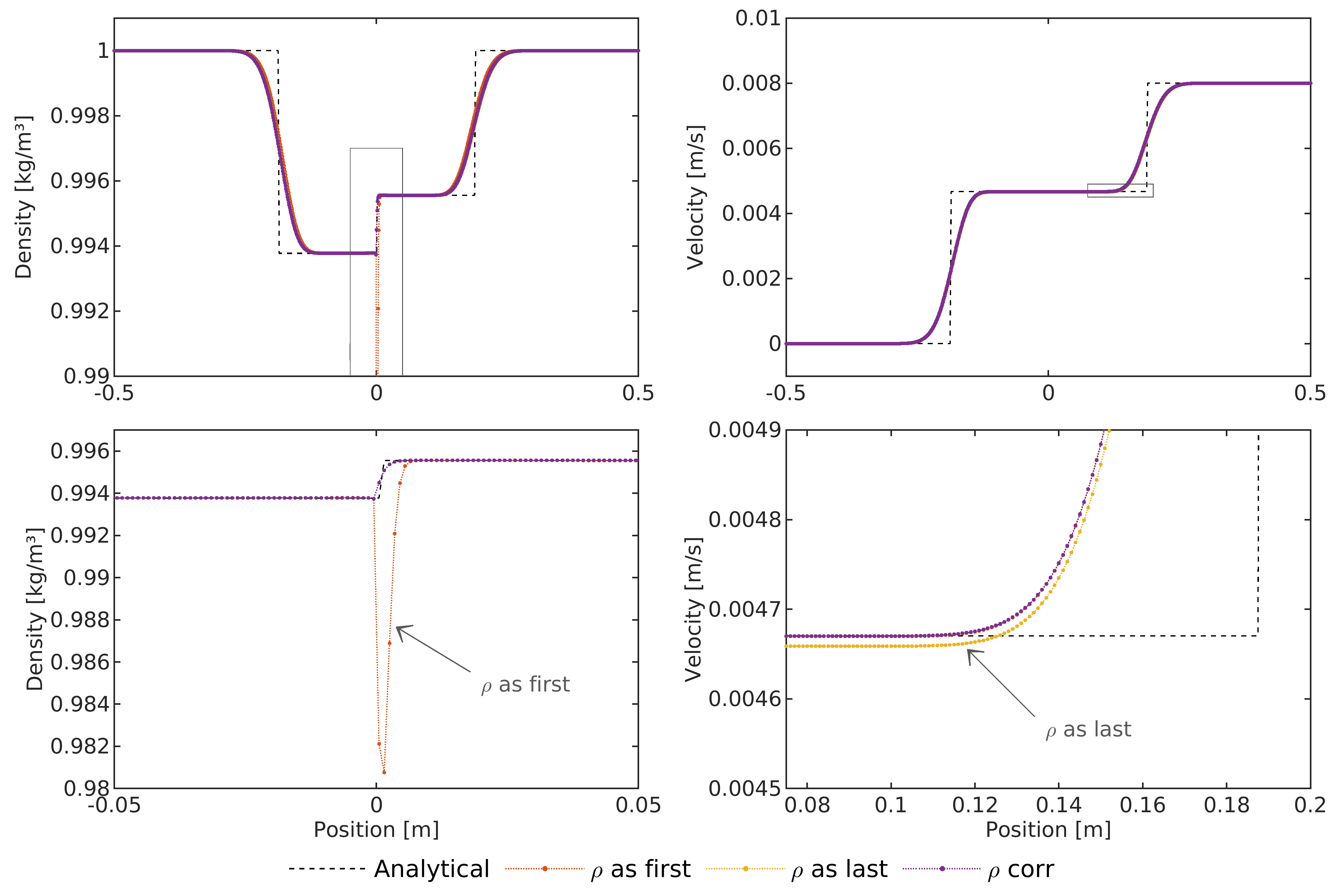}
\caption{Low Mach Air: comparison of different treatments of density equation, at final time $t_F$.
Initial conditions are given in Tab.~\ref{t:sp_initCond} and they are the same of Fig.~\ref{fsp:lmAir_mmrC}, and we use $N_t=50$. 
The two pictures in the second row show two details of density and velocity profiles, corresponding to the regions indicated by the rectangles in the plots of the first row.
In $\rho \;\mathsf{as \; first}$ and $\rho \;  \mathsf{as\;last}$, the mass equation is solved only once, at the beginning and at the end of the time step, respectively. $\rho \; \mathsf{corr}$ refers to the standard formulation with density correction. In all cases, the momentum correction equation is used.}
\label{fsp:lmAir_mmCfr}
\end{figure}

\begin{figure}
\centering
\includegraphics[width=0.75\textwidth]{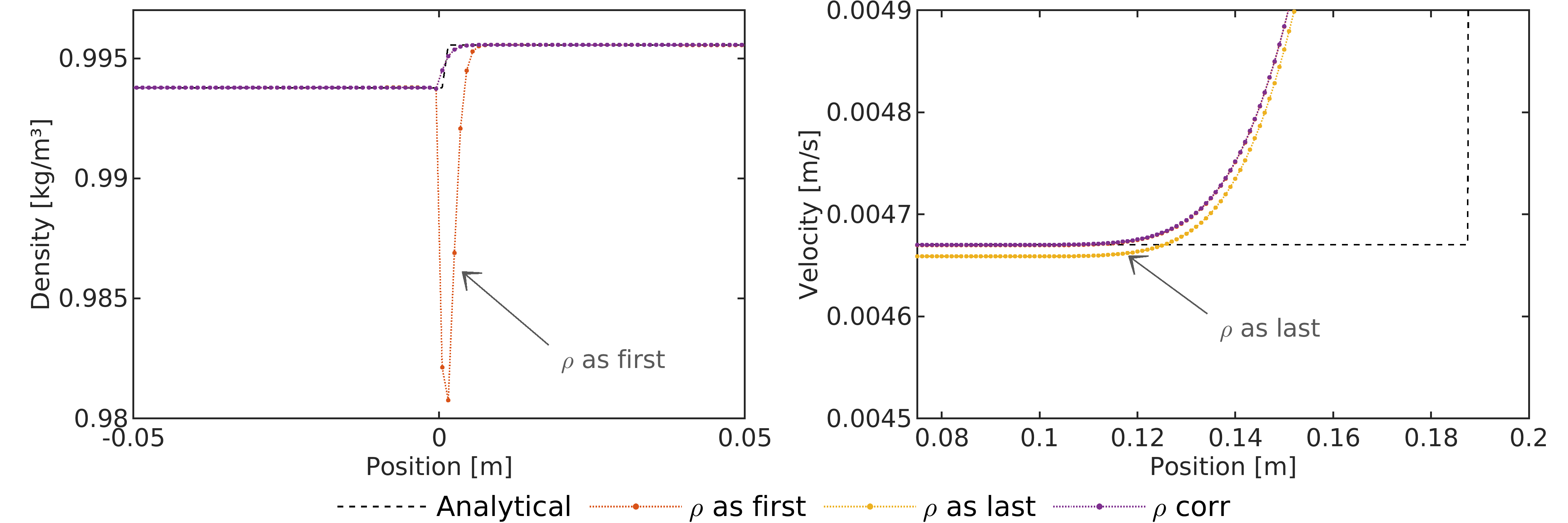}
\caption{Low Mach Air: results at $t_F$ obtained with the alternative formulation, involving velocity correction.
The test is the same of Fig.~\ref{fsp:lmAir_mmCfr}, i.e., \textsf{lmAir} test of Tab.~\ref{t:sp_initCond} with $N_t=50$. 
In $\rho \;\mathsf{as \; first}$ and $\rho \;  \mathsf{as\;last}$, the mass equation is solved only once, at the beginning and at the end of the time step, respectively. $\rho \; \mathsf{corr}$ refers to the standard formulation with density correction. 
Differently from Fig.~\ref{fsp:lmAir_mmCfr}, here the velocity correction equation~\eqref{ent:du} is used.
Since the results do not differ notably from the ones obtained with the momentum correction, we report here only the zoomed regions, highlighted by rectangles in the first row of Fig.~\ref{fsp:lmAir_mmCfr}.}
\label{fsp:lmAir_muCfr}
\end{figure}

A further open question in the development of the numerical method here proposed concerns the momentum or velocity correction, that is whether \eqref{ent:dm} can be substituted by \eqref{ent:du}.
For this reason, we have re-run the simulations presented in Figures~\ref{fsp:lmAir_mmrC} and \ref{fsp:lmAir_mmCfr} with the velocity correction, where \eqref{ent:dm} is solved instead of \eqref{ent:du}.
No notable differences are detected, and, in particular, the same conclusions about the density re-computation are drawn, as shown by Fig.~\ref{fsp:lmAir_muCfr}.

\begin{table}
\caption{Initial conditions for the single-phase tests presented in Sec.\ref{s:sphase}.
The coordinates $x_0$ and $x_N$ delimits the domain, which is split in $N$ primary cells. The initial position of the discontinuity is $x_\mathrm{d}$  and  the final time is $t_F$.
The six rightmost columns report the velocity, pressure, and density characterizing the left and right states of the Riemann problem, denoted by the subscripts $L$ and $R$, respectively.}
\label{t:sp_initCond}
\centering
\begin{tabular}{*{12}{c}}\toprule
   Test     & $x_0 $         &  $x_N$      &  $N$   & $x_\mathrm{d}$          &  $t_F$       
 &  $u_L$            & $u_R$              &  $P_L$            & $P_R$  &  $\rho_L$  & $\rho_R$\\
        & \small $[\mathrm{m}]$ &\small $[\mathrm{m}]$ & \small$[-]$  &\small $[\mathrm{m}]$ & \small$[\mathrm{s}]$
 & \small $[\mathrm{m/s}]$     &\small $[\mathrm{m/s}]$ &  \small$[\mathrm{Pa}]$ &\small $[\mathrm{Pa}]$
 & \small $[\mathrm{kg/m^3}]$  &  \small$[\mathrm{kg/m^3}]$ \\ \midrule
\textsf{lmAir}  & -0.5 & 0.5 & 1000 &  0 &   0.25  & 0 & 0.008 &  0.4& 0.399& 1.0 & 1.0\\  
\textsf{lmWater} & -0.5 & 0.5 & 1000 &  0 &   $10^{-4}$  & 0 & 15 &  $10^8$ & $0.98 \cdot 10^8$ & $10^3$ & $10^3$\\
\textsf{lmWaterLong} & -250 & 250 & 5000 &  0 &   0.095  & 0 & 15 &  $10^8$ & $0.98 \cdot 10^8$ & $10^3$ & $10^3$\\
\textsf{Lax}     & -0.5 & 0.5 & 1000 &  0 &   0.12  & 0.698 & 0 & 3.528 & 0.571 & 0.445 & 0.5 \\ \bottomrule
\end{tabular}
\end{table}

\subsection{Low Mach Riemann problem for a stiffened gas}\label{ss:lmwater}
\begin{figure}
\centering
\includegraphics[width=0.75\textwidth]{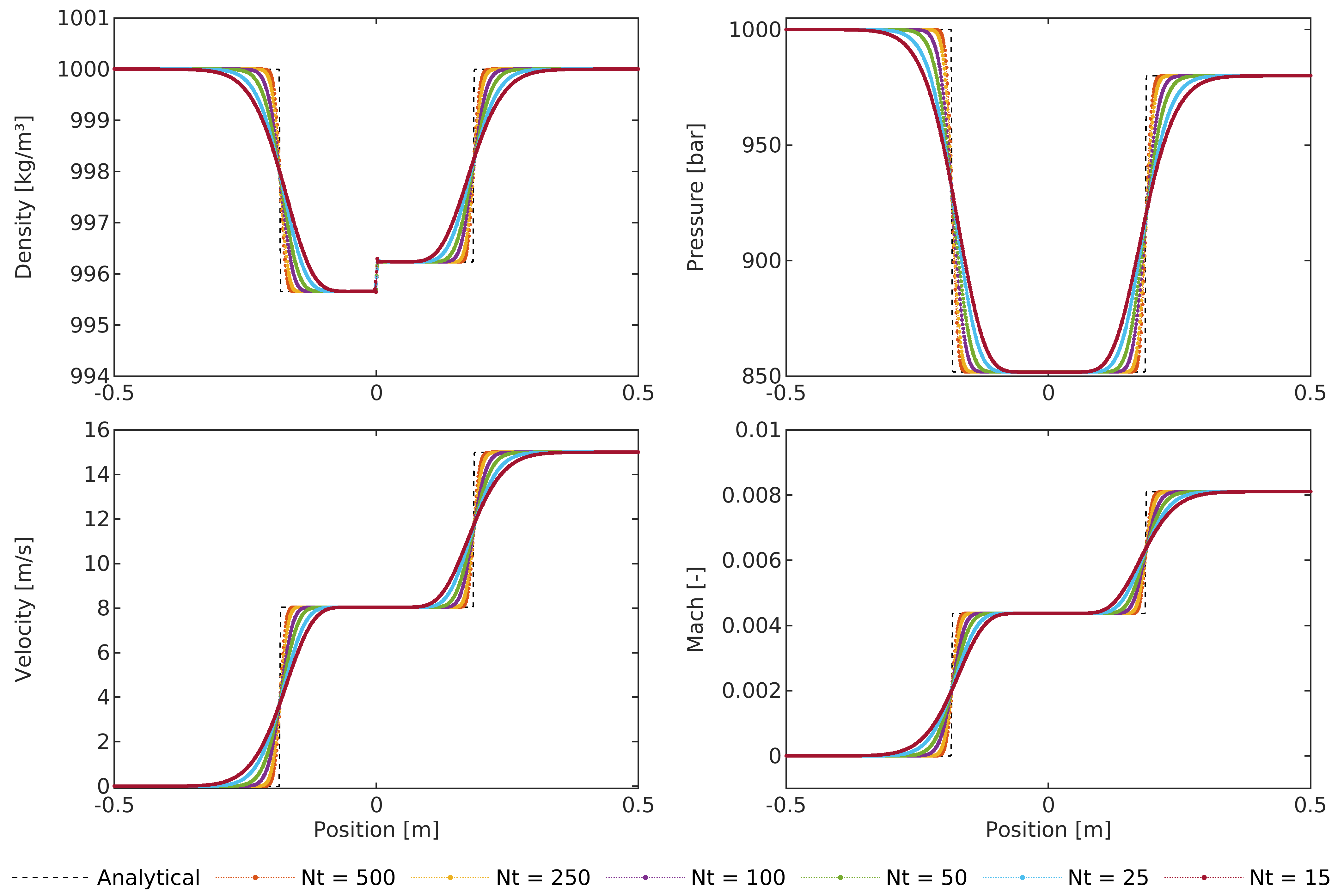}
\caption{Low Mach Water: results at $t_F=0.1~\mathrm{ms}$, obtained in six simulations, each one considering a different time step $\Delta t= t_F/N_t$, with $N_t$ the number of steps as indicated in the legend. The standard formulation with density re-computation and momentum correction is used.
The analytical solution of the Riemann problem (initial conditions given in Tab.~\ref{t:sp_initCond}) is shown as a dashed line.
The six numbers of steps $N_t$ correspond to the following CFL numbers.}\footnotesize
\vspace*{1ex}
\begin{tabular}{*{7}{l}}\toprule
$N_t$ &  500 & 250 & 100 & 50 & 25 & 15\\
$\max \mathrm{CFL}(|u|+c)$ &   0.4   &  0.7   &  1.9  &  3.7  &  7.5 & 12.4\\
$\max \mathrm{CFL}(|u|)$   &   0.003 &  0.006 &  0.015 &  0.03 & 0.06 &  0.1 \\ \bottomrule
\end{tabular}
\label{fsp:lmWater}
\end{figure}

In this section, we address the simulation of a water pipe flow under the stiffened gas model, proposed in~\cite{Abbate2017}. The thermodynamic parameters are given in Tab.~\ref{t:tmd_data} and the initial data are reported in the row \textsf{lmWater} in Tab.~\ref{t:sp_initCond}.
The Riemann problem is characterized by a weak pressure ratio and has the structure of the test presented in Sec.~\ref{ss:lmair}, with the contact discontinuity that moves at $u_s = 8.04~\mathrm{m/s}$.
Although we observe a higher speed in this test with respect to the test considering air, the Mach number is even lower, below 0.01, due to the high speed of sound.
Figure~\ref{fsp:lmWater} displays the results obtained by solving \eqref{ent:a}--\eqref{ent:dL} for one single phase.
We have considered different time steps, which correspond also to acoustic CFL numbers greater than one (up to 12), as indicated in the caption by $\mathrm{CFL}(|u|+c)$. The smaller is the time step, the better is the agreement with the analytical solution.  

\begin{figure}
\centering
\includegraphics[width=0.75\textwidth]{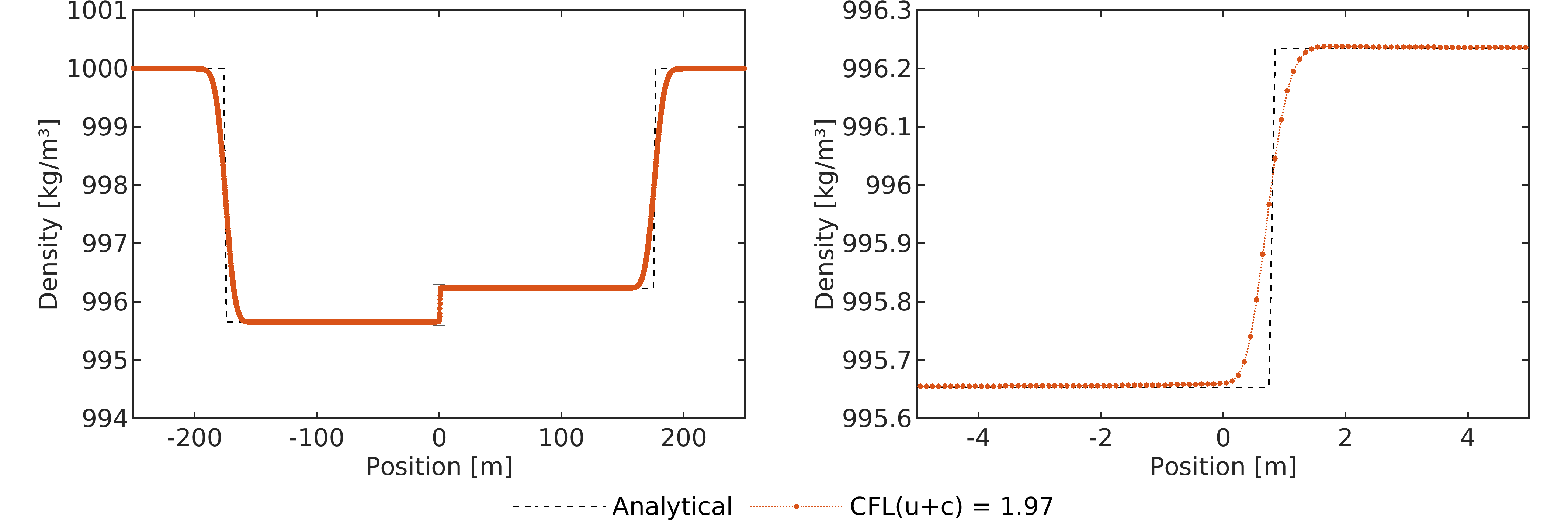}
\caption{Low Mach Water Long: the same Riemann problem test of Fig.~\ref{fsp:lmWater} is run over a longer time $t_F=0.095 ~\mathrm{s}$, using $N_t=900$, on a grid with $\Delta x= 0.1$. Initial conditions are given in Tab.~\ref{t:sp_initCond}. 
The analytical solution of the Riemann problem is shown as a dashed line.
The right picture shows the detail of the density field in proximity of the traveling contact discontinuity, i.e. the region enclosed in the rectangle in the left picture.}
\label{fsp:lmWater_long}
\end{figure}

Additionally, we test the capability to capture travelling material waves over a long simulation, by repeating the same Riemann problem test but over a longer time, that is $t_F=0.095~\mathrm{s}$, as proposed in~\cite{Abbate2017}.
All test information are reported in the row \textsf{lmWaterLong} in Tab.~\ref{t:sp_initCond}.
Figure~\ref{fsp:lmWater_long} displays the detail of the solution field  close to the contact discontinuity, which at the end of this test has reached $x_s = 0.7638~\mathrm{m}$, so it has crossed 7 grid cells.
The quality of our results compares well with the ones reported by Abbate et al.~\cite{Abbate2017} for their implicit scheme, so we can state that our scheme is able to correctly compute the position and the velocity of the moving material wave.

\subsection{Lax problem}\label{ss:lax}
\begin{figure}
\centering
\includegraphics[width=0.75\textwidth]{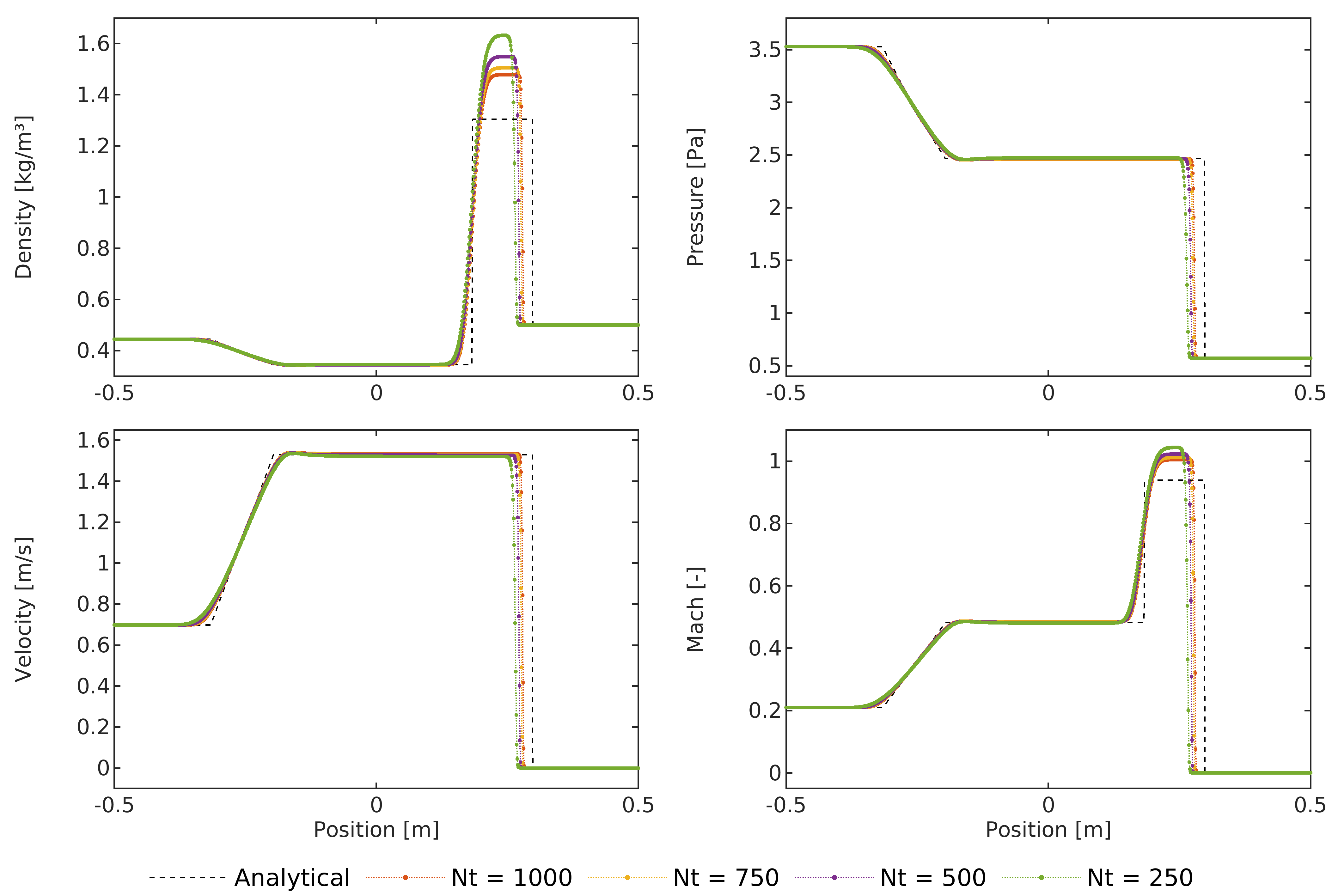}
\caption{Lax test: numerical results at $t_F=0.12~\mathrm{s}$, obtained in four simulations, each one considering a different time step $\Delta t= t_F/N_t$, with $N_t$ the number of steps as indicated in the legend. The standard formulation with density re-computation and momentum correction is used.
The analytical solution of the Riemann problem (initial conditions given in Tab.~\ref{t:sp_initCond}) is shown as a dashed line.
The four numbers of steps $N_t$ correspond to the following CFL numbers.}\footnotesize
\vspace*{1ex}
\begin{tabular}{*{5}{l}}\toprule
$N_t$ &  1000 & 750 & 500 & 250 \\
$\max \mathrm{CFL}(|u|+c)$ &   0.57 &  0.75 &  1.13 &  2.26  \\
$\max \mathrm{CFL}(|u|)$   &   0.18 &  0.25 &  0.37 &  0.74  \\ \bottomrule
\end{tabular}
\label{fsp:lax}
\end{figure}

Finally, we end this single-phase section presenting the results for the Lax shock-tube test~\cite{Lax1954}, routinely used to validate standard compressible schemes, to investigate the behavior of the proposed scheme at Mach numbers between 0.3 and 1, so not so low.
This shock test is characterized by an initial discontinuity also in the velocity, which is not null in the left chamber. Initial conditions and test data are given in Table~\ref{t:sp_initCond}, in the row \textsf{Lax}.
When the diaphragm bursts, the initial discontinuity evolves in a leftward moving rarefaction waves and a rightward moving shock wave, with a contact discontinuity in between.
The results at $t_F = 0.12~\mathrm{s}$ are shown in Fig.~\ref{fsp:lax} for different time steps $\Delta t$.
In the right part of the domain, where the Mach number is higher,
the numerical solution does not agree well with the analytical one.
However, this discrepancy is an expected manifestation of the non-conservation of the total energy, but beyond that, the numerical results of the proposed scheme show an acceptable agreement with the analytical ones although we are not operating within the target regime of weakly compressible flows.

\section{Numerical results for the hyperbolic operator for two-phase flows}\label{s:twophase}
In this section, we present two-phase flow results computed by using only the hyperbolic operator, without any relaxation process. Results of the complete numerical methods are shown in next section.
Taking into consideration the conclusions of the previous section, we use the standard formulation with density re-computation, i.e.~\eqref{ent:a}-\eqref{ent:dL}.
We organize our analysis in subsequent steps, starting from the numerical validation of two fundamental properties: the behavior of the hyperbolic operator without mixing in Sec.~\ref{ss:nomix}, and the fulfillment of the pressure non-disturbance condition in Sec.~\ref{ss:disc}.
Then, we present the results of the proposed method on some reference Riemann problems available in the literature about BN-type models, in Sec.~\ref{ss:riemann}, and, finally, on a water-air mixture problem in Sec.~\ref{ss:waterAir}.

\subsection{No mixing water-air test}\label{ss:nomix}
The first two-phase test involves liquid water and air governed by the stiffened gas model with the parameters listed in Tab.~\ref{t:tmd_data}.
As initial condition, the phases are uniformly dispersed with equal volume fraction $\alpha_1 = \alpha_2= 0.5$, in a shock-tube where a mild pressure jump is imposed between the two chambers:
$P_L =100\, \mathrm{bar}$ at the left, and $P_R =50\, \mathrm{bar}$ at the right. A null velocity and the temperature $T=270\, \mathrm{K}$ are applied uniformly in the domain.
The initial position of the discontinuity is at $x_\mathrm{d} = 0$ and the grid spacing is  $\Delta x = 0.001 \, \mathrm{m}$.

\begin{figure}
\centering
\includegraphics[width=0.75\textwidth]{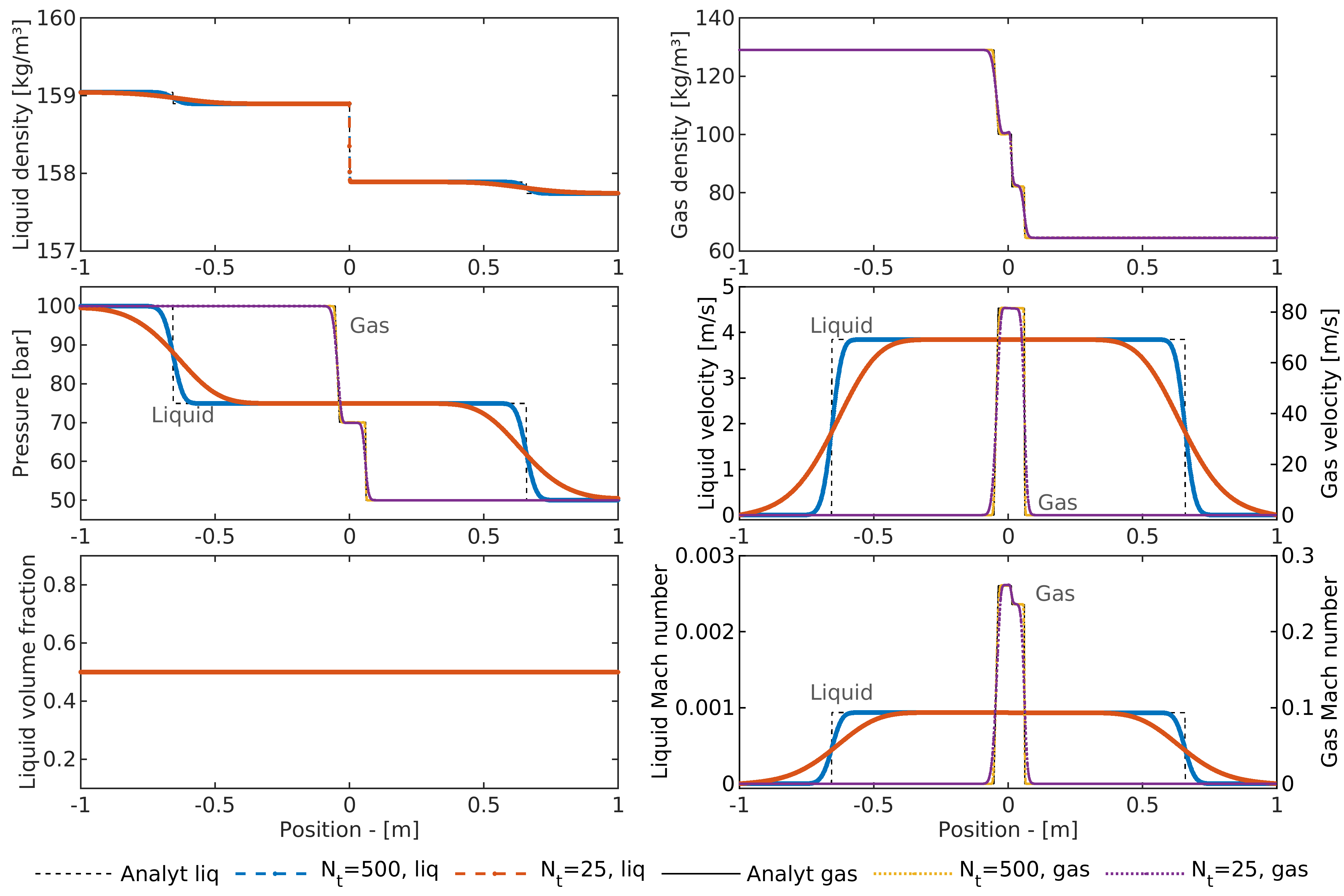}
\caption{No mixing test: results of the two-phase shock-tube Riemann problem with uniform volume fraction $\alpha_1=\alpha_2=0.5$ and with no relaxation, at $t_F=0.16\,\mathrm{ms}$ using $N_t=500$ and $N_t=25$ time steps. The absence of relaxation and the uniform volume fraction make the fluids (air and water) evolve separately, so the numerical results are compared to the analytical solutions of two single-phase Riemann problems. 
The acoustic CFL conditions are}\footnotesize
\vspace*{1ex}
\begin{tabular}{*{4}{l}}\toprule
for $N_t=500$,
   & liquid: &  $\max \mathrm{CFL}(|u|+c)=1.3$ & $\max \mathrm{CFL}(|u|)=0.001$ \\
   & gas:    &  $\max \mathrm{CFL}(|u|+c)=0.14$ & $\max \mathrm{CFL}(|u|)=0.03$ \\
for $N_t=25$,
   & liquid: &  $\max \mathrm{CFL}(|u|+c)=26.4$ & $\max \mathrm{CFL}(|u|)=0.02$ \\
   & gas:    &  $\max \mathrm{CFL}(|u|+c)=2.7$ & $\max \mathrm{CFL}(|u|)=0.5$ \\ \bottomrule
\end{tabular}
\label{ftp:nomix}
\end{figure}

Being the volume fraction uniform and given the absence of relaxation terms, in this test, each phase evolves independently from the other one. Thus, the exact solution can be computed by solving the Riemann problem for the Euler equations.
Figure~\ref{ftp:nomix} shows the results at the final time of $0.16 ~\mathrm{ms}$ computed with two different time steps: the smallest one corresponds to an acoustic CFL number slightly above 1 only for the liquid, while the biggest time step results in acoustic CFL number greater than 2 for both phases. 
Although the shock and the rarefaction waves appear smeared in liquid phase, the numerical results agree well with the analytical solutions, both in terms of position of the waves and of downstream conditions.

\subsection{Pure advection water-air problem}\label{ss:disc}
Here, we investigate a pure advection problem: a column of water-air mixture with a liquid volume fraction $\alpha_{1,\mathrm{c}}=0.9$ is transported at a velocity of $100~\mathrm{m/s}$ in a uniform pressure field at $P=1~\mathrm{bar}$, involving a mixture with
$\alpha_{1,\mathrm{L}}=\alpha_{1,\mathrm{R}}=0.1$. The initial temperature is $270\, \mathrm{K}$ for both phases.
The parameters of the stiffened gas model for the fluids are the same as in the previous test, and are listed in Tab.~\ref{t:tmd_data}.
Initially, the column is located at $0.2 < x < 0.4 $, within the domain $\Omega=[0,1]$.

\begin{figure}
\centering
\includegraphics[width=0.75\textwidth]{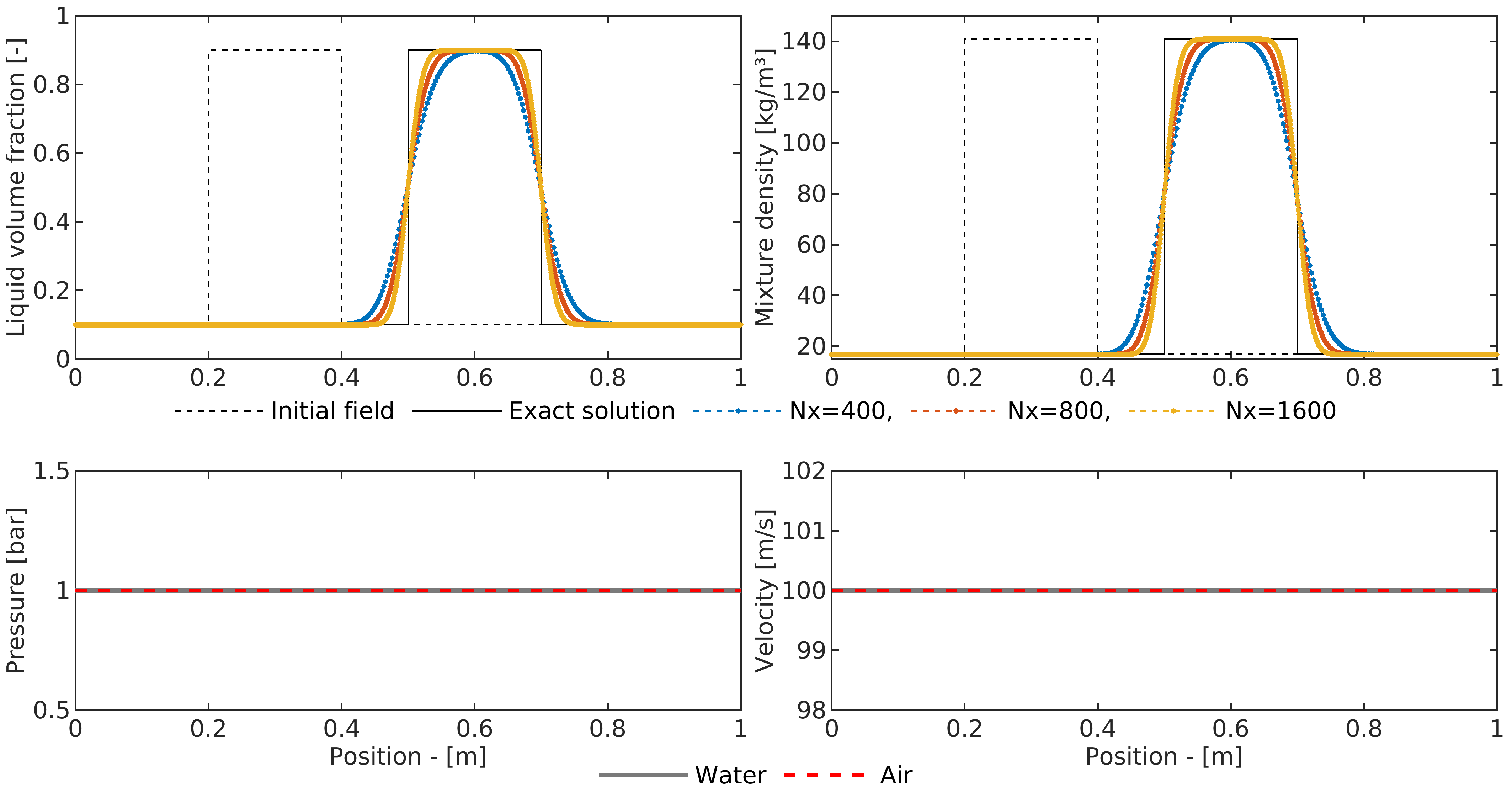}
\caption{Pure advetion test of a column of water-air mixture at uniform velocity, results at $t_F=3\,\mathrm{ms}$, with $\mathrm{CFL}(\vert u\vert)=0.5$.
The initial position of the column is shown as a dashed line in the top row, which displays also the water volume fraction $\alpha_1$ on the left, and the mixture density $\bar{\rho} = \arho_1 + \arho_2$ on the right;  the results at $t_F$ obtained over three different grids (with $\mathsf{Nx}$ the number of primary cells) are compared with the exact solution shown by a solid black line.
The second row presents the pressure and velocity fields at $t_F$ for both phases; only the results computed over the coarsest grid are shown for brevity, as no differences are observed using finer grids.}
\label{ftp:contdisc1}
\end{figure}

\begin{figure}
\centering
\includegraphics[width=0.4\textwidth]{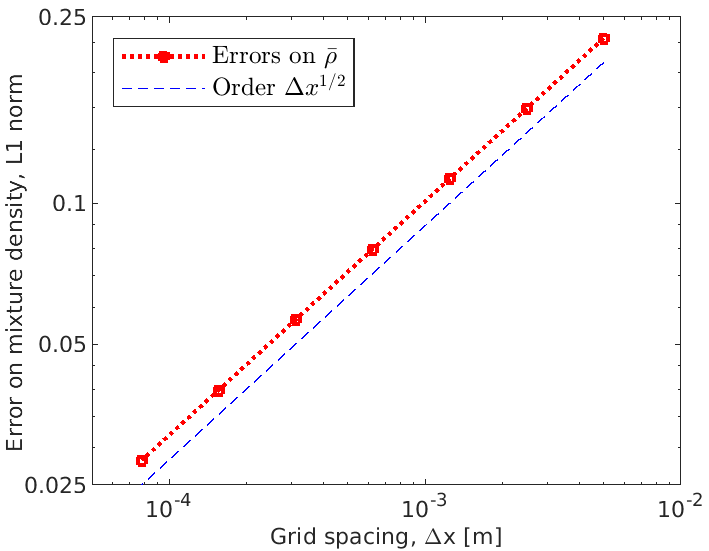}
\caption{Pure advetion test: grid convergence study. The considered grid spacings correspond to $100 \cdot 2^{[1:7]}$ points and the $L^1$ norm of the error on the mixture density is computed according to~\eqref{eres:errCD}. The dashed line displays the convergence rate of $\Delta x^{1/2}$.}
\label{ftp:contdisc2}
\end{figure}

This test is performed considering different discretizations, all imposing the convective $\mathrm{CFL}=0.5$. 
The results at time $t_F=3 ~\mathrm{ms}$ over three grids (with $400$, $800$, and $1600$ cells) are shown and compared to the exact solution in Fig.~\ref{ftp:contdisc1}.
From the second row of the picture, we can appreciate that the no  pressure or velocity oscillations arise and the initially uniform fields are correctly preserved during the time evolution. This achievement is crucial for a correct discretization of the non-conservative terms~\cite{Abgrall1996}.
Beyond pressure and velocity, a good agreement between the numerical and the exact solution is observed also for the volume fraction and mixture density variables, for which the smearing of the contact discontinuity decreases with the grid refinement.
To confirm this behavior, we have performed also a grid convergence study, presented in Fig.~\ref{ftp:contdisc2},
computing the discrete $L^1$ error between the numerical and the exact mixture density at the final time, normalized by the $L^1$ norm of the initial mixture density, as
\begin{equation}\label{eres:errCD}
E_{\bar{\rho}} (\Delta x) = 
\dfrac{ \int_\Omega \left\vert \bar{\rho}(\Delta x,t_F) - \bar{\rho}^\mathrm{ex}(x,t_F) \right\vert \mathrm{d}x}
{ \int_\Omega \left\vert \bar{\rho}^\mathrm{ex}(x,t_0) \right\vert \mathrm{d}x}
=\dfrac{ \sum_i \left\vert \bar{\rho}_i(t_F) - \bar{\rho}^\mathrm{ex}(x_i,t_F) \right\vert \Delta x}
{ \int_\Omega \left\vert \bar{\rho}^\mathrm{ex}(x,t_0) \right\vert \mathrm{d}x} \,.
\end{equation}
The numerical error converges with the order of $\Delta x^{1/2}$, as expected when using a first-order scheme for BN-type models~\cite{Herard2012}.
 
\subsection{Verification using a manufactured solution}
In this subsection, we perform a grid convergence study for a test with non-uniform pressure and velocity.
To do that, we switch off all relaxation terms and add to the right-hand side of the equations a source $\phi(x,t)$ to be determined by a manufactured solution. We follow the strategy proposed by Hennessey et al.~\cite{Hennessey2020}, and we express the exact solution in terms of the primitive variables
$\mathbf{W}=\left[ \alpha_1,\, \rho_1,\,  u_1,\, P_1,\, \rho_2,\,  u_2,\, P_2 \right]^\mathrm{T}$ as
\begin{equation}\label{e:manSol}
W^{\mathrm{ex}}_j(x,t) = \beta_j  + \delta_j \left(1 + a_{1,j} x + a_{2,j}x^2 \right)
 \left(1 + b_{1,j} t + b_{2,j} t^2 \right), \qquad j=1, \dots, 7
\end{equation}
where  $\left[ \beta_j,\, \delta_j ,\, a_{1,j} ,\, a_{2,j},\, b_{1,j} ,\, b_{2,j} \right] $
are constant.
The set of dimensionless values for $\beta_j$ is $\left[ 0.5 ,\, 0.5 ,\, 0 ,\, 0 ,\, 0.4,\, 0,\, -0.2 \right]$, while the remaining parameters are chosen randomly within the interval
$\left[0.1 , \, 0.2 \right]$.
These choices yield to a solution that varies smoothly and monotonically within the domain, starting from the constant states expressed by $\beta_j$ coefficients. The fluids are air and water, defined according to the stiffened gas model, with the parameters listed in Tab.~\ref{t:tmd_data}.

\begin{figure}
\centering
\includegraphics[width=0.4\textwidth]{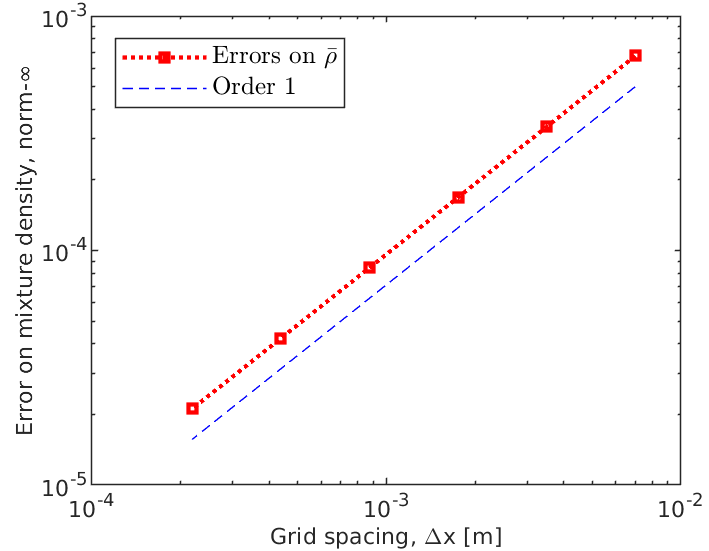}
\caption{Manufactured solution test: grid convergence study.
Infinity-norm of the error on the mixture density, normalized with respect to the integral of the density over the domain.
The final time step $t_F=0.5$ is reached in $N_T= 2400$ time steps, which correspond to a $\mathrm{CFL}(\vert u\vert)=0.25$ for air, on the smallest grid.
The dashed line displays the convergence rate of 1.
A least-square fit of the logarithmic values of the errors gives a slope of $1.0024$.}
\label{ftp:mansol}
\end{figure}

We run this test over different grids representing the domain $0\leq x\leq 1$, and we compute the final solution at time $t_F= 0.5$. The exact solution is given by evaluating Eq.~\eqref{e:manSol} at the final time.
The $L$-infinity norm of the error on the mixture density,  is shown in Fig.~\ref{ftp:mansol}.
The order of convergence is close to 1, as expected while using the Rusanov scheme, as here.

\subsection{Reference Riemann problems with perfect gases}\label{ss:riemann}
The goal of this section is to validate the proposed approach through some tests commonly used in the research community devoted to the development of one-dimensional numerical schemes for the BN-type models.
Neglecting tests involving strong shock waves or vanishing phases, 
we have selected from the literature three Riemann problems for which the analytical solution is given:
the first two, i.e., the \textit{sonic point} and the \textit{123-problem}, are taken from~\cite{Tokareva2010,Tokareva2016} (named there \textit{Test 3} and \textit{Test 4}, respectively); the third one reproduces the \textit{Test-case 1} in~\cite{Coquel2017}, and it is called \textit{solid contact} in the following.

Before describing each one, let us remark that these tests are not properly representative of low-Mach problems, but they provide anyhow an important contribution for the verification of the hyperbolic operator.
Moreover, in these three tests, both fluids follow the perfect gas model, with the air parameters given in Tab.~\ref{t:tmd_data}, so we prefer to use the notation \textit{phase 1} and \textit{phase 2}, rather than \textit{solid} and \textit{gas}.
Finally, for the sake of completeness, we report here the estimate for the shock speed~\cite{Tokareva2010} we have used to draw the analytical solutions:
\begin{equation}\label{eres:shockspeed}
u_\mathrm{s} = u_\mathrm{pre} \pm c_\mathrm{pre} \sqrt{1+ \frac{\gamma +1}{2 \gamma}\left(\frac{P_\mathrm{post}+ P_\infty}{P_\mathrm{pre} + P_\infty} -1 \right)}\,,
\end{equation}
where the subscript $\mathrm{pre}$ and $\mathrm{post}$ refer to the pre- and post-shock states and the plus and minus sign is used for a right and a left traveling shock, respectively.

\begin{figure}
\centering
\includegraphics[width=0.9\textwidth]{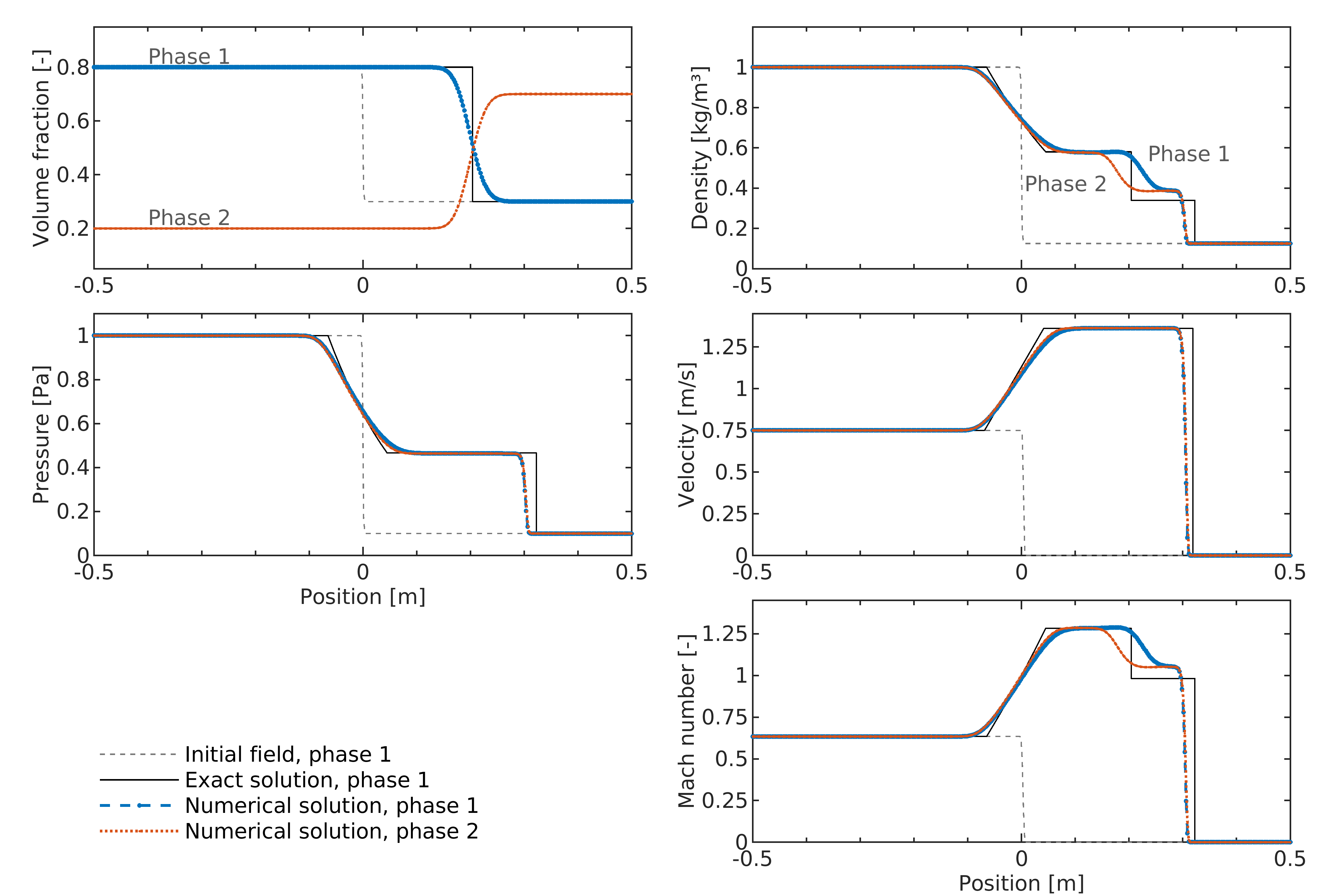}
\caption{Sonic point test: results at the time $t_F=0.15\,\mathrm{s}$. Both phases are air.
The initial solution for the phase 1 is shown by a dashed grey line; the initial pressure, density and velocity of phase 2 are the same.
The numerical results obtained over a grid with $\Delta x = 0.0025~\mathrm{m}$ and considering $N_t=500$ time steps are compared with the exact solution of this Riemann problem given in~\cite{Tokareva2010}. The convective CFL with respect to the shock speed, which is $V_s\approx 2~m/s$, is $\mathrm{CFL}(V_s)=0.24$, which is pretty similar to the acoustic one since $\mathrm{CFL}(|u|+c)=0.32$ for both phases.}
\label{ftp:sonicPoint}
\end{figure}

\paragraph{Sonic point} This test was presented in~\cite{Tokareva2010} to assess the correct resolution of a sonic rarefaction.
The two phases have initially the same pressure, density and velocity, but the mixture composition differs between the left and right states:
\begin{center}
\begin{tabular}{lllll}
\textsf{left:} & $P= 1.0 ~\mathrm{Pa}$ & $\rho=1.0~\mathrm{kg/m^3}$ & $u=0.75 ~\mathrm{m/s}$ & $\alpha_1=0.8$,\\
\textsf{right:} & $P= 0.1 ~\mathrm{Pa}$ & $\rho=0.125~\mathrm{kg/m^3}$ & $u=0.0 ~\mathrm{m/s}$ & $\alpha_1=0.3$.
\end{tabular}
\end{center}
Therefore, the solution of the two phases is the same except for the volume fraction, and it is composed by a shock wave and a contact discontinuity, both right-traveling, and a left sonic rarefaction wave.
The numerical results shown in Fig.~\ref{ftp:sonicPoint} at the time $t_F=0.15\,\mathrm{s}$ agree fairly well with the exact solution given in~\cite{Tokareva2010}, except for the intermediate state after the shock, where, however, some discrepancies are expected as we are using a  pressure-based, that is non-conservative, solver. 
Moreover, the asymmetry between phases in the density profiles is simply due to the smearing of the volume fraction discontinuity.
More important, in this test, is the correct resolution of the sonic rarefaction, without any non-physical entropy glitch at the sonic point.

\begin{figure}
\centering
\includegraphics[width=0.9\textwidth]{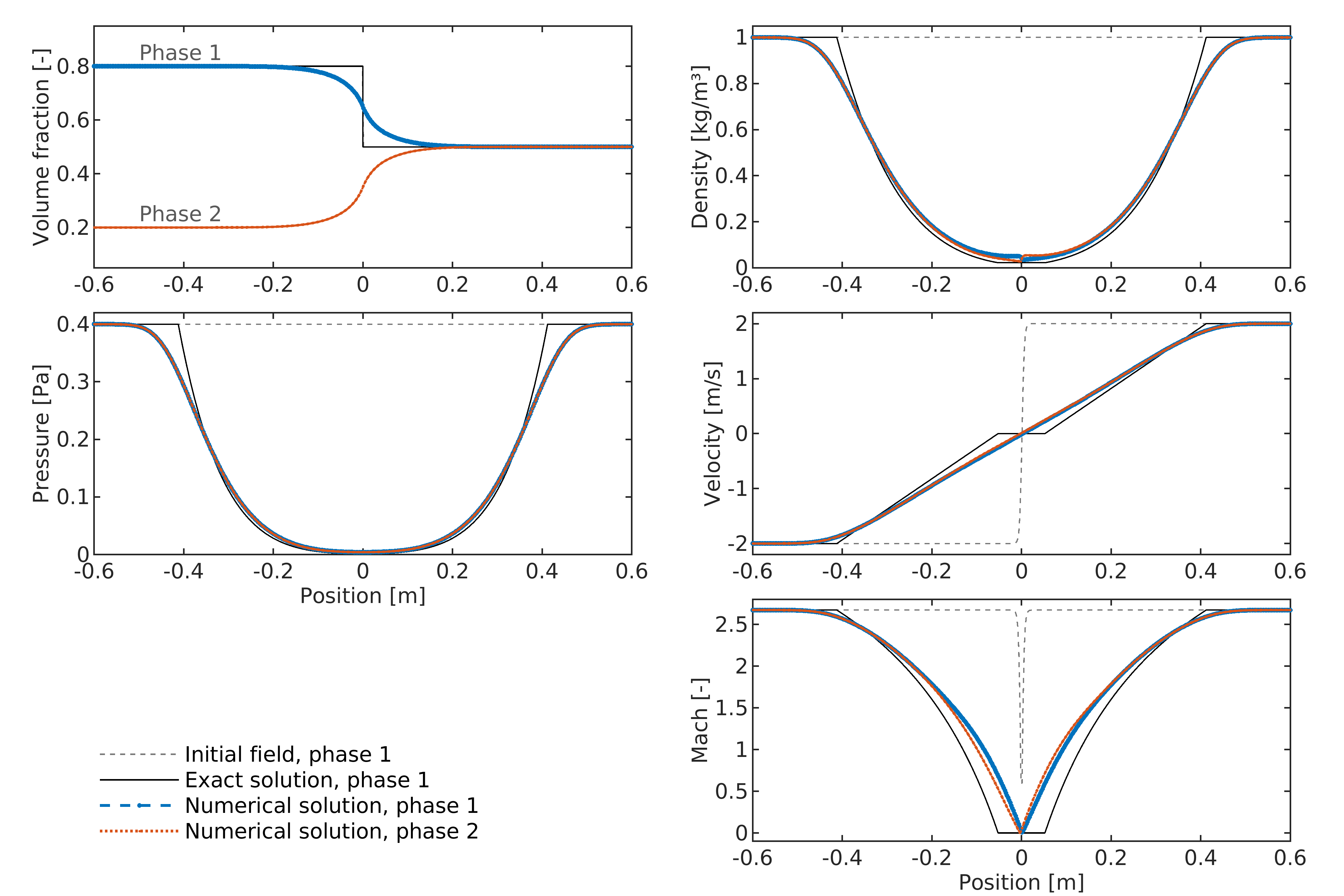}
\caption{Two-phase 123-problem: results at the time $t_F=0.15\,\mathrm{s}$. Both phases consist in air.
The initial solution for the phase 1 is shown by a dashed grey line; the initial pressure, density and velocity of phase 2 are the same.
The numerical results obtained over a grid with $\Delta x = 0.0024~\mathrm{m}$ and considering $N_t=250$ time steps are compared with the exact solution of this Riemann problem given in~\cite{Tokareva2010}. The convective and acoustic CFLs are $\max \mathrm{CFL}(|u|)=0.5$  and $\max \mathrm{CFL}(|u|+c)=0.69$, for both phases.}
\label{ftp:123probl}
\end{figure}

\paragraph{Two-phase 123-problem}
This test involves a region close to vacuum, so it is useful to assess the pressure positivity. 
Initially, the fluids are at uniform pressure $P=0.4~\mathrm{Pa}$ and density $\rho=1.0~\mathrm{kg/m^3}$. A discontinuity is imposed in the middle of the domain, $x_D=0$: on the left, the volume fraction is $\alpha_1=0.8$ and the velocity is $-2.0~\mathrm{m/s}$, on the right, the volume fraction is $\alpha_1=0.5$ and the velocity is $+2.0~\mathrm{m/s}$.
The solution consists in two symmetric rarefactions and a stationary contact discontinuity in between, where, at the final time $t_F=0.15~\mathrm{s}$, the pressure and density are extremely small: $P(x_D,t_F)=0.0019~\mathrm{Pa}$ and $\rho(x_D,t_F)=0.0219~\mathrm{kg/m^3}$~\cite{Tokareva2010}.
The numerical results are displayed in Fig.~\ref{ftp:123probl}. The pressure and the density are computed accurately, preserving the positivity.
However, the discontinuity in the volume fraction appears to be very diffused, but a similar behavior for the Rusanov's scheme is reported also by Coquel et al.~\cite{Coquel2017}.

\begin{figure}
\centering
\includegraphics[width=0.9\textwidth]{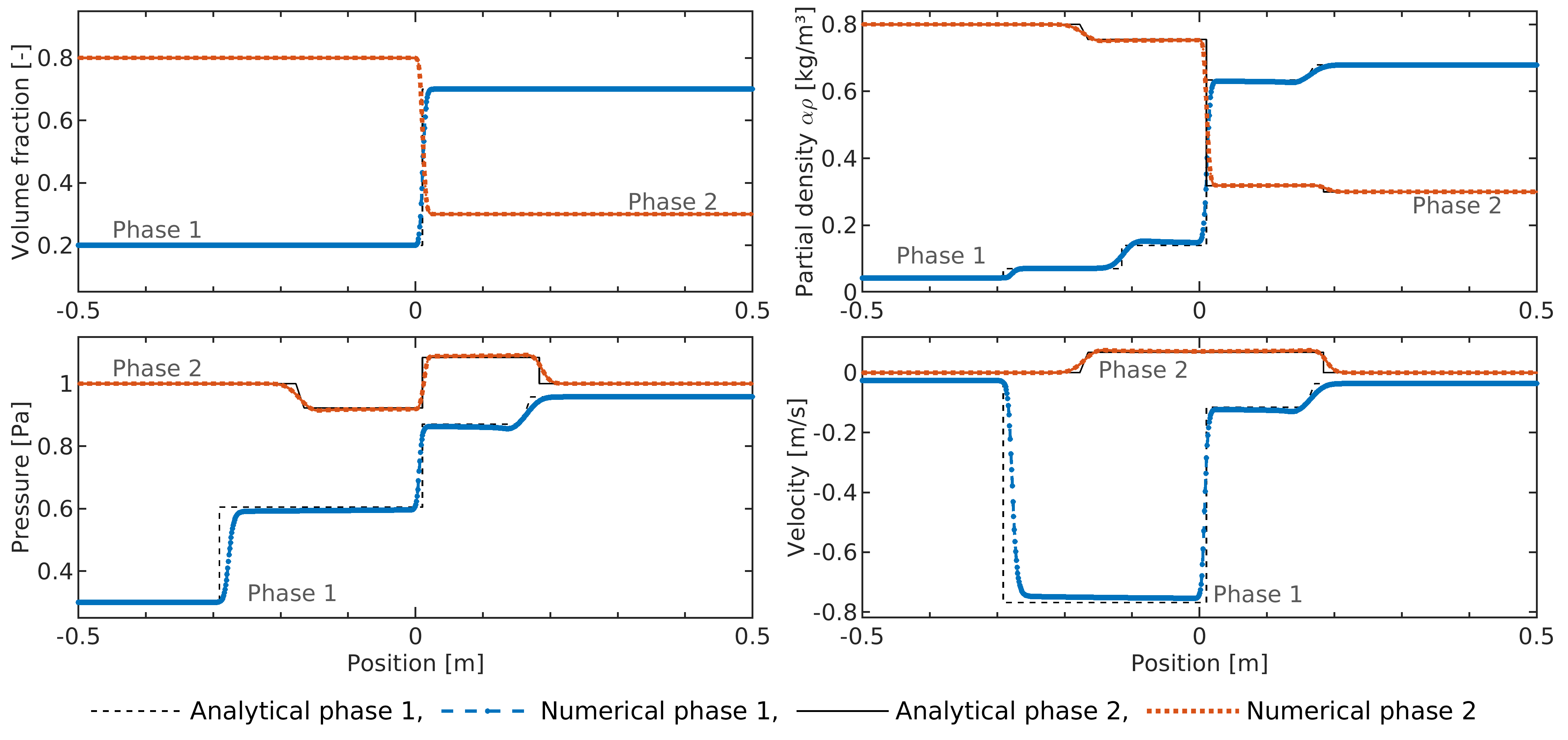}
\caption{Solid contact test: results at the time $t_F=0.15\,\mathrm{s}$. Both phases consist in air and the initial solution is given in Tab.~\ref{t:solidcont}.
The numerical results obtained over a grid with $\Delta x = 0.001~\mathrm{m}$ and considering $N_t=200$ time steps are compared with the analytical solution of this Riemann problem given in~\cite{Coquel2017}. 
The shock positions in the analytical solution are estimated with~\eqref{eres:shockspeed}.
On the top right, the solution variables $\arho_1$ and $\arho_2$ (called partial densities) are displayed.
The acoustic CFLs $\max \mathrm{CFL}(|u|+c)$ are $1.7$ for phase 1 and $0.95$ for phase 2. 
}
\label{ftp:solidcont}
\end{figure}

\paragraph{Solid contact}
The last Riemann problem we present was proposed by Coquel et. al.~\cite{Coquel2017} and, differently from the previous ones, it does not involve an initial symmetry between the two phases. 
The initial field is described in Tab.~\ref{t:solidcont}, and its evolution encompasses seven different types of waves: for phase 1, a left-traveling shock, a material contact discontinuity moving at velocity $u_1$, a phase fraction discontinuity moving with velocity $u_2$ and a right-traveling rarefaction wave; for phase 2, a left-traveling rarefaction fan, the phase fraction discontinuity, and a right-traveling shock.
To be able to compare our results with the analytical and numerical solution in~\cite{Coquel2017}, we define the interface velocity and pressure as $\ppi=P_1$ and $\ui=u_2$.
The solution computed after $0.15~\mathrm{s}$ is displayed in Fig.~\ref{ftp:solidcont}. A good agreement with the analytical solution both in terms of intermediate values and wave positions confirms the correctness of the numerical implementation of the hyperbolic operator.
This positive outcome is also justified by the fact that in this test, we have the lowest maximum Mach numbers among the three Riemann problems presented in this section, that is $\max M_1 =0.72$ for phase 1 and $\max M_2 =0.06$ for phase 2.

\begin{table}
\caption{Initial conditions for the \textit{Solid contact} Riemann problem in Sec.~\ref{ss:riemann}.
The subscripts $L$ and $R$ refer to the left and right state with respect to the initial position of the discontinuity $x_D=0$.}
\label{t:solidcont}
\centering 
\begin{tabular}{*{9}{c}}\toprule
       & $\alpha_L$   &  $u_L$    & $P_L$   &  $\rho_L$   & $\alpha_R $ &  $u_R$  &  $P_R$  & $\rho_R$\\
       &  \small$[-]$ & \small $[\mathrm{m/s}]$ & \small$[\mathrm{Pa}]$ &  \small$[\mathrm{kg/m^3}]$ 
       &  \small$[-]$ & \small $[\mathrm{m/s}]$ & \small$[\mathrm{Pa}]$ &  \small$[\mathrm{kg/m^3}]$  \\ \midrule
\textsf{Phase 1} & 0.2 & -0.02609   & 0.3   &  0.21430    &   0.7       & -0.03629  & 0.95776 & 0.96964\\  
\textsf{Phase 2} & 0.8 &  0.00007   & 1.0   &  1.00003    &   0.3       & -0.00004  & 1.0     & 0.99993 \\ \bottomrule
\end{tabular}
\end{table}

\subsection{Water-air mixture test}\label{ss:waterAir}
\begin{figure}
\centering
\includegraphics[width=0.9\textwidth]{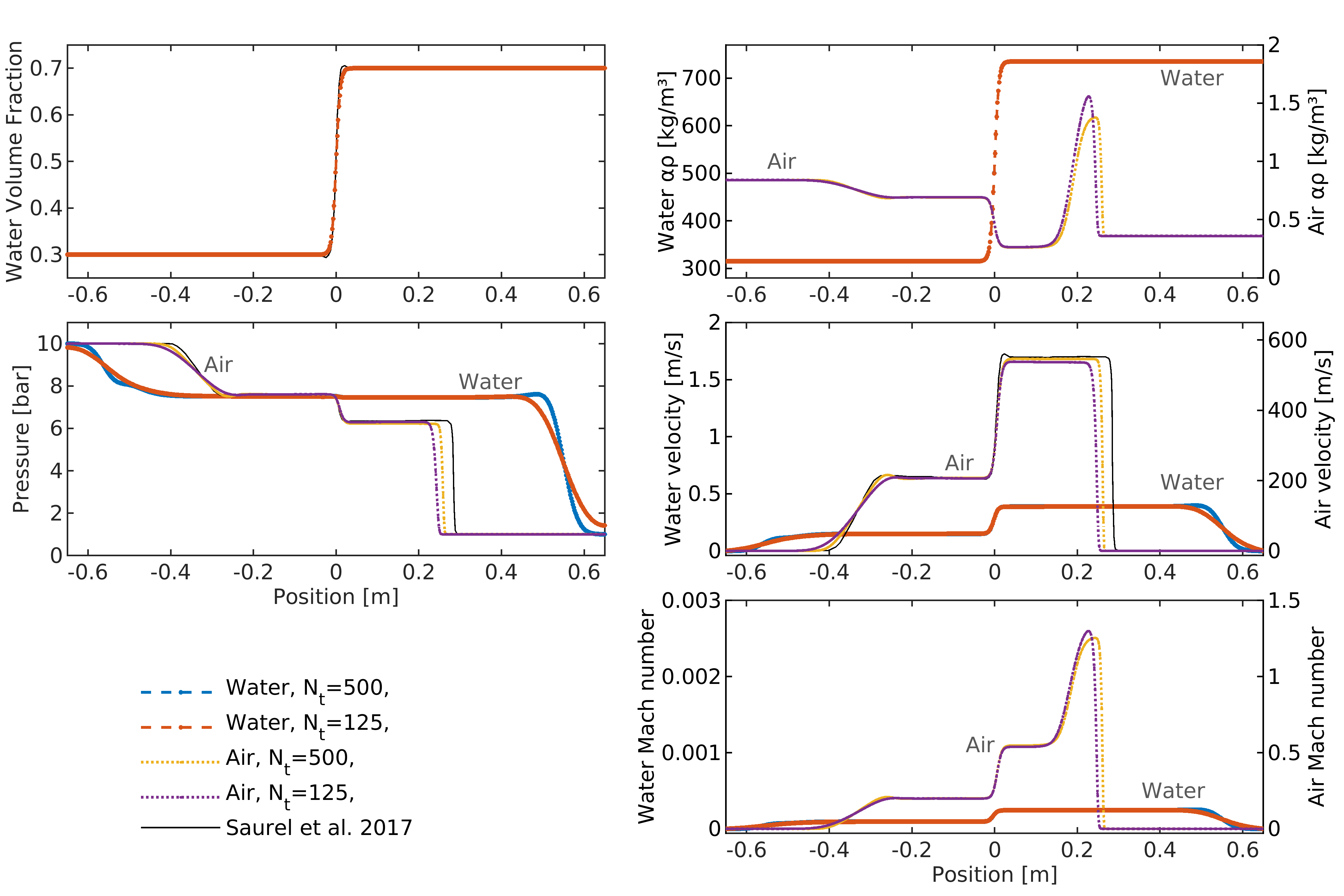}
\caption{Water-air mixture test: results at the time $t_F=350~\mathrm{\mu s}$, over a uniform grid with 650 cells, in absence of relaxation terms.
Two different time steps are considered: the smallest one (resulting from $N_T=500$ steps, and displayed in blue for water, and yellow for air) corresponds to $\max \mathrm{CFL}(|u|+c)\approx 0.5$ for both phases; the largest one (resulting from $N_T=125$ steps, and displayed in orange for water, and violet for air) corresponds to $\max \mathrm{CFL}(|u|+c)\approx 2.2$ for both phases.
The initial condition exhibits a pressure and volume fraction discontinuity at $x=0$, but its displacement (approximately $0.21~\mathrm{mm}$) is too small to be observed in these graphics.
The left column illustrates the water volume fraction and the pressures; the right column displays the partial densities $\arho$, the velocity and the Mach number of water (scales on left axis) and air (scales on the right axis).
The pressure and the velocity of the air, as well as the liquid volume fraction, given in Saurel et al.~\cite{Saurel2017} are displayed as reference.
}
\label{ftp:waterair}
\end{figure}

In this section, we reproduce the test proposed in~\cite{Saurel2017} under the name \textit{Smooth shock tube test case}.
The fluids are water and air but, differently from Sec.~\ref{ss:nomix}, an initial discontinuity is imposed also in the volume fraction.
The water is modeled under the stiffened gas model, using $P_\infty=6.0 \cdot 10^8~\mathrm{Pa}$ as in~\cite{Saurel2017}, while the remaining EOS parameters are the same given in Tab.~\ref{t:tmd_data}.
Initially, the fluids are at rest and the densities, $\rho_\mathrm{w}=1050~\mathrm{kg/m^3}$ for the water and $\rho_\mathrm{a}=1.2~\mathrm{kg/m^3}$ for the air, are uniform along the tube. 
Pressure and volume fraction are different: in the left chamber ($x<0$), $P= 10^6 ~\mathrm{Pa}$ and $\alpha_\mathrm{w}=0.3$, whereas in the right chamber ($x>0$), $P= 10^5 ~\mathrm{Pa}$ and $\alpha_\mathrm{w}=0.7$.
The domain $\Omega=[-0.65,0.65]~\mathrm{m}$ is divided in 650 primary cells.
The results computed at $t_F = 350 ~\mathrm{\mu s}$, with two different numbers of time steps ($N_t=500$ and $N_t=125$), are shown in Fig.~\ref{ftp:waterair}.
In the first case, the maximum acoustic CFL is smaller than one ($\max CFL(u+|c|) \approx 0.5$ for both phases), whereas in the second case, the acoustic CFL is greater than 2 for both phases, with convective CFL reaching $\max_\mathrm{a} \mathrm{CFL}(|u|)=0.75$ for air and  $\max_\mathrm{w} \mathrm{CFL}(|u|)=0.0005$ for water.
An estimation based on the values of the solution variables across the volume fraction discontinuity leads to a value for the interface velocity $u_I \approx 0.6~\mathrm{m/s}$; since at the final time step it has moved only about $\Delta x/10$, its displacement cannot be distinguished in Fig.~\ref{ftp:waterair}.
As a reference, Fig.~\ref{ftp:waterair} displays also the results for the air reported in~\cite{Saurel2017}.
The match is reasonably good in proximity to the rarefaction wave and the contact discontinuity.
On the contrary, the shock position is not captured correctly by the present method. This is an inherent limitation of the adopted pressure-based formulation and we are aware that the introduced error increases with the shock strength, however this test is on the boundary of the target application area, as the maximum Mach number of the air is well above one.
Concerning the water results, we cannot compare them with the results in~\cite{Saurel2017}, as their model is based on different assumptions which make the dispersed phase---water in this test---invariant across the shock. Nevertheless, as pointed out also in~\cite{Saurel2017}, such a large pressure disequilibrium between water and air in this test is not much physically reasonable.
Indeed, this test serves mainly as a further validation for the hyperbolic operator, especially when dealing with different scale velocities between the phases and acoustic CFLs greater than one: even with the largest time step, the scheme is stable and no spurious oscillations appear in the solution.

\section{Numerical results for two-phase flows with relaxation}\label{s:relax}
In this section, we finally present the results obtained with the full numerical scheme, that is including velocity and pressure relaxation.
In Sec.~\ref{ss:waterAlu}, the results of the BN-type model with pressure and velocity relaxation are compared to the analytical results of Kapila's model for two-phase flows in mechanical equilibrium, and the role of the finite relaxation parameters is investigated. 
Then, the water-air mixture test of Sec.~\ref{ss:waterAir} is re-run in Sec.~\ref{ss:waterAirRelax} with pressure relaxation to compare the results of the proposed model to the ones achieved through a different numerical method for a BN-type model. The last three subsections refers to specific features:
a strong rarefaction that generates a gas pocket in Sec.~\ref{ss:waterExpansion},
the simulation of almost-pure fluids in Sec.~\ref{ss:almostPure},
and the use of cubic equation of states in Sec.~\ref{ss:coo}.

\begin{table}
\caption{Stiffened gas parameters for the pure fluids used in the  two-phase numerical tests with relaxation involving a water-aluminum mixture (Sec.~\ref{ss:waterAlu}), a water-air mixture (Sec.~\ref{ss:waterAirRelax}), and almost-pure water and air flows (Sec.~\ref{ss:almostPure}).}
\label{t:tmd_data2}
\centering
\begin{tabular}{l*{4}{c}}\toprule
        &  $\gamma$ \small $[-]$& $P_\infty$ \small $[\mathrm{Pa}]$     &  $c_v$ \small $[\mathrm{J/kg\,K} ]$ & $q$ \small $[\mathrm{J/kg} ]$  \\ \midrule
 Aluminum: &  3.4      &  $21.5 \cdot 10^9$ &  897.0  & 0 \\
 Water: &     4.4      &  $6.0 \cdot 10^8$ &  4178.0  & 0 \\
 Air:   &     1.4      &  0          &  717.6  & 0 \\ \bottomrule
\end{tabular}
\end{table}

\subsection{Water-aluminum mixture test}\label{ss:waterAlu}
The test presented in this section was proposed by Furfaro et al.~\cite{Furfaro2015} and involves a mixture of two condensed phases, water and aluminum. 
The parameters of the stiffened gas models used in this test are summarized in Tab.~\ref{t:tmd_data2}.
Initially, the phases are uniformly dispersed with equal volume fraction $\alpha_1 = \alpha_2= 0.5$ in a shock-tube with two chambers: the left one at high pressure ($P_L =10^9\, \mathrm{Pa}$), and the right one at low pressure ($P_R =10^5\, \mathrm{Pa}$). The densities are uniform: $\rho_\mathrm{w}=1000~\mathrm{kg/m^3}$ for water and $\rho_\mathrm{al}=2700~\mathrm{kg/m^3}$ for aluminum. The initial velocity is zero everywhere.
Due to relaxation processes, as the time evolves, the volume fraction changes across the expansion and compression waves, besides the contact discontinuity originating from the initial pressure jump.
To build a reference solution, we consider the mechanical equilibrium model of Kapila~\cite{Kapila2001}, which is the limit model of BN model considering instantaneous pressure and mechanical relaxation, and the exact Riemann solver proposed in~\cite{Petitpas2007}.
The computed reference solution is given in Tab.~\ref{t:wateralu_sol}.
The maximum Mach number is $0.06$ (achieved by water) and this motivates the choice of this test to illustrate the capabilities of the proposed pressure-based method.

\begin{table}
\caption{Reference solution for the water-aluminum mixture test, computed according to the Riemann solver proposed by Petitpas et al. in~\cite{Petitpas2007} for the Kapila's model, which assumes mechanical equilibrium between phases.
The solution consists of a right-traveling rarefaction, a contact surface, and a left-traveling shock, and it is characterized by the four constant states given in this table. The two intermediate states, labeled  \textsf{left*} and  \textsf{right*}, are separated by the contact surface, across which the pressure $P$ and velocity $u$ are continuous. In addition to them, the densities of each fluid and the volume fraction of the water are given in the table.}
\label{t:wateralu_sol}
\centering
\begin{tabular}{l@{~}l*{4}{c}}\toprule
       &                          &  \textsf{left} &  \textsf{left*}   &  \textsf{right*}  & \textsf{right}\\  \midrule
 $P$   & \small $[\mathrm{Pa}]$   &     $10^9$     & $4.583\cdot10^8$  & $4.583 \cdot 10^8$ &    $10^5$ \\
 $u$   & \small $[\mathrm{m/s}]$  &     $0$        &    $124.1 $       & $124.1$            &   $0$       \\
 $\rho_\mathrm{w}$  & \small $[\mathrm{kg/m^3}]$  &   $1000$     &   $910.3$     &  $1134.0$     &   $1000$    \\
 $\rho_\mathrm{al}$ & \small $[\mathrm{kg/m^3}]$  &   $2700$     &   $2680.7$    &  $2716.8$     &   $2700$    \\
 $\alpha_\mathrm{w}$ & \small $[-]$  &    $ 0.5$   &     $0.5217$     &  $0.4701$           &   $0.5$       \\\bottomrule
\end{tabular}
\end{table}

\begin{figure}
\centering
\includegraphics[width=0.9\textwidth]{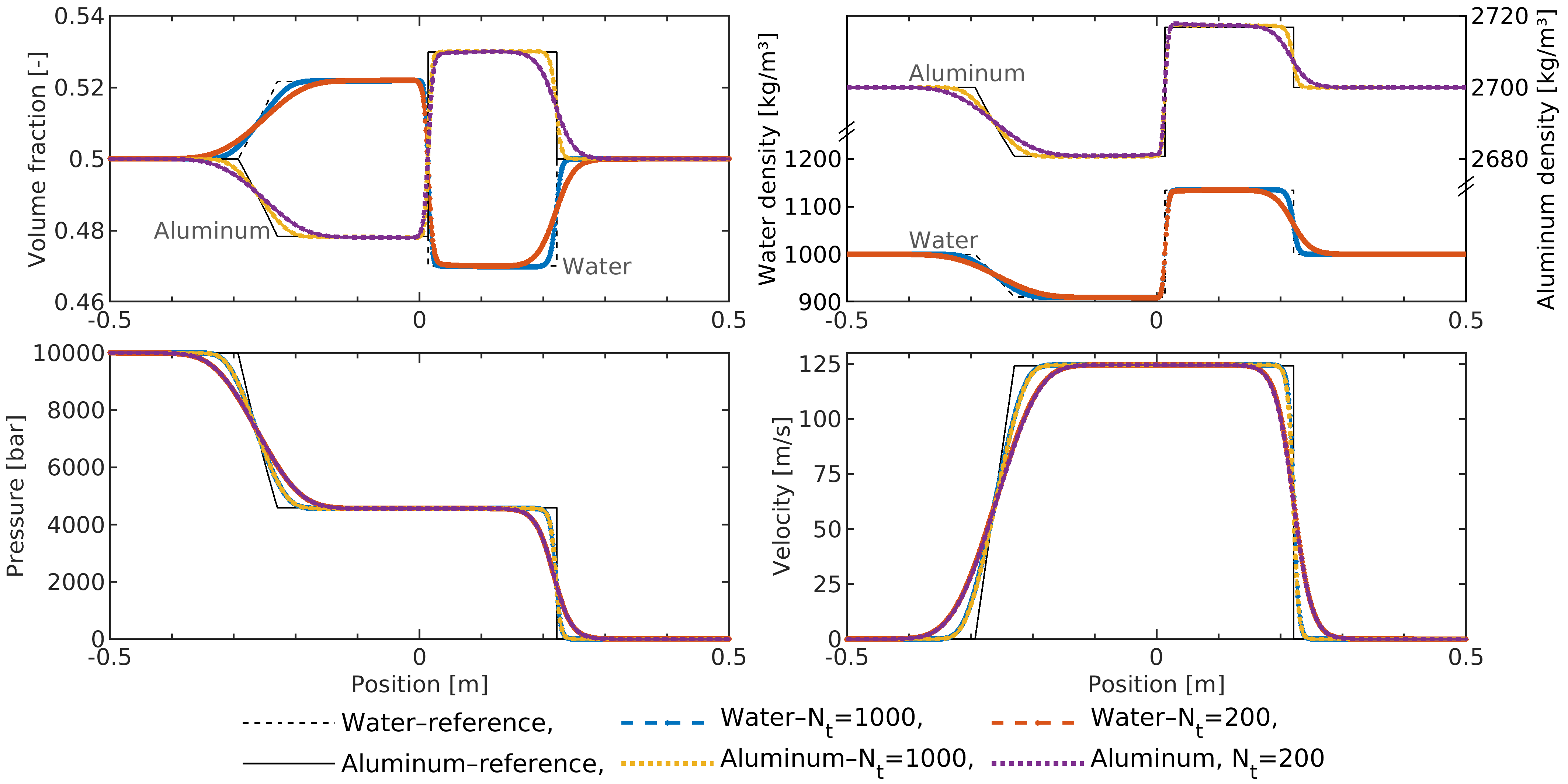}
\caption{Water-aluminum mixture test, with  $\lambda=10^9~\mathrm{kg/(m^3s)}$ and $\mu=10^5~\mathrm{m \, s/kg}$: results at $t_F=111~\mathrm{\mu s}$ obtained in two simulations characterized by a different number of time steps $N_t$, over a uniform mesh with $N_x=1000$ cells.
The choice $N_t=1000$ corresponds to $\max \mathrm{CFL}(|u|+c)$ of $0.3$ for water and of $0.6$ for aluminum, while $N_t=200$ leads to $1.5$ for water and $3.0$ for aluminum.
The results are compared with the analytical solution of Kapila's model given in Tab.~\ref{t:wateralu_sol}.
The top-left panel shows how the initially uniform volume fractions change due to the relaxation processes.
The top-right panel displays the phasic densities: please, note the different scales for water (left axis) and aluminum (right axis).
In the bottom line, we have the pressure and the velocity of each phase: for each simulation, it is impossible to distinguish between water and aluminum as the relaxation processes drive them toward the equilibrium. In these plots, the reference pressure and velocity are only one, displayed as solid line.}
\label{fr:waAl_cfrDT}
\end{figure}

\begin{figure}
\centering
\includegraphics[width=0.8\textwidth]{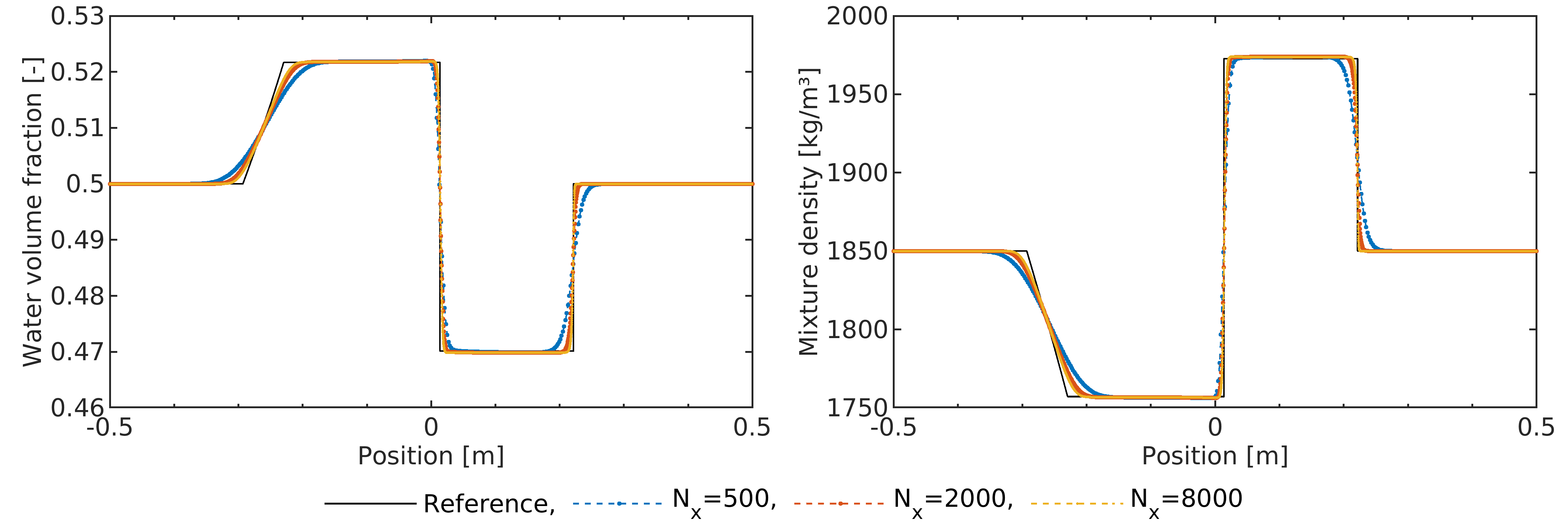}
\caption{Water-aluminum mixture test, with  $\lambda=10^9~\mathrm{kg/(m^3s)}$ and $\mu=10^5~\mathrm{m \, s/kg}$: grid convergence study at $t_F=111~\mathrm{\mu s}$ considering $\max \mathrm{CFL}(|u|+c_\mathrm{w})=0.3$  and $\max \mathrm{CFL}(|u|+c_\mathrm{al})=0.6$.
The results of three simulations over uniform grids with $N_x=\{500, 2000, 8000\}$ cells are displayed, along with the reference solution of Tab.~\ref{t:wateralu_sol}.
The variables are the water volume fraction (on the left) and the mixture density $\bar{\rho}=\arho_\mathrm{w} + \arho_\mathrm{al}$ (on the right).
}
\label{fr:waAl_gridconv}
\end{figure}

We start the illustration of the results considering a set of relaxation parameters sufficiently large to drive water and aluminum toward mechanical equilibrium, namely $\lambda=10^9~\mathrm{kg/(m^3s)}$ and $\mu=10^5~\mathrm{m \, s/kg}$.
The results at the time $t_F=111~\mathrm{\mu s}$ obtained with the proposed method correlate well with the reference solution, as Fig.~\ref{fr:waAl_cfrDT} shows.
A good match is reached also considering acoustic CFLs higher than one, indeed in the simulation carried out using $N_t=200$ time steps we have $\max \mathrm{CFL}(|u|+c)_\mathrm{w}=1.5$ and $\max \mathrm{CFL}(|u|+c)_\mathrm{al}=3.0$.
Considering these parameters, we conducted also a grid convergence study, illustrated in Fig.~\ref{fr:waAl_gridconv}. As expected, a coarse grid leads to smoother results, but with grid refinement, the solution converges toward the reference one.

\begin{figure}
\centering
\includegraphics[width=0.9\textwidth]{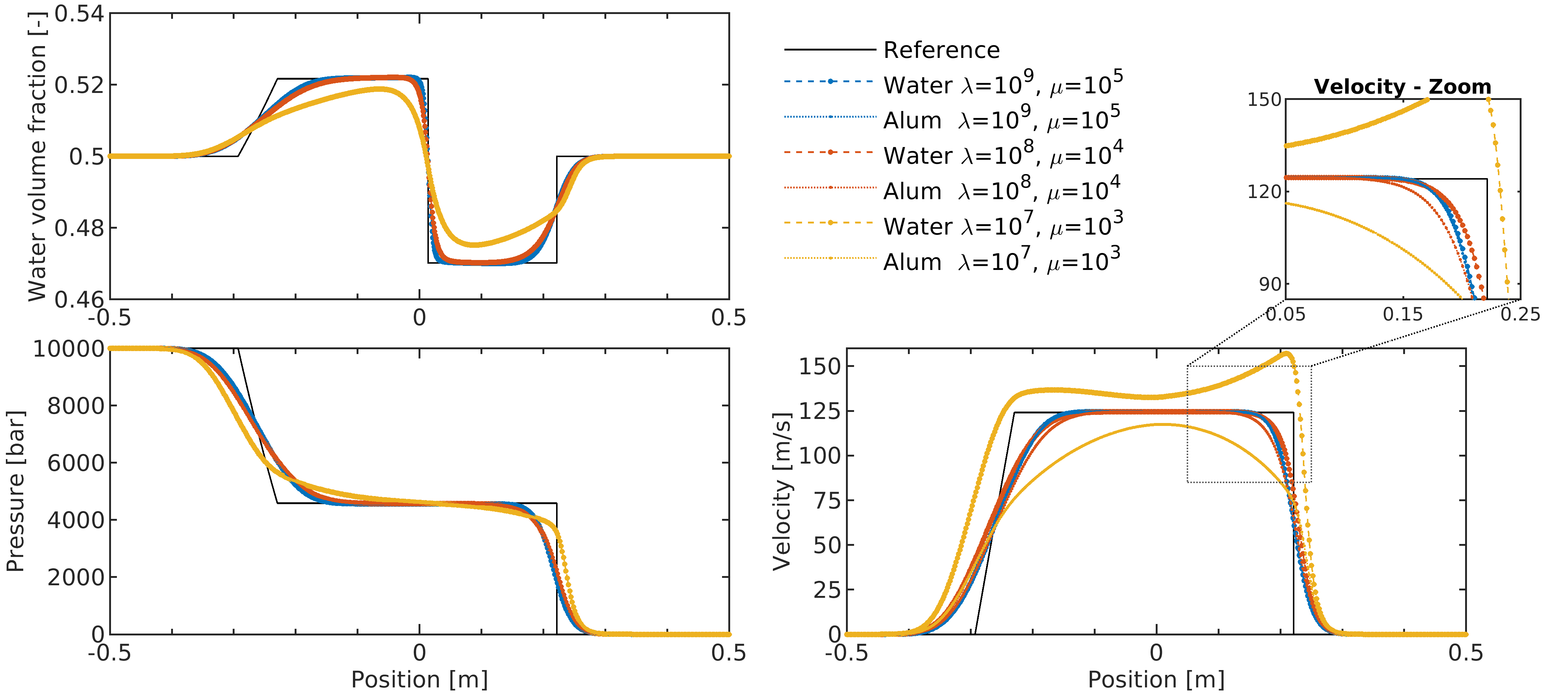}
\caption{Water-aluminum mixture test: effects of different finite relaxation parameters.
Results are computed at $t_F=111~\mathrm{\mu s}$, using $1000$ cells and $N_t=200$ time steps (i.e., approximately $\max \mathrm{CFL}(|u|+c_\mathrm{w})=1.5$  and $\max \mathrm{CFL}(|u|+c_\mathrm{al})=3.0$ in all tests).
The results obtained with three different sets of relaxation parameters $\lambda$ and $\mu$ are displayed, along with the reference solution of Tab.~\ref{t:wateralu_sol}.
The units of the $\lambda$ and $\mu$ (omitted in the legend for brevity) are, respectively, $\mathrm{kg/(m^3s)}$ and $\mathrm{m \, s/kg}$.
On the top-left, only the volume fraction of water is displayed for clarity.
On the bottom-left, the pressure profiles of water and aluminum of each single test are indistinguishable, but they differ among tests.
On the top-right, a zoomed view of the velocities near and after the shock is displayed to highlight the dis-equilibrium.
}
\label{fr:waAl_cfrRelPU}
\end{figure}

After the preliminary verification of the results, we use this test to illustrate now the effects of finite relaxation parameters.
In the previous tests, we have seen how finite, large values can successfully replicate the mechanical equilibrium.
Intuitively, smaller relaxation parameters leave a higher level of dis-equilibrium between phases.
This behavior is confirmed by Fig.~\ref{fr:waAl_cfrRelPU}, which displays the results obtained with different sets of relaxation parameters:
\begin{itemize}
\item the \textit{strong set} with $\lambda=10^9~\mathrm{kg/(m^3s)}$ and $\mu=10^5~\mathrm{m \, s/kg}$ used also in the previous tests,
\item an \textit{intermediate set} with $\lambda=10^8~\mathrm{kg/(m^3s)}$ and $\mu=10^4~\mathrm{m \, s/kg}$,
\item a \textit{mild set} with  $\lambda=10^7~\mathrm{kg/(m^3s)}$ and $\mu=10^3~\mathrm{m \, s/kg}$.
\end{itemize}
These tests are performed over a grid with $\Delta x=0.001~\mathrm{m}$ at acoustic CFLs higher than one.
From Fig~\ref{fr:waAl_cfrRelPU}, we note a substantial dis-equilibrium in the phasic velocity, in particular while using the \textit{mild set}, but it can be observed also for the \textit{intermediate set}. To have a quantitative idea, the maximum velocity difference in these sets is
$$ \mathit{intermediate}: \; \max(u_\mathrm{w}-u_\mathrm{al}) = 15.8~\mathrm{m/s}, \qquad
 \mathit{mild}:\; \max(u_\mathrm{w}-u_\mathrm{al}) = 78.4~\mathrm{m/s} \,. $$
Conversely, no disequilibrium in the pressure can be distinguished, but the effect of decreasing relaxation parameters is a smoothing of the expansion and compression waves, which causes the disappearance of the intermediate uniform regions between them (that is the regions called \textsf{left*} and  \textsf{right*} in the mechanical equilibrium solution in Tab.~\ref{t:wateralu_sol}).
This behavior is reflected also by the volume fraction profile, where this smoothing diminishes also the degree of the mixing, as with the \textit{mild set} the water volume fraction $\alpha_\mathrm{w}$ reaches only a maximum of $0.519$ after the rarefaction and a minimum of $0.475$ after the compression, instead of $\max(\alpha_\mathrm{w})=0.522$ and $\min(\alpha_\mathrm{w})=0.470$ as in other cases.

\begin{figure}
\centering
\includegraphics[width=0.9\textwidth]{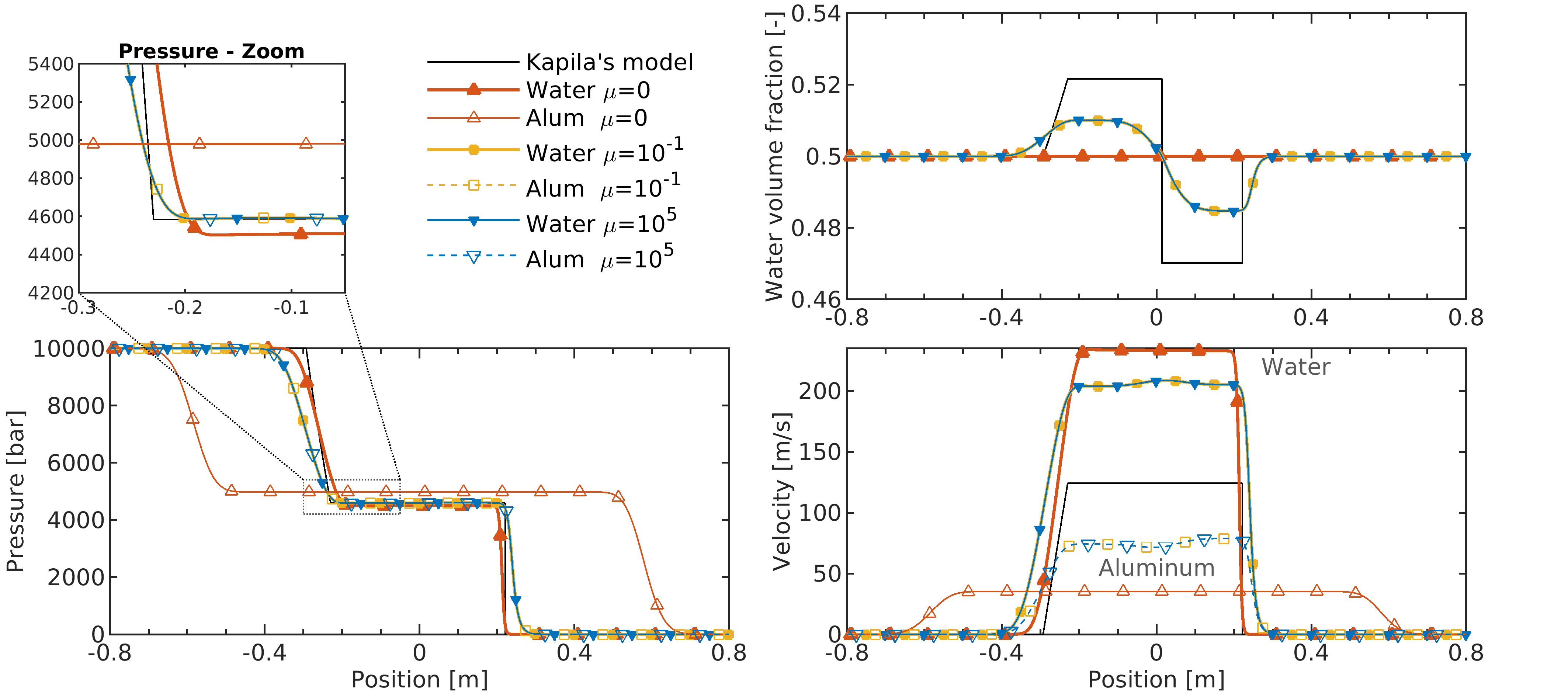}
\caption{Water-aluminum mixture test: effects of pressure relaxation, without velocity relaxation ($\lambda =0$).
Results are computed at $t_F=111~\mathrm{\mu s}$, using $1600$ cells and $N_t=200$ time steps (i.e., approximately $\max \mathrm{CFL}(|u|+c_\mathrm{w})=1.5$  and $\max \mathrm{CFL}(|u|+c_\mathrm{al})=3.0$ in all tests).
A first test is run without pressure relaxation ($\mu=0$), then two tests are run using $\mu\in \{ 10^{-1}, \, 10^5\}~\mathrm{m \, s/kg}$ (units are omitted in the legend for brevity).
The solution of Tab.~\ref{t:wateralu_sol} is also displayed for comparison with previous figures, but here it is labeled as ``Kapila's model'' as it is not a reference solution for this non-equilibrium test.
Markers distinguish lines which, especially for pressure, overlap between water and aluminum (filled vs. empty symbols) and between the two tests with pressure relaxation (differentiated  by yellow squares vs blue downward-pointing triangles).
A zoomed view of the pressure field close to the rarefaction end is shown in the top-left corner. 
}
\label{fr:waAl_cfrRelP}
\end{figure}

Finally, we better investigate the effects of pressure relaxation only, without relaxing the velocities (i.e., using $\lambda=0$).
Figure~\ref{fr:waAl_cfrRelP} compares the results obtained without pressure relaxation (so full disequilibrium, considering the hyperbolic operator only), with a \textit{weak} relaxation parameter $\mu=10^{-1}~\mathrm{m \, s/kg}$ and with a \textit{strong} one $\mu=10^{5}~\mathrm{m \, s/kg}$.
To allow for the faster waves in the first test, we consider a longer domain, $\Omega=[-0.8, \,  0.8]~\mathrm{m}$, but the same grid spacing and time step as before. 
The first observation we can make from Fig.~\ref{fr:waAl_cfrRelP} is that the \textit{weak} parameter is sufficient to drive the phasic pressures to the equilibrium, as the lines for water and aluminum appear to be overlapped also in the zoomed view. These lines overlap also to the ones corresponding to the \textit{strong} relaxation and the Kapila's model, confirming that the pressure equilibrium is achieved in both tests.
Similar observations between the two different relaxation parameters can be drawn also for the volume fraction and the velocity fields, which, conversely, differ significant from the Kapila's model because of the velocity dis-equilibrium.
Furthermore, comparing the results with and without pressure relaxation, we can observe that wave speeds in the relaxed condition are similar to the wave speeds in non-equilibrium water, that is much slower than the aluminum ones. The equilibrium pressure is also pretty similar to the non-equilibrium water, whereas the equilibrium velocity is an intermediate value between the ones of water and aluminum. 

\subsection{Water-air mixture test with pressure relaxation}\label{ss:waterAirRelax}
\begin{figure}
\centering
\includegraphics[width=0.9\textwidth]{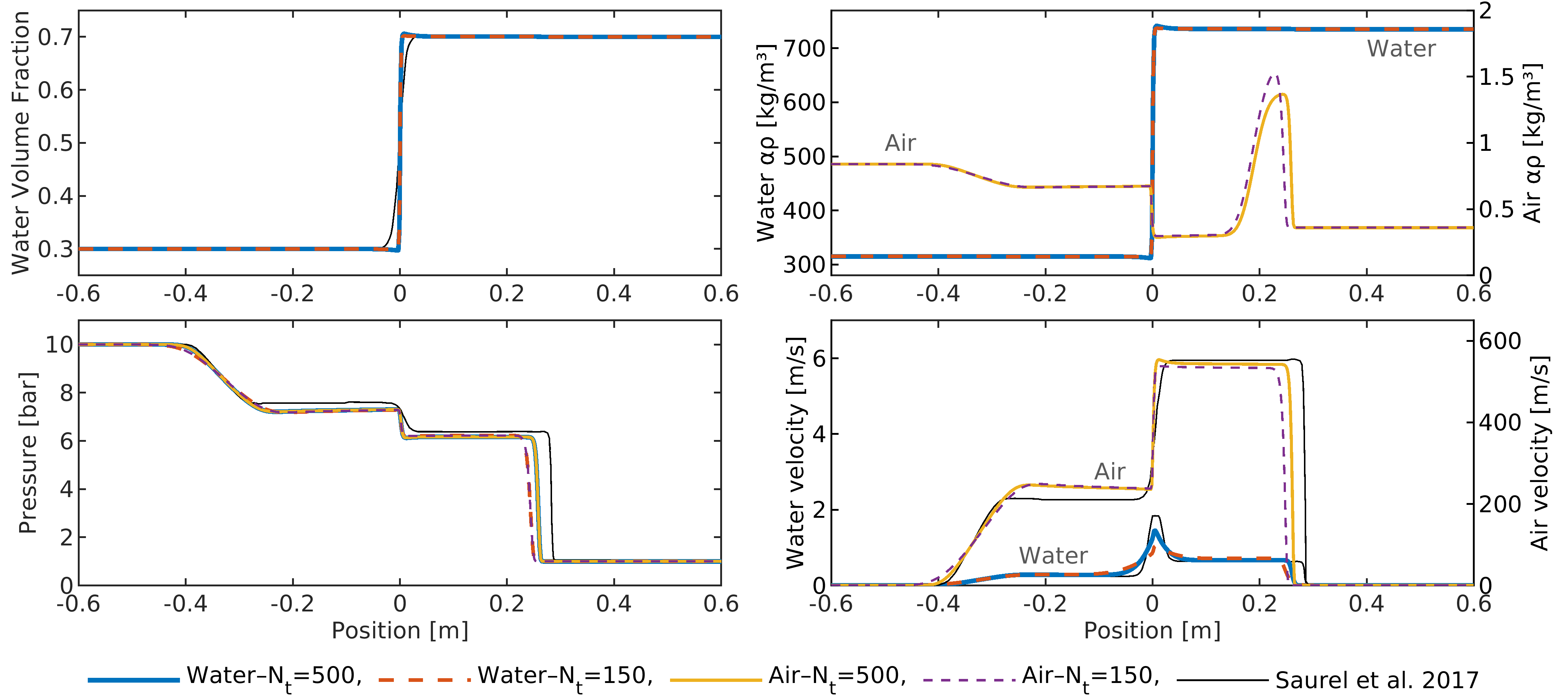}
\caption{Water-air mixture test with pressure relaxation ($\mu=10^5~\mathrm{m \, s/kg}$): results at the time $t_F=350~\mathrm{\mu s}$, over a uniform grid with 600 cells, in absence of velocity relaxation ($\lambda=0$).
Two different time steps are considered: the smallest one (resulting from $N_T=500$ steps, and displayed in blue and yellow solid lines) corresponds to $\max \mathrm{CFL}(|u|+c)\approx 0.55$ for both phases; the largest one (resulting from $N_T=150$ steps, and displayed in orange and violet dashed lines) corresponds to $\max \mathrm{CFL}(|u|+c)\approx 1.85$ for both phases.
In the left column, we see the water volume fraction and the pressures, which overlap between phases because of relaxation; in the right column, we have the partial densities $\arho$ and the velocity of each phase (note the different scales: for water on left axis, for air on the right axis).
Thin, black lines display the pressure, the velocity and the volume fraction in the limit of stiff pressure relaxation given by Saurel et al.~\cite{Saurel2017}.
}
\label{fr:waterairRelax}
\end{figure}

In this section, we reconsider the \textit{Smooth shock tube test case} proposed in~\cite{Saurel2017}, but differently from Sec.~\ref{ss:waterAir}, we use now pressure relaxation, and we compare our results again with the ones shown in~\cite{Saurel2017} under stiff pressure relaxation.
The initial conditions and the stiffened gas parameters of air and water are those given in Sec.~\ref{ss:waterAir} and in Tab.~\ref{t:tmd_data2}. In summary:  $\rho_\mathrm{w}(x)=1050~\mathrm{kg/m^3}$ and $\rho_\mathrm{a}(x)=1.2~\mathrm{kg/m^3}$, with 
$$\mathsf{left} \, (x<0):\;
P= 10^6 ~\mathrm{Pa}\,,\;\; \alpha_\mathrm{w}=0.3\,;
\qquad
\mathsf{right} \, (x>0):\;
P= 10^5 ~\mathrm{Pa}\,,\;\; \alpha_\mathrm{w}=0.7\,.$$ 
The domain $\Omega=[-0.6,0.6]~\mathrm{m}$ is divided in 600 primary cells.
Figure~\ref{fr:waterairRelax} shows the results computed at $t_F = 350 ~\mathrm{\mu s}$, with two different time steps---the larger one resulting in an acoustic CFL of about $1.85$ for both phases---and a relaxation parameter $\mu=10^5~\mathrm{m \, s/kg}$. 
The agreement with the reference results is not excellent, probably because the maximum Mach number for air is higher than $1.2$, so our non-conservative scheme is not able to correctly capture the shock speed and this inaccuracy affects also the pressure profiles, where we have a difference of about $0.3~\mathrm{bar}$ (i.e., $4\%$) after the rarefaction and of $0.2~\mathrm{bar}$ (i.e., $3\%$) after the shock. 
We also remind that the model used in~\cite{Saurel2017} is not symmetric and it is based on different modeling hypotheses for the water and we are not able to estimate whether and how this variation in the models may justify the discrepancy in the results.
Nevertheless, the general behavior of the pressure and velocity of both phases is consistent with the reference results, and this serves as a confirmation of the correctness of the pressure relaxation scheme.

\begin{figure}
\centering
\includegraphics[width=0.9\textwidth]{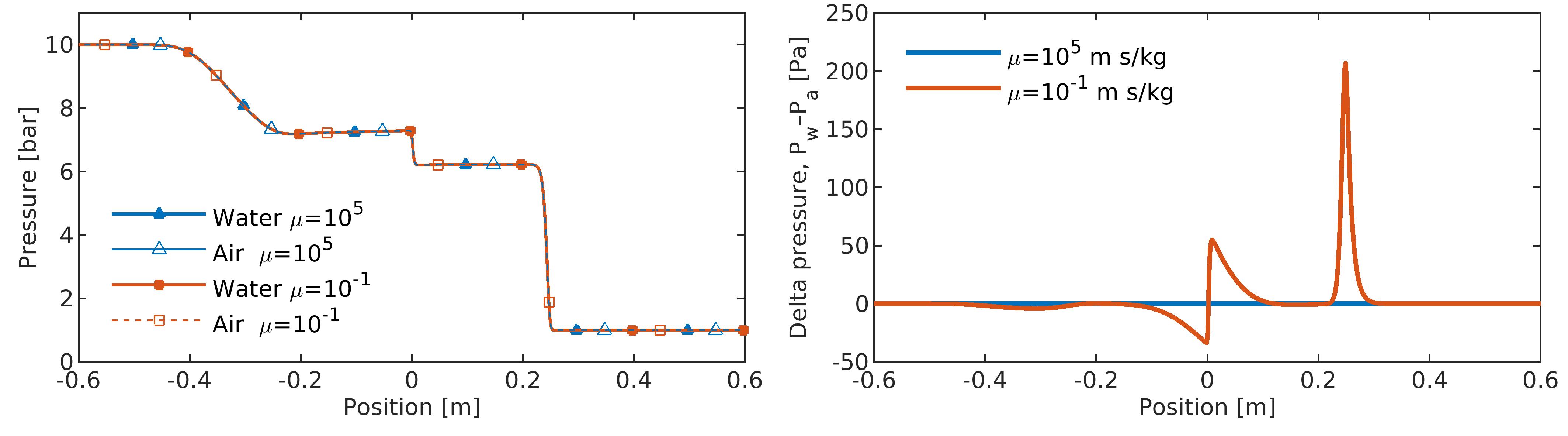}
\caption{Water-air mixture test with different pressure relaxation parameters: results at the time $t_F=350~\mathrm{\mu s}$, over a uniform grid with 600 cells, in absence of velocity relaxation ($\lambda=0$), considering $N_T=150$ time steps.
The compared $\mu$ values are $10^5$ (blue) and $10^{-1}~\mathrm{m \, s/kg}$ (orange).
On the left, the water and air pressure profiles (in $\mathrm{bar}$) are displayed, but it is impossible to distinguish any difference.
On the right, the pressure disequilibrium between water and air (in $\mathrm{Pa}$) is plotted over the domain.
}
\label{fr:waterairCfrMu}
\end{figure}

To better investigate the effect of the finite pressure relaxation parameter, we have re-run this test considering the largest time step and different values of $\mu$. Figure~\ref{fr:waterairCfrMu} compares the pressure profiles and the differences in the phasic pressure obtained with two values of $\mu$. The value $\mu=10^5~\mathrm{m \, s/kg}$ is large enough to drive the phasic pressures toward the equilibrium, and the phase difference is null everywhere.
On the other hand, the small value $\mu=10^{-1}~\mathrm{m \, s/kg}$ allows a certain degree of phase disequilibrium, especially across the shock, where it reaches the maximum ($206.7~\mathrm{Pa}$), and across the material discontinuity, where there is also a sign change. Indeed, due to the smearing of material discontinuity, the water on its left undergoes an expansion (the increase of the volume fraction has an effect similar to an expanding nozzle) and the water on its right is slightly compressed, and the weak pressure relaxation is not sufficient to overcome this phenomenon.

\subsection{Two-phase water expansion tube}\label{ss:waterExpansion}
The benchmark presented in this section was originally proposed in~\cite{Saurel2008} and it has been investigated also in~\cite{Zein2010,Han2017}, with and without phase transition. Here, we consider it without phase transition.
This test involves a tube filled with liquid water at pressure $P=1~\mathrm{bar}$ and density $\rho_\mathrm{liq}=1150~\mathrm{kg/m^3}$, to which a weak volume fraction of vapor $\alpha_\mathrm{vap}=0.01$ is added.
As in~\cite{Han2017}, we compute the vapor initial conditions from the pressure and temperature of the liquid, considering the stiffened gas parameters given in Tab.~\ref{t:tmd_dataWater}.
An initial velocity discontinuity is located at the center of the tube ($x=0$): the left velocity is  $-2~\mathrm{m/s}$, the right one is   $2~\mathrm{m/s}$.

\begin{figure}
\centering
\includegraphics[width=0.9\textwidth]{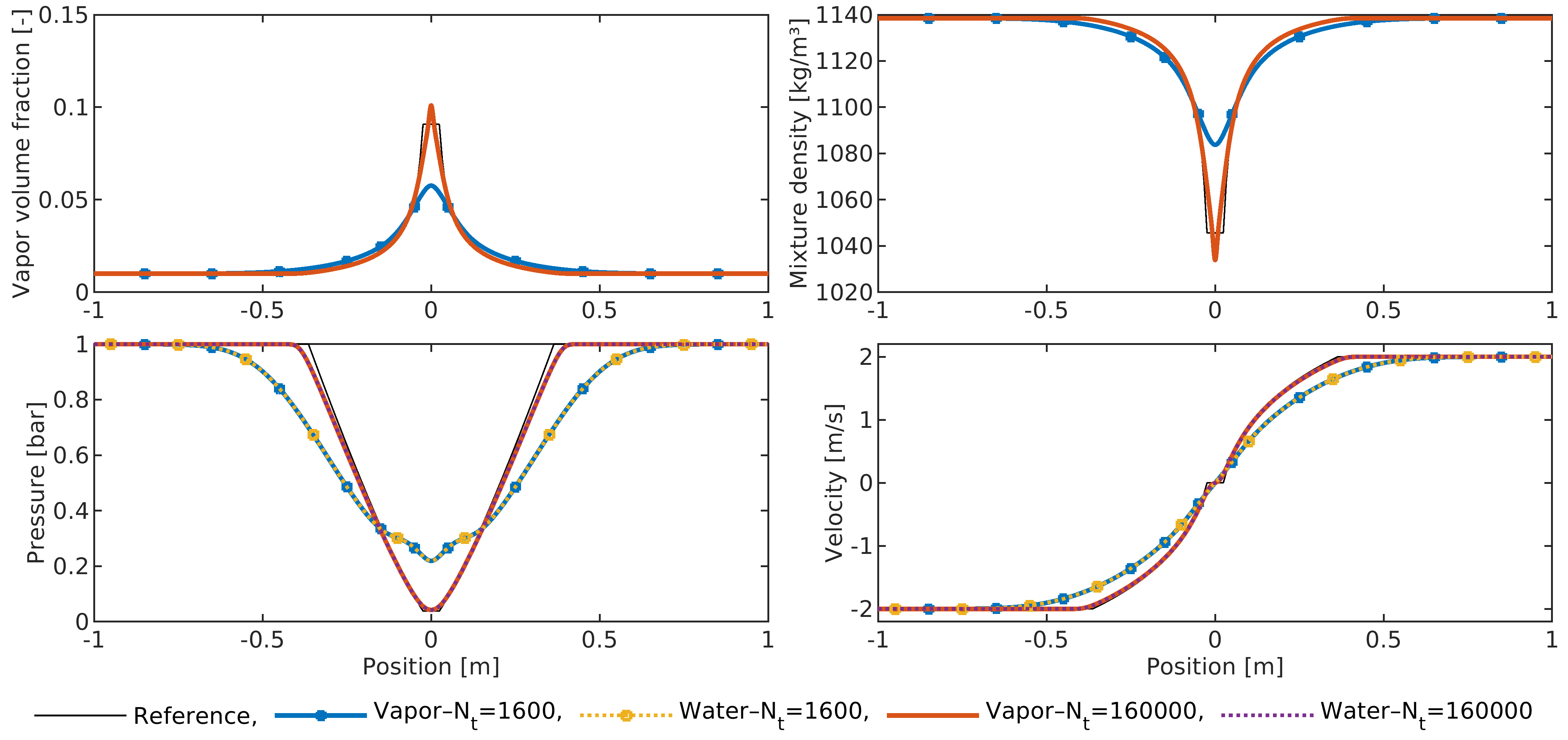}
\caption{Water expansion tube test: results at the time $t_F=3.2~\mathrm{ms}$, over a uniform grid with 20000 cells, with pressure and velocity relaxation ($\lambda=10^7~\mathrm{kg/(m^3s)}$ and $\mu=10^5~\mathrm{m \, s/kg}$).
The results obtained with $N_T=160000$ and with $N_T=1600$ time steps are displayed, along with the reference solution, computed assuming mechanical equilibrium between phases according to~\cite{Petitpas2007}.
The displayed variables are: on the top, vapor volume fraction (initially, $\alpha_\mathrm{vap}=0.01$ everywhere) and mixture density $\bar{\rho}=\arho_\mathrm{liq} + \arho_\mathrm{vap}$ (it follows the color of \textit{Vapor} in the legend);
on the bottom, pressure and velocity of each phases (\textit{Water} in the legend refers to liquid water).}
\label{fr:symRaref}
\end{figure}

The solution is computed at time $t_F=3.2~\mathrm{ms}$ and it consists of two symmetric rarefaction waves moving outwards. At the center of the domain, the volume fraction increases because of the vapor mechanical expansion and a gas pocket is dynamically generated~\cite{Saurel2008}. 
As for the water-aluminum test in Sec.~\ref{ss:waterAlu}, a reference solution is computed in the limit of stiff mechanic relaxation according to the Riemann solver proposed by Petitpas et al~\cite{Petitpas2007}.
The numerical results computed using a grid spacing $\Delta x = 10^{-4}~\mathrm{m}$ are shown in Fig.~\ref{fr:symRaref}.
First, we compute the solution using $N_T=160000$ time steps, corresponding to a maximal acoustic CFL of $0.3$ (for the liquid). With this choice,  we achieve a good match with the reference solution, especially in terms of pressure, capturing correctly the constant region between the rarefaction waves, while a little overshoot affects both the density and the volume fraction, but this behavior has been exhibited also in the previous works.
Considering the small velocities and the severe rarefactions involved in this test, the capability of the proposed low-Mach scheme to correctly compute the solution at mild CFL numbers is a notable result.
Remarkably, we perform a second test enforcing a 100 times bigger time step, leading to maximal acoustic CFL number of $28.6$ for the liquid and $15.8$ for the vapor.
As we can see from Fig.~\ref{fr:symRaref}, the dissipation at this level prevents to fully capture the intermediate state, being all profiles smeared with respect to the reference ones, but the general behavior of the fluids is well captured.
By comparison, Zein et al.~\cite{Zein2010} used a CFL equal to $0.03$ to obtain a stable solution, while Han et al.~\cite{Han2017} used a maximum CFL of $0.9$.

\begin{table}
\caption{Stiffened gas parameters for the two-phase water expansion tube, presented in Sec.~\ref{ss:waterExpansion}, as given in~\cite{Saurel2008}.}
\label{t:tmd_dataWater}
\centering
\begin{tabular}{l*{4}{c}}\toprule
        &  $\gamma$ \small $[-]$                  & $P_\infty$ \small $[\mathrm{Pa}]$   
        &  $c_v$    \small $[\mathrm{J/kg\,K} ]$  &  $q$       \small $[\mathrm{J/kg} ]$  \\ \midrule
 Liquid water:   &   $2.35$      &  $10^9$   &  $1816$  &  $-1167 \cdot 10^3$ \\
 Vapor water:    &   $1.43$      &  $0$      &  $1040$  &  $ 2030 \cdot 10^3$ \\\bottomrule
\end{tabular}
\end{table}

\subsection{Almost pure fluids test}\label{ss:almostPure}
Here, we present a test involving almost pure fluids, water and air, with material parameters given in Tab.~\ref{t:tmd_data2}.
The configuration involves a shock-tube with a left chamber ($x<0$) filled with air at $P_L =100\, \mathrm{bar}$ and the right chamber ($x>0$) filled with water at $P_R =50\, \mathrm{bar}$.
Since the proposed method is not able to deal with null volume fractions, we define the air volume fraction in the left chamber as $\alpha_\mathrm{a}=1-\varepsilon$ and the water one in the right chamber as  $\alpha_\mathrm{w}=1-\varepsilon$, where $\varepsilon=10^{-4}$.
The air and water densities are  $\rho_\mathrm{a}=100~\mathrm{kg/m^3}$ and $\rho_\mathrm{w}=1000~\mathrm{kg/m^3}$, respectively.
Starting from a state of rest, the almost pure liquid on the right is set into motion by the almost pure gas at higher pressure on the left.
Considering pure fluids separated by a material discontinuity, it is possible to compute the analytical solution according to the Euler equations. It comprises a left-traveling rarefaction for air and a right-traveling shock wave for water, while the intermediate state is characterized by a pressure $P^\star=98.887\, \mathrm{bar}$ and a velocity $u^\star = 2.989~\mathrm{m/s}$.
This velocity confirms the low-Mach regime, as the maximum Mach number is $0.008$, and the shock speed is $1636~\mathrm{m/s}$ ($1.0025$ times the pre-shock speed of sound).

\begin{figure}
\centering
\includegraphics[width=0.9\textwidth]{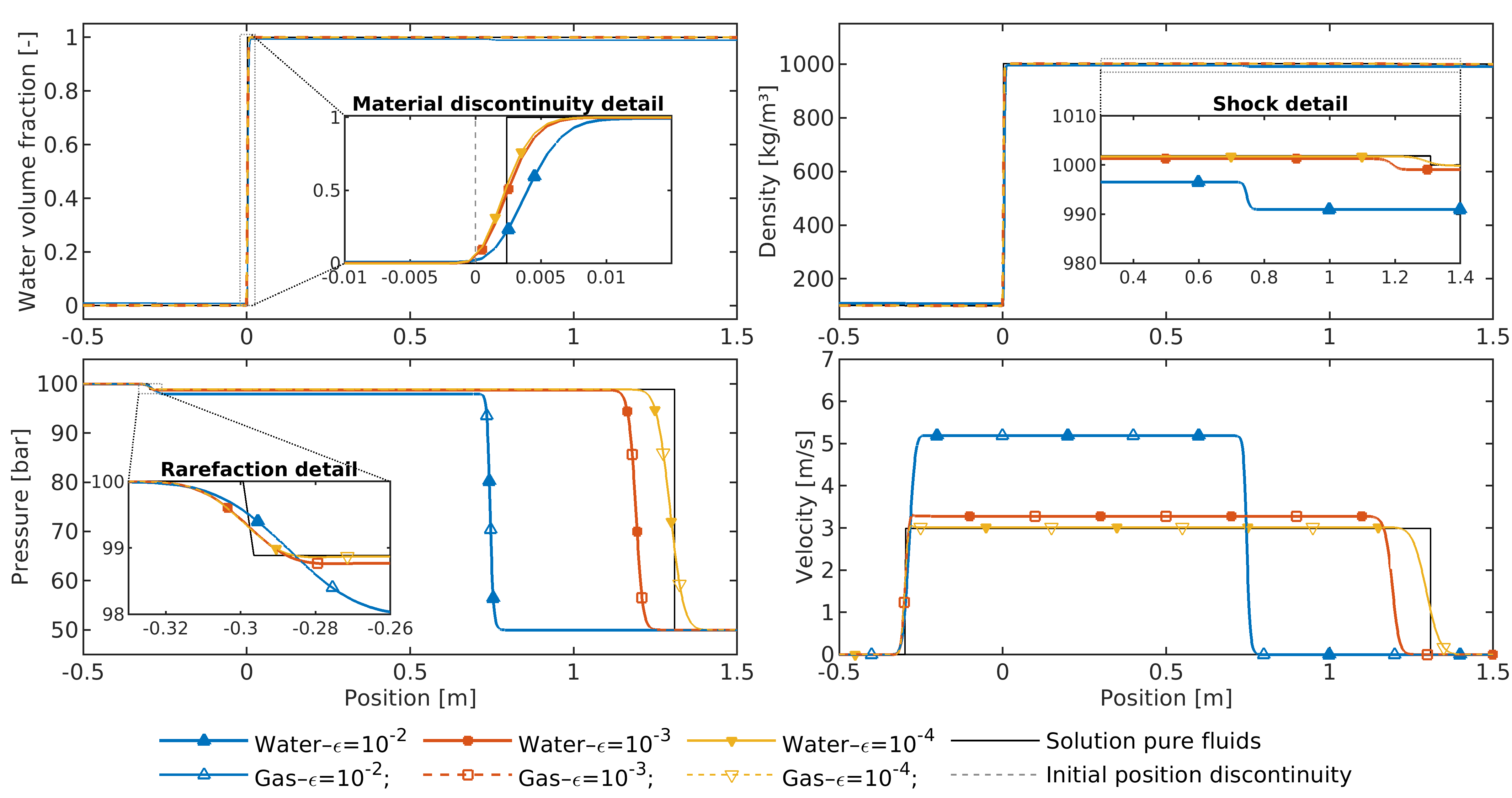}
\caption{Almost pure fluids test: results at the time $t_F=0.8~\mathrm{ms}$, over a uniform grid with 2000 cells ($\Delta x=10^{-3}~\mathrm{m}$), considering $N_t=1200$ time steps, and imposing pressure and velocity relaxation.
On the top-left, the water volume fraction is displayed, along with a detailed view close to the material discontinuity, where the dashed vertical line illustrates the initial position of the material discontinuity. On the top-right, the density, or more precisely the mixture density $\bar{\rho}$, is displayed, with a zoom of the region close to the shock; the disagreement among the initial densities (clearly visible for the blue line at $x=1.4$) is due to the different values of $\varepsilon$, which lead to different weights in the average of the densities.
On the bottom line, we have the pressure profiles, with a zoomed view of the rarefaction, and the velocities.
In all frames, the solid black line displays the pure-fluid solution computed analytically.
}
\label{fr:apureShort}
\end{figure}

The numerical results at $t_F=0.8~\mathrm{ms}$, obtained with pressure and velocity relaxation, are displayed in Fig.~\ref{fr:apureShort}  and they reach an excellent agreement with the analytical, pure-fluid solution, except the shock wave which is considerably smeared.
Figure~\ref{fr:apureShort} shows also how the solution varies for higher values of $\varepsilon$, that is for higher levels of mixing. The figure contains some zoomed frames to highlight the details close to the most important flow structures.
The level of mixture given by $\varepsilon=10^{-2}$, that is the same as the one considered for the liquid-water expansion test in sec.~\ref{ss:waterExpansion}, is already enough to significantly depart from the pure-fluid behavior. Indeed, the intermediate velocity is higher, resulting in a faster material discontinuity and a slower shock.
On the other hand, $\varepsilon=10^{-3}$ is already sufficient to capture qualitatively the behavior of the pure fluids, but it leads to some discrepancies in the values of the intermediate state. These are overcome by using the value $\varepsilon=10^{-4}$.

\begin{figure}
\centering
\includegraphics[width=0.9\textwidth]{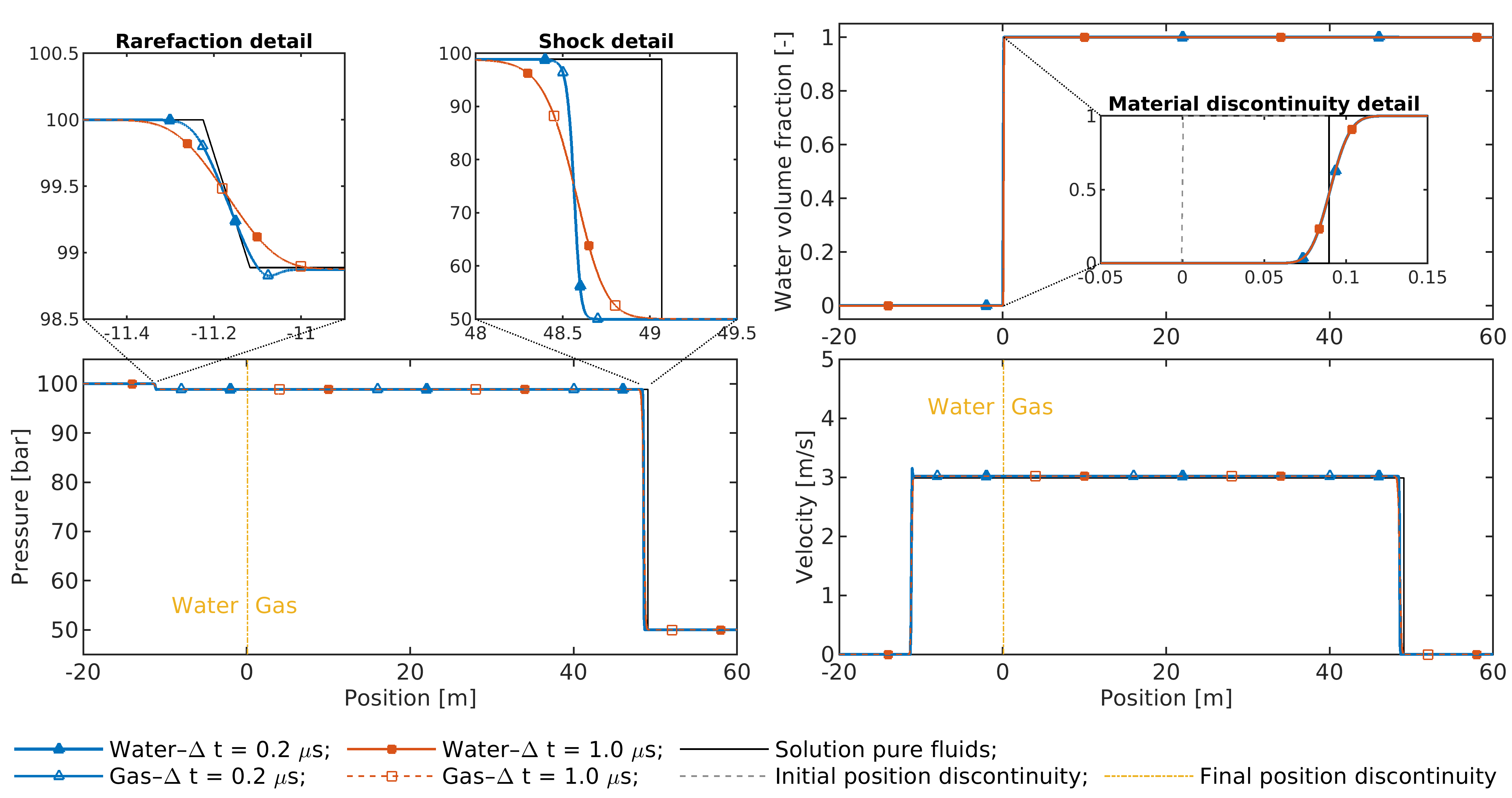}
\caption{Almost pure fluids test: results at the time $t_F=0.03~\mathrm{s}$, over a uniform grid with 80000 cells ($\Delta x=10^{-3}~\mathrm{m}$), with pressure and velocity relaxation. The results obtained with two time steps (leading to $\max(|u|+c)_\mathrm{w}$ of $0.33$ and $1.64$) are compared.
On the top-right, the water volume fraction is displayed, along with an highlight of the displacement of the material discontinuity (here, the grey dashed line illustrates its initial position, the solid lines the final one).
On the bottom, the pressure and velocity are displayed and the final position of the material discontinuity is shown by a yellow vertical line.
On the top-left corner, two zoomed views of the pressure fields in proximity of the rarefaction fan and of the shock wave are shown.
In all frames, the solid black line displays the pure-fluid solution computed analytically.
}
\label{fr:apureLong}
\end{figure}

To investigate better the capability to correctly capture the interface between almost pure fluids, we repeat this simulation, for a long time. We consider a longer domain, $\Omega=[-20, 60]~\mathrm{m}$, with the same grid spacing ($\Delta x = 10^{-3}~\mathrm{m}$) and we compute the solution at $t_F=0.03~\mathrm{s}$, using the time steps $\Delta t = 0.2~\mu s$ and $\Delta t = 1.0~\mu s$, corresponding to a maximal acoustic CFL for water of $0.33$  and $1.64$, respectively.
The results are displayed in Fig.~\ref{fr:apureLong}.
The final position of the material interface  is $x^\star=0.0897~\mathrm{m}$ and it is computed correctly, as it can be observed in the zoomed view in the top-right corner of Fig.~\ref{fr:apureLong}. Similarly, also the intermediate state and the rarefaction fan are in excellent agreement with the analytical solution. On the contrary, the position of the shock is not computed accurately, because the non-conservativeness of the scheme introduces an error in the shock velocity, which, after a long time, results visible in the position.
However, the relative error in the shock position is about $1.0\%$, so it is acceptable.

\subsection{Two-phase carbon dioxide test in saturation conditions, with accurate EOS}\label{ss:coo}
In this final section, we compare the results computed under the stiffened gas approximation with the ones computed using a more accurate thermodynamic model for CO\textsubscript{2}, based on the Peng-Robinson (PG) EOS~\cite{PengRobinson1976}.
In particular, we take advantage of the implementation of this EOS  tailor-made for CO\textsubscript{2} flows, provided by an in-house thermodynamic library developed at SINTEF Energy~\cite{Wilhelmsen2017}, that exploits the concept of the corresponding states to enhance the accuracy of specific properties, such as the density and the speed of sound, for the liquid phase.

\begin{figure}
\centering
\includegraphics[width=0.5\textwidth]{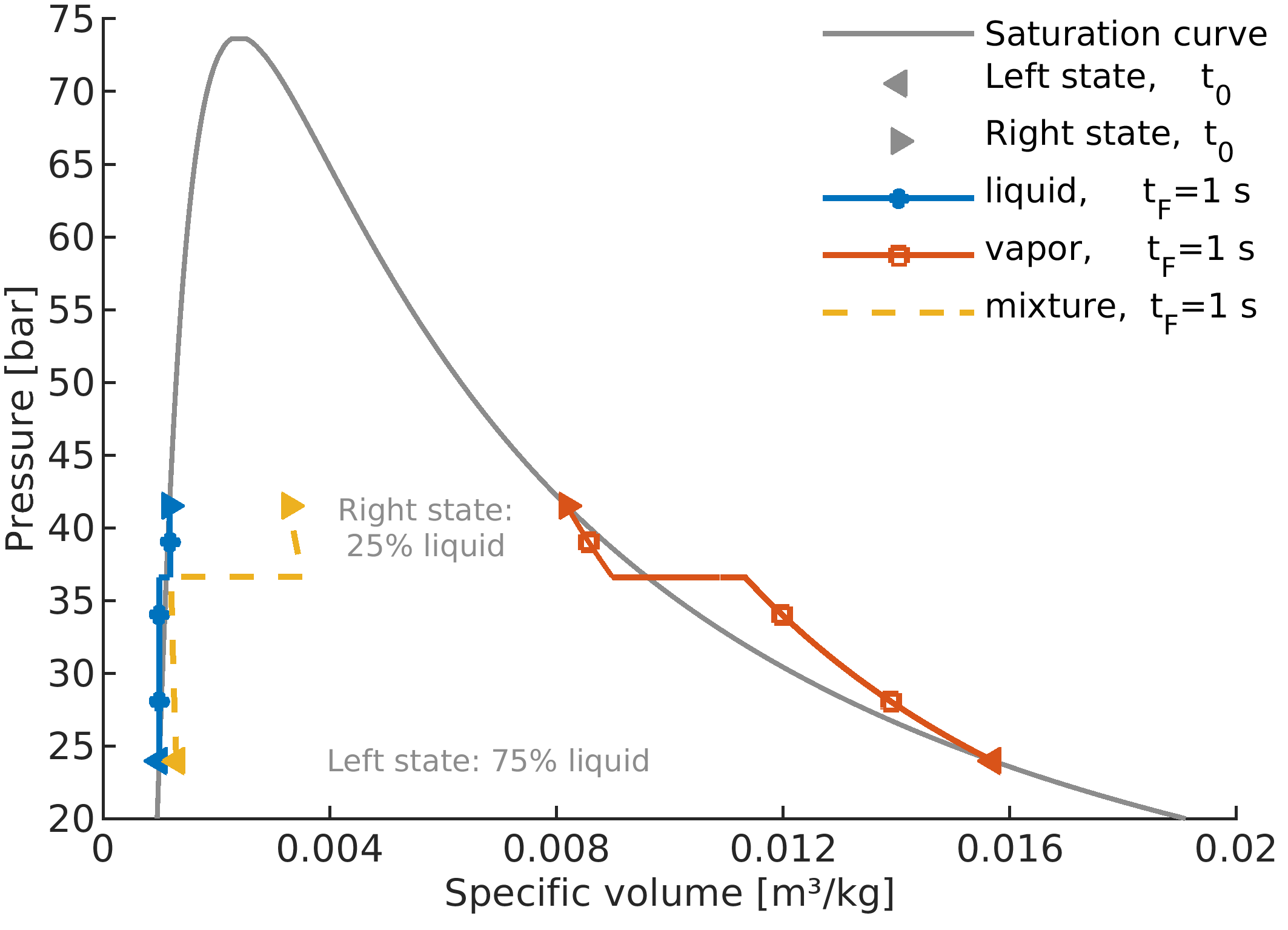}
\caption{Two-phase CO\textsubscript{2} test: visualization of the flow field computed using the Peng-Robinson~EOS in the thermodynamic plane pressure--specific volume. The saturation curve is plotted on the background with a grey line. The initial states for the liquid (blue) and vapor (orange) phase, as well as the CO\textsubscript{2} mixture (yellow), are displayed by full triangles: left-pointing triangles refer to the left state (saturated temperature $T_L=260~\mathrm{K}$) and right-pointing ones refer to the right state (saturated temperature $T_R=280~\mathrm{K}$).
The results at $t_F=1.0~\mathrm{s}$ are displayed by lines.
The evolution of the mixture (dashed line) is plotted as $1/\bar{\rho}$ versus the mixture pressure $\bar{P}=\alpha P_\mathrm{liq} + \alpha P_\mathrm{vap}$, where however the phasic pressures $P_\mathrm{liq}$ and $P_\mathrm{vap}$ are equal thanks to relaxation.
}
\label{fr:co2VLE}
\end{figure}

The set-up in which this comparison is carried out consists in a pipe, $600~\mathrm{m}$ long, filled with two-phase CO\textsubscript{2} mixture at saturation conditions, that is the initial conditions of the fluids, represented in the thermodynamic plane, lie along the saturation, or vapor-liquid equilibrium, curve.\footnote{A representation of the initial conditions in the pressure-volume plane is given in Fig.~\ref{fr:co2VLE}; for illustrations of the saturation curve of CO\textsubscript{2} with different EOSs, see~\cite{Hammer2013}.}
In the left part of the domain, the CO\textsubscript{2} mixture is composed by $75\%$ of liquid ($\alpha_\mathrm{liq}=0.75$) and $25\%$ of vapor  ($\alpha_\mathrm{vap}=0.25$), at the saturation temperature $T_L=260~\mathrm{K}$.
In the right part, the composition is inverted and the saturation temperature is $20~\mathrm{K}$ higher, that is $\alpha_\mathrm{liq}=0.25$, $\alpha_\mathrm{vap}=0.75$ and  $T_R=280~\mathrm{K}$.
We remind that, at saturation conditions, defining the temperature unequivocally defines also the pressure and the density of each phase.
Hence, in the test with the Peng-Robinson EOS, no other inputs are defined.
Conversely, in the test with the stiffened gas model, as we do not have a saturation model, we do need to specify also the initial pressure, for which we use the saturated values computed according to the Peng-Robinson EOS, that are  $P_L= 23.98~\mathrm{bar}$ and $P_R= 41.5~\mathrm{bar}$. The parameters of the stiffened gas model for CO\textsubscript{2} are taken from~\cite{LundAursand2012} and reported in Tab.~\ref{t:tmd_coo}.

\begin{figure}
\centering
\includegraphics[width=0.9\textwidth]{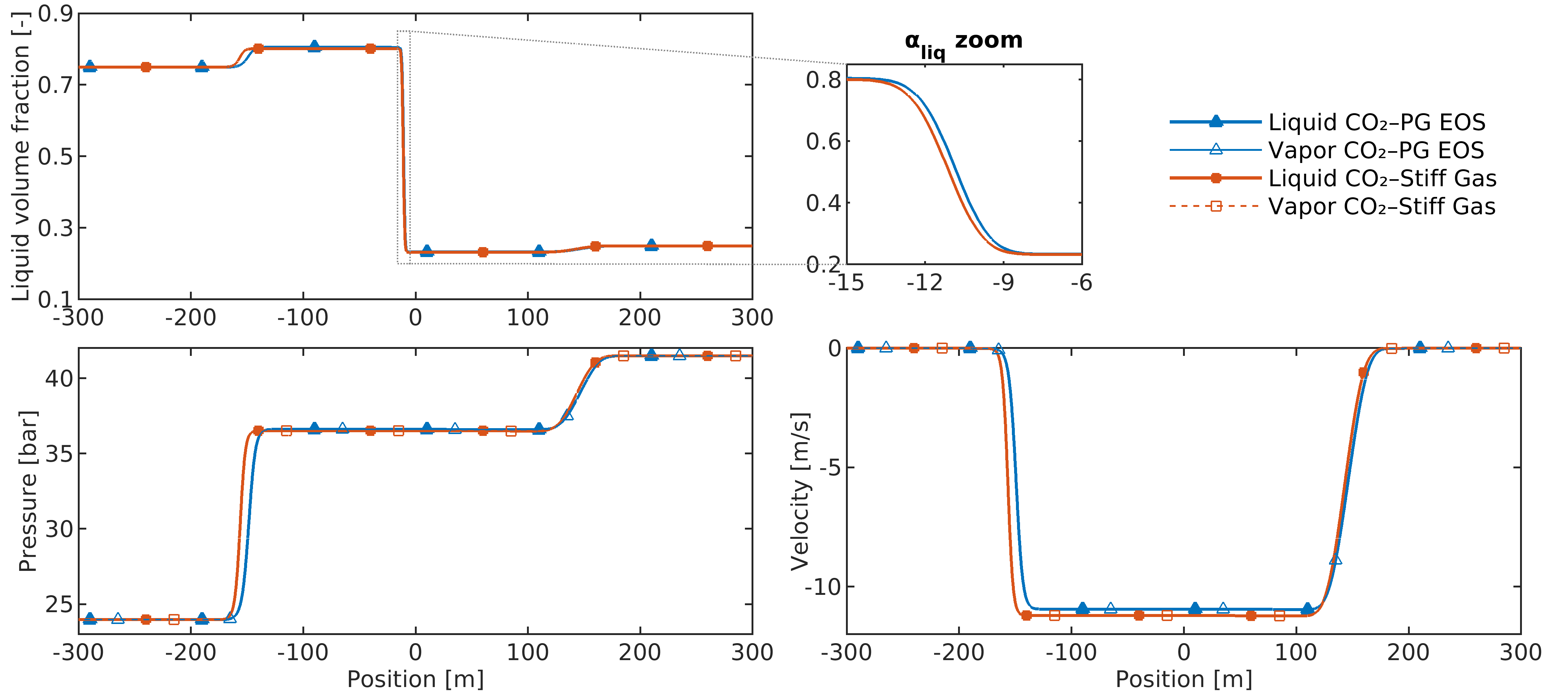}
\caption{Two-phase CO\textsubscript{2} test: results at the time $t_F=1.0~\mathrm{s}$, over a uniform grid with 6000 cells ($\Delta x=0.1~\mathrm{m}$), considering $N_t=2000$ time steps.
Initially, the left ($x<0$) and right ($x>0$) parts of the tube contain a saturated mixture with different composition and at different temperatures.
The results obtained with the Peng-Robinson EOS (blue and triangular markers, \textsf{PG EOS} in the legend) are compared to the ones obtained with the stiffened gas model (orange and square markers, \textsf{Stiff Gas} in the legend).
On the top, the liquid volume fraction is displayed, with a detailed view close to the material discontinuity in the box on the right. 
On the bottom, the pressure and the velocities are displayed: the liquid (full markers) and vapor (empty markers) CO\textsubscript{2} are driven to the equilibrium by means of relaxation.
}
\label{fr:co2Profiles}
\end{figure}

We compute the flow field for $t_F=1~\mathrm{s}$ from the moment the diaphragm separating the two mixtures, initially at rest, is removed. We consider $N_T=2000$ time steps and a grid spacing $\Delta x= 0.1~\mathrm{m}$, which result in $\max CFL(\vert u\vert +c)_\mathrm{liq}=2.3$ and $\max CFL(\vert u\vert +c)_\mathrm{vap}=1.2$. Higher values of CFL could be enforced without preventing stability, but we found that these values leads to a good compromise between accuracy and efficiency.
Figure~\ref{fr:co2Profiles} illustrates the profiles of the liquid volume fraction, the pressure and the velocity, which are driven toward the equilibrium by relaxation processes with $\lambda=10^7~\mathrm{kg/(m^3s)}$ and $\mu=10^3~\mathrm{m \, s/kg}$.
The different thermodynamic models lead to a difference in the minimum velocity reached in the intermediate region between the shock and the rarefaction wave. This is reflected on the position of the material discontinuity: as highlighted by the zoomed view on the top-right box in Fig.~\ref{fr:co2Profiles}, the stiffened gas predicts a larger displacement (in the negative $x$-direction) of the contact discontinuity due to faster fluids velocities.
Quantitatively, the position of the contact discontinuity is $-11~\mathrm{m}$ using the Peng-Robinson EOS, while it is $-11.3~\mathrm{m}$ using the stiffened gas.

\begin{figure}
\centering
\includegraphics[width=0.4\textwidth]{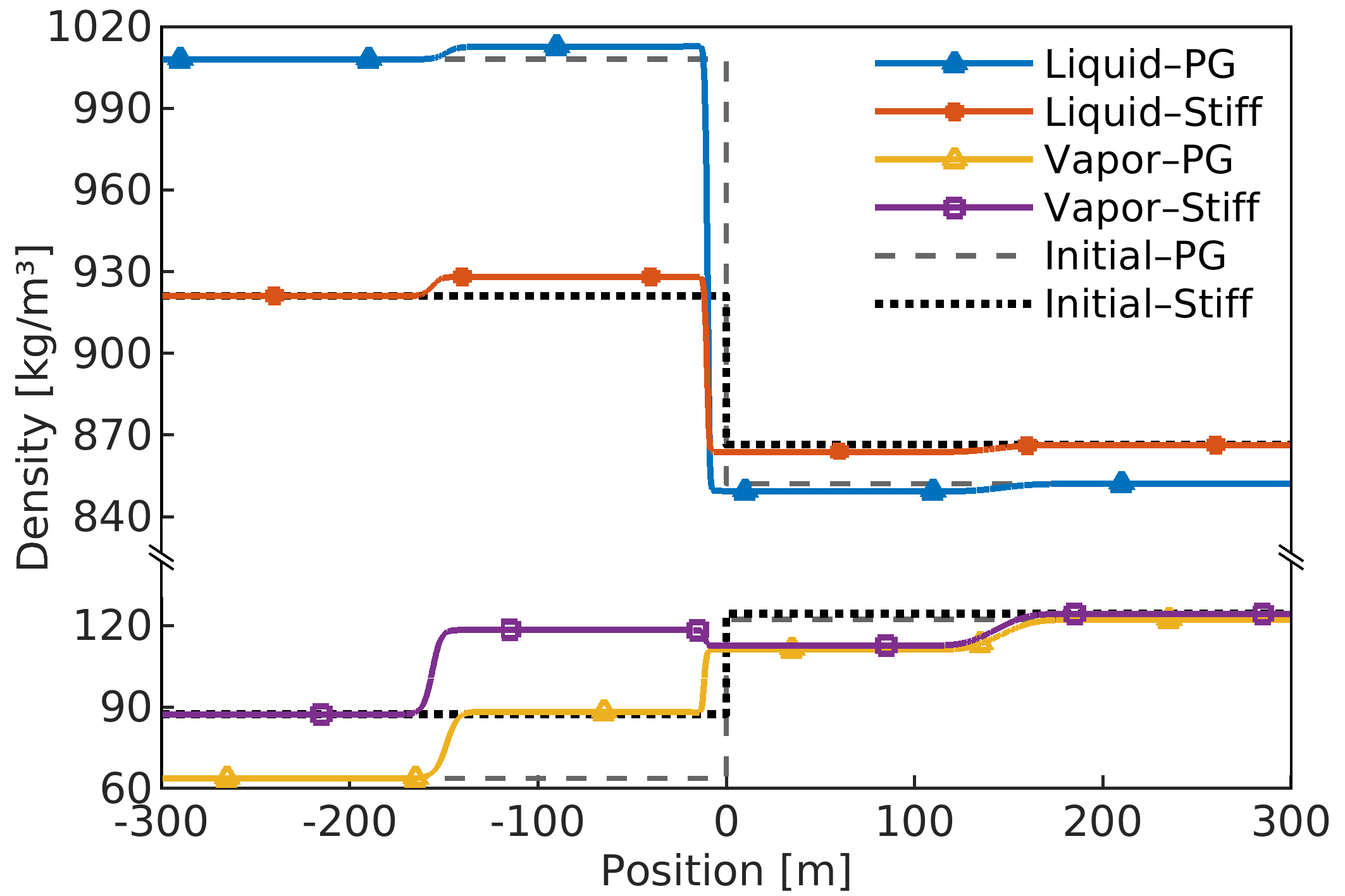}
\caption{Two-phase CO\textsubscript{2} test: comparison of the  density ($\rho_\mathrm{liq}$ and $\rho_\mathrm{vap}$) at the initial time (black dashed and dotted lines) and at $t_F=1.0~\mathrm{s}$ (colored lines with markers) computed with Peng-Robinson EOS (black dashed; blue and yellows lines with triangular markers) and with stiffened gas model (black dotted; orange and violet lines with square markers). The upper part of the plot refers to the liquid CO\textsubscript{2}, while the bottom part refers to the vapor CO\textsubscript{2}. Note that the $y$-axis is discontinuous, but the tick spacing is preserved.
}
\label{fr:co2Density}
\end{figure}

The differences introduced by the thermodynamic models are clearer in the density profiles, displayed in Fig.~\ref{fr:co2Density}.
Given as input the same temperatures and pressures, the initial conditions entail already discrepancies in the density:
$$\text{for liquid:}\;\; \rho_{L,\mathrm{liq}}^\mathrm{PG}> \rho_{L,\mathrm{liq}}^\mathrm{Stiff} \quad \text{and}
 \quad \rho_{R,\mathrm{liq}}^\mathrm{PG}< \rho_{R,\mathrm{liq}}^\mathrm{Stiff}  \; ;
\qquad 
\text{for vapor:}\;\; \rho_{L,\mathrm{vap}}^\mathrm{PG}< \rho_{L,\mathrm{vap}}^\mathrm{Stiff} \quad \text{and} 
\quad \rho_{R,\mathrm{vap}}^\mathrm{PG}< \rho_{R,\mathrm{vap}}^\mathrm{Stiff}\,.
$$
Moreover, Peng-Robinson~EOS predicts stronger initial density jumps for both phases.
This leads to a notable disagreement in the behavior of the CO\textsubscript{2} vapor after the diaphragm rupture, when the two thermodynamic models predict opposite (in sign) density jumps across the material discontinuity.

Finally, we plot the results obtained with the Peng-Robinson~EOS in the pressure-volume thermodynamic plane, in Fig.~\ref{fr:co2VLE}. This illustration gives a clear idea of how  the proposed full non-equilibrium BN-type model allows each phase to evolve independently according to its own thermodynamic model, although the phasic pressures (and velocity) are immediately driven toward the equilibrium. 
The resulting flow states lie in close proximity to one side (liquid on the left, vapor on the right) of the dome, and only the evolution of the mixture density develops in the fully two-phase region. However, no special treatment, such as the definition of a speed of sound for the mixture, is required for this situation.

\begin{table}
\caption{Stiffened gas parameters for the carbon dioxide tests presented in Sec.~\ref{ss:coo}, as given in~\cite{LundAursand2012}}
\label{t:tmd_coo}
\centering
\begin{tabular}{l*{4}{c}}\toprule
        &  $\gamma$ \small $[-]$& $P_\infty$ \small $[\mathrm{Pa}]$   &  $c_v$ \small $[\mathrm{J/kg\,K} ]$ & $q$       \small $[\mathrm{J/kg} ]$ \\ \midrule
 Liquid CO\textsubscript{2}: &     1.23      &  $1.32 \cdot 10^8$ &   $2440$ & $-6.23 \cdot 10^5$\\
 Vapor  CO\textsubscript{2}: &     1.06      &  $8.86 \cdot 10^5$ &  $2410$  &  $-3.01 \cdot 10^5$ \\\bottomrule
\end{tabular}
\end{table}

\section{Summary and conclusions}\label{s:conclusion}
Starting from the symmetric variant proposed by Saurel and Abgrall~\cite{Saurel1999}, we derived a pressure-based BN-type model for weakly compressible two-phase flows, which, as illustrated by Remark~\ref{rmk:incompressibility}, encompasses the divergence-free condition for multiphase flows in the zero-Mach limit, recovering the correct scaling of the pressure.
Although the single components of the finite-volume solver we built to solve the 1D pressure-based BN-type model are not novel by themselves, the resulting numerical method is unique and includes notable features.
For instance, it preserves by construction the non-disturbance condition on pressure and velocity, it overcomes the stringent limitation on the time step imposed by the acoustics, it allows for different thermodynamic models, and it provides a wide variety of modeling choices thanks to the relaxation terms with finite parameters, which can be used to control whether and how the phasic pressure and/or velocity are driven toward the equilibrium.

The main drawback of the proposed method relates to the non-conservative character of the pressure-based formulation, which prevents to resolve accurately shock waves.
However, this behavior was expected and numerical tests showed that, for  moderate intensities, the error in the position of the shock is between $1$ and $2\%$, consistently with the predictions made by Karni~\cite{Karni1996}.
Notwithstanding the acceptable impact on weakly compressible flows, a remedy for this limitation is imperative to move toward an all-speed scheme, so this matter will be given a high priority in the future development of the method.
To mitigate the effects of non-conservativeness, we will consider strategies already proposed for single-phase flows---for instance, an additional scheme-dependent viscous term that provides a correction at the leading order of the discrete approximation~\cite{Karni1992}, or the use of the pressure equation only as a prediction for variables in the total energy equation~\cite{Kwatra2009}---and for reduced two-phase flows, e.g., considering a supplementary, explicit correction step~\cite{Zhang2019} or moving to residual distribution schemes~\cite{Abgrall2018}.

As told in the introduction, this work describes the first stage of a longer project, which aims to develop a reliable and robust tool for multi-phase simulations of weakly-compressible flows. To reach this aim, the numerical scheme underlying the solution strategy described in this work should be enhanced: higher-order discretization techniques will be considered, in addition to the previously mentioned correction for non-conservativeness.
Afterwards, we will extend the proposed numerical tool to multi-dimensional grids. In particular, to account for complex geometries, we will look at unstructured grids, for which a face-based staggered discretization has been recently proposed by Berm\'{u}dez et al.~\cite{Bermudez2020} in the framework of weakly compressible single-phase flows.

In conclusion, we propose here the first numerical tool based on the BN-type model that includes simultaneously all the features outlined in the first paragraph of this conclusive section.
As identified above, the proposed method is not free from defects, but the main limitations could be overcome by adapting the solution strategy to include some of the techniques already proposed in the literature for other models.
Given these points, we think this work could pave the way for a wider application of BN-type models, especially to investigate two-phase flows involving liquid components, which generally exhibit low Mach number and require specific, accurate thermodynamic models. The ingredients of the solution strategy, taken individually, are not complex, so the main ideas presented in this work could be applicable also in software used in applied research and industry, where simplicity and efficiency are of utmost importance.
Similarly, the modular nature of the presented method could facilitate its integration in other BN solvers, mainly used in academic research, so far.
Furthermore, the use of finite relaxation parameters may enhance the modeling capabilities of BN--type models. Indeed, if experimental data were available, parametric analyses could be carried out to define the best values (or range of values) for each multiphase regimes, contributing to increase the fidelity of numerical results.

\section*{Acknowledgement}
This publication has been produced with support from the \textit{NCCS Centre}, performed under the Norwegian research program Centres for Environment-friendly Energy Research (FME). The authors acknowledge the following partners for their contributions: Aker Solutions, Ansaldo Energia, CoorsTek Membrane Sciences, EMGS, Equinor, Gassco, Krohne, Larvik Shipping, Norcem, Norwegian Oil and Gas, Quad Geometrics, Shell, Total, V\aa r Energi, and the Research Council of Norway (257579/E20).

This work was initiated while B. Re was a post-doctoral researcher at University of Z\"{u}rich (UZH). The author gratefully acknowledges the financial support received under the grant \textit{Forschungskredit} of the University of Zurich, grant no. [FK-20-121].

The authors would like to thank Svend Tollak Munkejord and Morten Hammer (SINTEF, Norway) for the constructive discussions about multiphase CO\textsubscript{2} flows and the help with the implementation of the thermodynamic library.

\bibliography{Biblio2020}

\appendix

\section{Derivation of the pressure formulation for the BN-type model}
\label{a:pform}
In this appendix, we show step by step how we have achieved the pressure formulation of Eq.~\eqref{emp:peq}, starting from the equation of the total energy in~\eqref{em:bn}.

For a generic fluid satisfying the EOS $e=e(\rho,P)$, we can express the partial derivative of $E=e + \frac{m^2}{2\rho}$ with respect to a generic variable $\xi$ as
\begin{equation*}
\pder{E}{\xi} =  
\left[ \tmdder{e}{P}{\rho} \pder{P}{\xi} + \tmdder{e}{\rho}{P} \pder{\rho}{\xi} \right] + u  \pder{m}{\xi} - \frac{u^2}{2} \, \pder{\rho}{\xi}
\end{equation*}
where the terms in square brackets are the partial derivative $\pder{e}{\xi}$. Reminding definitions~\eqref{emt:tmdder} and the triple product rule, i.e., 
$ \tmdder{e}{P}{\rho} \tmdder{P}{\rho}{e} \tmdder{\rho}{e}{P} = -1$, we can re-write the derivative of $E$ as
\begin{equation}
\label{eaa:dEtdXi}
\pder{E}{\xi} =  
\left[ \frac{1}{\kappa} \pder{P}{\xi} -\frac{\chi}{\kappa} \pder{\rho}{\xi} \right] + u  \pder{m}{\xi} - \frac{u^2}{2} \, \pder{\rho}{\xi} \,.
\end{equation}

Now, we insert Eq.~\eqref{eaa:dEtdXi} into the total energy equation for phase $\phSymbol$ in \eqref{em:bn}, re-formulating it as
\begin{equation}
\label{eaa:tmp1}
\begin{split}
\aph \left[
\frac{1}{\kappa\ph}\pder{\Pph}{t} -\frac{\chi\ph}{\kappa\ph} \pder{\rhoph}{t} + \uph  \pder{\mph}{t} -  \frac{\uph^2}{2} \, \pder{\rhoph}{t} \right]  
&+\left( e\ph + \frac{\mph^2}{2\rhoph} \right) \left[ \pder{\aph}{t}  + \pder{\alpha \uph}{x} \right]\\
+ \alpha \uph \left[
\frac{1}{\kappa\ph}\pder{\Pph}{x} -\frac{\chi\ph}{\kappa\ph} \pder{\rhoph}{x} + \uph  \pder{\mph}{x} -  \frac{\uph^2}{2} \, \pder{\rhoph}{x} \right]
&+ \pder{(\alpha \Pph \uph)}{x} - \ppi \ui\pder{\aph}{x} =  \ppi \mu\Delta\ph P - \ui \lambda\Delta\ph u
\end{split}\end{equation}
where we have introduced the operator $\Delta\ph$ which takes the difference between the phase $\phSymbol$ and the opposite one, i.e., $\Delta_1 P = P_1 - P_2$ and $\Delta_2 u = u_2 - u_1$.

Terms in Eq.~\eqref{eaa:tmp1} can be re-arranged as
\begin{equation*}
\begin{split}
\frac{1}{\kappa\ph} \left[\aph\pder{\Pph}{t}+\alpha\uph \pder{\Pph}{x}\right]  
&-\left(\frac{\chi\ph}{\kappa\ph} +\frac{\uph^2}{2} \right) \left[\aph \pder{\rhoph}{t} + \alpha \uph \pder{\rhoph}{x}\right]  \\
+\uph \left[ \aph \pder{\mph}{t} + \alpha \uph  \pder{\mph}{x}  \right]  
&+\left( e\ph + \frac{\mph^2}{2\rhoph} \right) \left[ \pder{\aph}{t}  + \pder{\alpha \uph}{x} \right]\\
+ \pder{(\alpha \Pph \uph)}{x} - \ppi \ui\pder{\aph}{x} &= -\mu \ppi \Delta\ph P - \lambda \ui \Delta\ph u \, .
\end{split}\end{equation*}
Now, we use the density and momentum equations in \eqref{em:bn} to replace the derivative of $\rhoph$ and $\mph$:
\begin{equation}
\begin{split}
\frac{1}{\kappa\ph} \left[\aph\pder{\Pph}{t}+\alpha\uph \pder{\Pph}{x}\right]  
&+\left(\frac{\chi\ph}{\kappa\ph} +\frac{\uph^2}{2} \right) \rhoph \left[\pder{\aph}{t} + \pder{\alpha\uph }{x} \right]  \\
-\uph \left[\mph \pder{\aph}{t} + \mph  \pder{\alpha\uph}{x} + \pder{\alpha \Pph}{x} - \ppi \pder{\aph}{x} + \lambda \Delta\ph u \right]  
&+\left( e\ph + \frac{\mph^2}{2\rhoph} \right) \left[ \pder{\aph}{t}  + \pder{\alpha \uph}{x} \right]\\
+ \pder{(\alpha \Pph \uph)}{x} - \ppi \ui\pder{\aph}{x} &= -\mu \ppi \Delta\ph P - \lambda \ui \Delta\ph u \, .
\end{split}\end{equation}
Expanding the derivative of $\alpha \Pph \uph$ and noting that $\frac{\uph^2}{2} \rhoph  -\uph \mph+\frac{\mph^2}{2\rhoph} =0$, the previous equation reads
\begin{equation}
\begin{split}
\frac{1}{\kappa\ph} \left[\aph\pder{\Pph}{t}+\alpha\uph \pder{\Pph}{x}\right]  
&+\left(e\ph  + \frac{\chi\ph \rhoph}{\kappa\ph} \right)  \left[\pder{\aph}{t} + \pder{\alpha\uph }{x}\right]  \\
 +\left[  \uph\ppi \pder{\aph}{x} - \uph \lambda \Delta\ph u \right]  
&+  \alpha \Pph\pder{\uph}{x} - \ppi \ui\pder{\aph}{x} = -\mu \ppi \Delta\ph P - \lambda \ui \Delta\ph u \, .
\end{split}\end{equation}
Now, we use also the volume fraction equation in  \eqref{em:bn} to replace the temporal derivative of $\aph$, and we re-arrange the terms, to have
\begin{equation}
\begin{split}
\frac{1}{\kappa\ph} \left[\aph\pder{\Pph}{t}+\alpha\uph \pder{\Pph}{x}\right]  
&+\left(\frac{\chi\ph \rhoph}{\kappa\ph}+ \Pph + e\ph \right) \aph \pder{\uph}{x} - \left(\frac{\chi\ph \rhoph}{\kappa\ph} + \ppi + e\ph \right)(\ui - \uph)\pder{\aph }{x} \\ 
& = -\left(\frac{\chi\ph \rhoph}{\kappa\ph} +\ppi + e\ph \right)  \mu \Delta\ph P -(\ui- \uph ) \lambda  \Delta\ph u \, .
\end{split}\end{equation}

Finally, recalling the definitions of the speed of sound \eqref{emt:spsound} and \eqref{emt:spsoundInt}, we arrive to 
\begin{equation}
\aph\pder{\Pph}{t}+\alpha\uph \pder{\Pph}{x} 
+\alpha\rhoph c\ph^2 \pder{\uph}{x} - \rhoph \cci{\phSymbol}(\ui - \uph)\pder{\aph }{x} 
 = -\rhoph \cci{\phSymbol}\mu \Delta\ph P - \kappa\ph (\ui- \uph ) \lambda  \Delta\ph u \, ,
\end{equation}
which is the pressure formulation of the BN-type model given in Eq.~\eqref{emp:peq}.

\section{Definition of non-conservative operator in momentum equations}
\label{a:Hp}
We show here the process that has led to the definition of the operator $ H_P(\aph\tnn, \ppi\tn)_k \approx \int_{\zeta_k} \ppi\tn \pder{\aph\tnn}{x} \mathrm{d}x $ in the momentum equations.

Assuming  $(\uph)_k = (\uph)\kpo = (\uph)\kmo = u$ and $(\Pph)_i = (\Pph)\ipo = P$
and taking into account that that $(\amph)_k\tns = (\arhoph)_k\tns (\uph)_k\tns$, the momentum equation~\eqref{ens:m} reads
\begin{equation} \begin{split}\label{ens:nc1}
\frac{\volz{k}}{\Delta t} \bigg[(\arhoph)_k\tns (u)_k\tns - (\arhoph)_k\tn (u)_k\tn \bigg]
 = &- (u)\tns \left[ F\Rus\kph\left( \arhoph\tns, u\tn \right) - F\Rus\kmh\left( \arhoph\tns, u\tn \right)\right]\\
   &- (P)\tn \left[(\aph)\ipo\tnn - (\aph)_i\tnn \right] + H_P(\aph\tnn, \ppi\tn)_k \,.
\end{split} \end{equation}
We recall the mapping from the primary to the staggered (see Remark~\ref{rmk:mapping}):  $(\arhoph)_k = 0.5\left[ (\arhoph)_i + (\arhoph)\ipo \right]$. Therefore, the Rusanov fluxes in Eq.~\eqref{ens:nc1} are
\begin{align*}\begin{split}
F\Rus\kph
  &=  \frac{1}{2} u\tn \left[ (\arhoph)\kpo\tns + (\arhoph)_k\tns \right]
    - \frac{1}{2} \vert u\tn \vert  \left[ (\arhoph)\kpo\tns -  (\arhoph)_k\tns  \right]   \\
  &=  \frac{1}{2} u\tn \left[ \frac{(\arhoph)_{i+2}\tns +(\arhoph)\ipo\tns}{2}  
                            + \frac{(\arhoph)\ipo\tns +(\arhoph)_i\tns}{2} \right]\\
    &- \frac{1}{2} \big\vert u\tn \big\vert  \left[ \frac{(\arhoph)_{i+2}\tns +(\arhoph)\ipo\tns}{2} - \frac{(\arhoph)\ipo\tns +(\arhoph)_i\tns}{2} \right]  \\
  &= \frac{1}{2} \left[F\Rus_{i+\frac{3}{2}}(\arhoph, u) + F\Rus\iph(\arhoph, u)\right]\,; \end{split}\\
F\Rus\kmh
  &=\frac{1}{2} \left[F\Rus\iph(\arhoph, u) + F\Rus\imh(\arhoph, u)\right] \,,
\end{align*}
where  $F\Rus\imh(\arhoph, u)$, $F\Rus\iph(\arhoph, u)$ and  $F\Rus_{i+\frac{3}{2}}(\arhoph, u)$ are the density flux at primary cell interfaces, as defined in Eq.~\eqref{ens:rus_ad}.

Now, we substitute the previous flux forms in the right hand side of Eq.~\eqref{ens:nc1} and the expressions for $(\arhoph)_k\tnn$ and $(\arhoph)_k\tn$ in the left hand side, obtaining
\begin{multline} \label{ens:nc2}
\frac{\volz{k}}{\Delta t} \frac{(\arhoph)_i\tns + (\arhoph)\ipo\tns}{2} u\tns
 - \frac{\volz{k}}{\Delta t}\frac{(\arhoph)_i\tn + (\arhoph)\ipo\tn}{2}  u\tn \\
 = - \frac{1}{2} (u)\tns \left[F\Rus_{i+\frac{3}{2}}(\arhoph\tns, u\tn) + F\Rus\iph(\arhoph\tns, u\tn)
  - F\Rus\iph(\arhoph\tns, u\tn) - F\Rus\imh(\arhoph\tns, u\tn)\right]\\
   - (P)\tn \left[(\aph)\ipo\tnn - (\aph)_i\tnn \right] + H_P(\aph\tnn, \ppi\tn)_k \,.
\end{multline}
Finally, we substitute in the left hand side also the expressions for $(\arhoph)_i\tns$ and $(\arhoph)\ipo\tns$ given by Eq.~\eqref{ens:ad_uunif}).
Considering an internal cell of a uniform grid\footnote{
Considering only an internal cell is motivated by the different computation of momentum on the boundary cells, which does not require the solution of Eqs.~\eqref{ens:m} and \eqref{ens:dm}, as explained in Sec.~\ref{s:boundarycondition}.
}, for which $\volz{k}=\volc{i}=\volc{i+1}=\Delta x$, we have
\begin{equation} \label{ens:nc3} \begin{split}
\frac{\volz{k}}{\Delta t}\frac{1}{2} \bigg[
 &  \left[ (\arhoph)_i\tn  + (\arhoph)\ipo\tn \right] u\tns 
   -\left[ (\arhoph)_i\tn + (\arhoph)\ipo\tn \right] u\tn \bigg] \\
 & -\frac{1}{2} u\tns \left[
    F\Rus\iph\left( \arhoph\tns, u\tn)\right) - F\Rus\imh\left(\arhoph\tns, u\tn)\right)
  + F\Rus_{i+\frac{3}{2}}\left( \arhoph\tns, u\tn)\right) - F\Rus\iph\left(\arhoph\tns, u\tn)\right) \right]\\
=& -\frac{1}{2} u\tns \left[F\Rus_{i+\frac{3}{2}}(\arhoph\tns, u\tn) + F\Rus\iph(\arhoph\tns, u\tn)
  - F\Rus\iph(\arhoph\tns, u\tn) - F\Rus\imh(\arhoph\tns, u\tn)\right]\\
 &- (P)\tn \left[(\aph)\ipo\tnn - (\aph)_i\tnn \right] + H_P(\aph\tnn, \ppi\tn)_k \,.
\end{split}\end{equation}
All the flux terms cancel out, so, to ensure that no velocity variations arise, it is requires that
\[ -(P)\tn \left[(\aph)\ipo\tnn - (\aph)_i\tnn \right] + H_P(\aph\tnn, \ppi\tn)_k =0 \,.\]
In light of this, we define the non-conservative operator $H_P$ as
\[ H_P(\aph\tnn, \ppi\tn)_k = (\ppi)_k\tn \left[(\aph)\ipo\tnn - (\aph)_i\tnn \right]\]
where $(\ppi)_k\tn = \frac{1}{2}\left[ (\ppi)_i\tn + (\ppi)\ipo\tn \right]$ is the interface pressure mapped at the staggered cell $\zeta{k}$.

The same reasoning applies to the equation of the momentum update~\eqref{ens:dm}.  In this case, we obtain intermediate equations very similar to Eqs.~\eqref{ens:nc1}, \eqref{ens:nc2}, \eqref{ens:nc3}, but with the term $\delta u\tnn$ instead of $u\tn$. The constraint to have $\delta u\tnn = 0$ is 
\[ -(\delta P)\tnn \left[(\aph)\ipo\tnn - (\aph)_i\tnn \right] + H_P(\aph\tnn, \delta \ppi\tnn)_k =0 \,.\]
so, we define 
\[ H_P(\aph\tnn, \delta \ppi\tnn)_k = \left[(\ppi )_k\tnn - (\ppi)_k\tn \right] \left[(\aph)\ipo\tnn - (\aph)_i\tnn \right] \,,\]
which, clearly, ensures also that, if $P\tnn=P\tn$, $H_P(\aph, \delta \ppi) =0 $.
Indeed, if we have a uniform pressure and velocity field, the equations in the correction step, that is the pressure and the update momentum ones, should be identically null.

\end{document}